\documentclass{article}

\usepackage[main, preprint]{neurips_2026}

\usepackage[utf8]{inputenc} 
\usepackage[T1]{fontenc}    
\usepackage{hyperref}       
\usepackage{url}            
\usepackage{booktabs}       
\usepackage{amsfonts}       
\usepackage{nicefrac}       
\usepackage{microtype}      
\usepackage{xcolor}         
\usepackage{tikz}
\usepackage{import}
\usepackage{bm}
\usepackage{amsmath}
\usepackage{comment}
\usepackage{gensymb}

\usepackage{titlesec}
\titlespacing{\section}{0pt}{6pt}{4pt}
\titlespacing{\subsection}{0pt}{4pt}{2pt}

\DeclareMathOperator*{\argmin}{argmin}

\title{Posterior Geometry and Identifiability in Multi-Response Bayesian Calibration}

\author{%
 Anton van Beek\\
  University College Dublin\\ Dublin, Belfield 4 \\
  \texttt{anton.vanbeek@ucd.ie} \\
  \And
  Adam. M Boyce \\
  University College Dublin\\ Dublin, Belfield 4 \\
  \texttt{adam.boyce@ucd.ie} \\
  \AND
  Will J. Dawson \\
  University College London\\ Gower Street, London \\
  \texttt{will.dawson.20@ucl.ac.uk} \\
  \And
  James B. Robinson \\
  University College London\\ Gower Street, London \\
  \texttt{j.b.robinson@ucl.ac.uk}\\
}

\begin{document}  
\tikzset{
  mynode/.style={execute at begin node=\setlength{\baselineskip}{-1mm}
  }
}

\maketitle

\begin{abstract}
Calibration under model misspecification is inherently ill-posed because calibration parameters and structural discrepancy are statistically confounded without additional assumptions. Bayesian formulations address this ambiguity through prior and covariance modeling choices, including multi-response observations and cross-source alignment. However, how these choices, together with output-dependent uncertainty, shape posterior geometry and practical parameter identifiability remains poorly understood. We present a latent-variable multi-response calibration model that combines empirical Bayes estimation of nuisance hyperparameters with a Fisher-information-based local Gaussian approximation of the calibration-parameter posterior. Closed-form expressions are derived for squared-exponential and Mat\'ern covariance functions, revealing how multi-response correlations, latent alignment, and output-specific uncertainty influence local posterior geometry. Controlled numerical experiments show that additional responses provide complementary geometric constraints on calibration parameters, while noise modeling can improve uncertainty quantification by reducing overconfidence without necessarily improving parameter localization. A lithium-ion battery study further illustrates how the model distinguishes strongly from weakly constrained parameters. Together, these results establish expected-Fisher posterior geometry as a scalable diagnostic of local parameter identifiability in multi-response calibration.

\end{abstract}
\section{Introduction}
\label{introduction}
Simulation models play a central role in scientific discovery and engineering, enabling researchers to study systems that are difficult or impossible to probe through physical experiments \cite{koermer2024}. Before such models can inform downstream decisions, they must typically be calibrated to physical observations by tuning modeling parameters that can be physically meaningful but difficult to measure directly \cite{wei2022,higdon2004}. In practice, however, simulation models are almost invariably misspecified due to incomplete physics, numerical approximations, or simplifying assumptions \cite{kennedy2001}. Under model misspecification, calibration requires the joint inference of model parameters and a discrepancy/bias stemming from model misspecification. This joint inference problem is fundamentally ill-posed: without additional structure, calibration parameters and discrepancy are not statistically identifiable from observational data alone. As shown by Tuo and Wu \cite{tuo2015}, procedures that prioritize predictive accuracy can compromise the physical interpretability of inferred parameters, highlighting the need for principled methods that jointly address model discrepancy and parameter estimation. A principled statistical formulation of this problem was introduced in the Bayesian calibration method of Kennedy and O’Hagan \cite{kennedy2001}, which provides a probabilistic decomposition of model output, bias, and observational noise.

The probabilistic decomposition of the calibration problem in a Bayesian formulation makes the ill-posedness of joint parameter and discrepancy inference explicit and addresses it through prior assumptions on calibration parameters and structural bias. Extensions such as Co-Kriging jointly model multiple experimental fidelities through cross-covariance structures \cite{LeGratiet2014}, increasing representational flexibility but substantially expanding the hyperparameter space, which can weaken identifiability and degrade numerical stability. Related approaches improve computational tractability through adaptive sampling \cite{oliveira2024}, low-rank covariance approximations \cite{spantini2015}, or modular calibration strategies \cite{bayarri2007}, though these approximations may themselves affect posterior inference and identifiability \cite{yousefpour2024}. More recently, latent space models have been proposed to represent multiple data sources through shared low-dimensional embeddings \cite{comlek2025}, avoiding explicit fidelity hierarchies while improving empirical performance, yet the statistical mechanisms by which latent representations influence calibration, discrepancy modeling, and posterior curvature remain poorly characterized.

Identifiability can also improve when calibration is performed jointly across multiple outputs, since multi-response observations impose cross-output consistency constraints that reduce confounding between parameter effects and structural discrepancy \cite{arendt2012}. Recent work suggests that latent representations can further strengthen these constraints \cite{yousefpour2024}, but it remains unclear whether such improvements arise from reduced model flexibility, implicit geometric regularization, or constraints induced by shared embeddings across data sources.

At present, there is no clear characterization of how multi-output structure, latent embeddings, and output-specific noise collectively determine the information available for calibration parameter inference. Without such understanding, improvements in predictive performance cannot be reliably translated into guarantees of parameter identifiability or physical interpretability.

We present a latent-variable calibration and bias-correction model that provides a computationally tractable characterization of local posterior geometry in multi-response Bayesian calibration. The proposed model incorporates output-specific experimental uncertainty and latent representations of multiple data sources, allowing cross-output correlations and measurement uncertainty to influence posterior curvature. Inference is performed using an empirical Bayes approximation, in which nuisance hyperparameters, including the covariance, latent embedding, and nugget parameters, are estimated by maximum likelihood. Conditional on these estimates, closed-form expressions for the expected Fisher information under squared-exponential and Mat\'ern covariance functions yield a local Gaussian approximation of the calibration-parameter posterior. Through controlled numerical experiments, we demonstrate how latent alignment, multiple responses, and output-dependent uncertainty reshape local posterior geometry and parameter uncertainty. Finally, a lithium-ion battery calibration study illustrates how the proposed framework distinguishes strongly and weakly constrained calibration parameters while accounting for model discrepancy.

\section{Practical Calibration and Bias Correction of Multi-Response Simulations}
\label{problem_form}


\subsection{Model Formulation and Assumptions}
\label{modeling_settings}
We consider a setting with two data sources: a set of \(d_s\) simulation models
\begin{equation}
f^{(s)}_i:\chi\times \Theta\rightarrow \mathbb{R},\quad i=1,\ldots,d_s,
\end{equation}
parametrized by design variables \(\mathbf{x}\in\chi\subset\mathbb{R}^d\) and calibration parameters \(\bm{\theta}\in\Theta\subset\mathbb{R}^p\). The calibration parameters represent physically meaningful but unobservable quantities that govern the system’s behavior. In contrast, the second data source is a set of \(d_e\) physical experiments
\begin{equation}
f^{(e)}_j:\chi \rightarrow \mathbb{R},\quad j = 1,\ldots,d_e,
\end{equation}
producing outputs without explicit calibration inputs. To correlate these data sources, we introduce calibration parameters \(\bm{\theta}_e \in \mathbb{R}^p\), which are the same for each simulation model \( f_j^{(e)}(\cdot)\). These latent parameters account for unknown but shared physical conditions influencing experimental outputs. 

Each output is indexed by \(\eta \in \mathcal{H} \subset \mathbb{N}\) to identify its source that may vary in predictive fidelity and in the physical properties they represent. We denote by \(\mathcal{S}_s\), \(\mathcal{S}_e\) the index sets corresponding to simulation and experimental outputs, respectively. As the calibration parameters of the physical experiments are unobservable, we introduce a vector of parameters for observations \(k = 1,\ldots n\) as
\begin{equation}
    \tilde{\bm{\theta}}_k = \begin{cases}
        \bm{\theta}_k, & \eta_k \in \mathcal{S}_{s}, \\
        \bm{\theta}_e, & \eta_k \in \mathcal{S}_{e}.
    \end{cases}
    \label{theta_con}
\end{equation}
Thus, a training data set with \(n\) samples, \(D_n = \cup_{k=1}^n \{\mathbf{X}_k, Y_k\}\) consists of inputs \(\mathbf{X}_k = \left\{\mathbf{x}_k^T, \tilde{\bm{\theta}}_k^T, \eta_k\right\}^T\) and corresponding outputs \(Y_k, \quad k = 1,\ldots,n\). It should be noted that \(\eta_k\) is a categorical index for which embedding strategies will be discussed.

Next, we need to jointly model the outputs from all data sources as a Gaussian process (GP). Specifically, we assume that the observations \(\mathbf{Y}\) at inputs \(\mathbf{X}\) follow a multivariate normal distribution
\begin{equation}
    \mathbf{Y}\sim\mathcal{N}\left(\mathcal{M}\left(\mathbf{X}\right), \text{cov}\left(\mathbf{X},\mathbf{X}^{\prime} \right) \right),
    \label{priorGP}
\end{equation}
where \(\mathcal{M}\left(\cdot\right)\) is the prior mean function and \(\mathrm{cov}(\cdot,\cdot)\) is the covariance function. The latent calibration parameters \(\bm{\theta}_e\) enter this model through their influence on the covariance, capturing uncertainty and correlations across different data sources and outputs. They modulate the similarity between inputs by influencing the covariance function, allowing the model to capture correlations induced by shared calibration effects across outputs.

The prior mean function \(\mathcal{M}(\mathbf{X})\) can be flexibly specified to incorporate prior knowledge. Common choices include polynomial expansions and feed forward neural networks \cite{yousefpour2024}. While neural network-based prior mean functions can capture complex trends, they often reduce interpretability and complicate inference. Here, we adopt a linear basis expansion of the form \(\mathcal{M}(\mathbf{X}; \bm{\beta}) = \mathbf{M} \bm{\beta}\) where \(\mathbf{M}\) is an \(n \times \rho\) matrix of basis functions evaluated at the inputs, with the \(i^{th}\) row given as \(\mathbf{M}_k = \left\{ m_1(\mathbf{X}_k), \ldots, m_\rho(\mathbf{X}_k) \right\}\), and \(\bm{\beta} \in \mathbb{R}^\rho\) are associated weights to be estimated.

The covariance is given as \(\text{cov}\left(\cdot,\cdot \right) = \sigma^2 r(\cdot,\cdot)\) where \(\sigma^2\) denotes the prior variance and \(r\left(\cdot,\cdot\right)\) a correlation function. When the design space consists solely of continuous variables, the correlation function \(r:\chi\times\chi\rightarrow \mathbb{R}\) encodes the inductive bias that nearby inputs produce similar outputs, with correlations decaying as inputs become more distant. While many alternative correlation functions exist \cite{xu2020,Plumlee2016}, common choices include the stationary Squared-exponential and Mat\'ern correlations, given respectively by
\begin{align}
    r\left(\mathbf{x},\mathbf{x}^{\prime} ; \bm{\omega}\right) & =  \exp\left( - \left( \mathbf{x} - \mathbf{x}^{\prime}\right)^T\bm{\Omega}_{\chi}\left( \mathbf{x} - \mathbf{x}^{\prime}\right) \right),\label{cont_gauss}\\
    r\left(\mathbf{x},\mathbf{x}^{\prime} ; \bm{\omega}\right) & =\frac{2^{1-\nu}}{\Gamma(\nu)}\left(\sqrt{2\nu\left( \mathbf{x} - \mathbf{x}^{\prime}\right)^T\bm{\Omega}_{\chi}\left( \mathbf{x} - \mathbf{x}^{\prime}\right)} \right)^{\nu}\mathcal{K}_{\nu}\left(\sqrt{2\nu\left( \mathbf{x} - \mathbf{x}^{\prime}\right)^T\bm{\Omega}_{\chi}\left( \mathbf{x} - \mathbf{x}^{\prime}\right)} \right),
    \label{cont_matern}
\end{align}
Here \(\mathbf{\Omega}_{\chi}=\mathrm{diag}(\left\{10^{\omega_1},\ldots,10^{\omega_d}\right\})\) parameterizes anisotropic inverse length scales with \(\omega_i\in\mathbb{R}\) and \(\bm{\omega}=\left\{ \omega_1,\ldots,\omega_d\right\}^T\). We use log-scale parameters to improve numerical stability. The smoothness parameter satisfies \(\nu>0\); in our experiments we restrict \(\nu\in\{1/2,3/2,5/2\}\) to obtain closed-form expressions with stable gradients. Finally, \(\Gamma(\cdot)\) denotes the Gamma function and \(\mathcal{K}_{\nu}(\cdot)\) the modified Bessel function of the second kind.

When the input space additionally contains categorical variables \( \eta\) , we introduce a mapping \(\mathbf{z}:\mathcal{H}\rightarrow\mathbb{R}^{d_z}\) that embeds each categorical source indicator as an input into a continuous latent space (i.e., \(\mathbf{z}(\eta)=\left\{z_1(\eta),\ldots,z_{d_z}(\eta)\right\}\)). While these embeddings could be learned directly, prior work has shown that a linear mapping of the form \(\mathbf{z}(\eta)=\mathbf{g}(\eta)\mathbf{A}\) achieves strong performance \cite{oune2021}. Here, \(q=d_s+d_e\) is the total number of data sources, \(\mathbf{A}\in \mathbb{R}^{q\times d_z}\) is a matrix of learnable hyperparameters, and \(\mathbf{g}(\eta)=\left\{g_1(\eta),\ldots,g_{d_z}(\eta) \right\}\) is a fixed vector representation of the categorical input. In this work, we adopt a two-dimensional latent space (\(d_z=2\)), which has been shown to provide adequate predictive performance in non-calibration settings \cite{zhang2020} and for \(\mathbf{g}(\cdot)\) we use one-hot encoding. Let \(\mathbf{x}_0 = \left\{\mathbf{x}^T, \bm{\theta}^T,\eta \right\}^T\) denote a generic input. The correlation functions in Eqs.~\ref{cont_gauss} and~\ref{cont_matern} are then extended to
\begin{align}
    r\left(\mathbf{x}_0,\mathbf{x}_0^{\prime} ; \bm{\omega}\right)  = &  \exp\left( - \Delta\left(\mathbf{x}_0,\mathbf{x}_0^{\prime}\right)\right),\label{Gauss_corr2} \\
    r\left(\mathbf{x}_0,\mathbf{x}_0^{\prime}; \bm{\omega}\right) =\ &\dfrac{2^{1-\nu}}{\Gamma(\nu)}\left(\sqrt{2\nu\Delta\left(\mathbf{x}_0,\mathbf{x}_0^{\prime}\right)} \right)^{\nu}\mathcal{K}_{\nu}\left(\sqrt{2\nu\Delta\left(\mathbf{x}_0,\mathbf{x}_0^{\prime}\right)} \right), \label{Matern_corr2}   
\end{align}
where \(\Delta\left(\mathbf{x}_0,\mathbf{x}_0^{\prime}\right) =(\mathbf{x}-\mathbf{x}^{\prime})^T\mathbf{\Omega}_{\chi}(\mathbf{x}-\mathbf{x}^{\prime}) +(\bm{\theta}-\bm{\theta}^{\prime})^T\bm{\Omega}_{\Theta}(\bm{\theta}-\bm{\theta}^{\prime})+\left\|\mathbf{z}(\eta)-\mathbf{z}(\eta^{\prime})\right\|_2^2\) and \(\mathbf{\Omega}_{\Theta}=\mathrm{diag}(\left\{10^{\omega_{d+1}},\ldots,10^{\omega_{d+p}}\right\})\) parameterizes inverse length scales associated with the calibration parameters. No additional length-scale parameters are introduced for the latent variables, as relative distances in the latent space are directly controlled by the matrix \(\mathbf{A}\). This additive construction preserves positive definiteness of the resulting covariance matrix. Additive covariance structures across design variables, calibration parameters, and categorical embeddings are widely used in GPs \cite{williams2006, duvenaud2011}. Such decompositions provide a favorable trade-off between expressivity, interpretability, and statistical robustness.

\subsection{Modeling Output Dependent Uncertainty}
\label{output_noise}

Experimental observations often exhibit uncertainty arising from measurement error, unobserved inputs, or intrinsic variability in the physical process. For example, physical experiments may be corrupted by sensor noise, while certain simulation models exhibit stochastic variability due to random initial conditions \cite{vanbeek2020c} (e.g., molecular dynamics or agent-based simulations). Such uncertainty can be modeled through additive Gaussian noise, where each observation may be associated with an output-dependent noise level (i.e., \(\varepsilon_i \sim \mathcal{N}(0,\tau_{\eta_i}^2)\) with \(\eta_i\in\left\{1,\ldots,d_e+d_s \right\}\) denoting the output or data source corresponding to observation \(i\)). For clarity, we first describe the homoscedastic case, and later extend the formulation to allow output-specific noise levels. In GP models, observation uncertainty is commonly incorporated via a nugget (or jitter) term added to the correlation matrix, yielding \(\mathbf{R}_{\delta} = \mathbf{R} + 10^{\lambda}\mathbf{I}\) where \(R_{ij} = r(\mathbf{X}_i,\mathbf{X}_j)\), \(\lambda\) is a log-scale nugget hyperparameter, and \(\mathbf{I}\) is an \(n \times n\) identity matrix. This formulation accounts for observation noise while also improving numerical conditioning during likelihood-based hyperparameter estimation (see Sec.~\ref{parm_est}).

While a single nugget parameter is appropriate for single-output GP regression, this assumption is restrictive when experimental variance varies by data source. To accommodate source-dependent noise, we model observation uncertainty as \(\mathbf{R}_{\delta}=\mathbf{R}+\bm{\pi}\) where \(\bm{\pi}\) is an \(n \times n\) diagonal matrix with entries \(\pi(\eta_i)\) defined by a mapping \(\pi:\mathcal{H} \rightarrow \mathbb{R}\), where \(\eta_i\) indexes the data source of observation \(i\). Assuming deterministic simulations, we focus on modeling noise in physical experiments.

One approach is to use a common variance for all physical experiments,
\begin{equation}
\pi(\eta;\lambda)=\begin{cases}
    0, & \text{if } \eta\in\mathcal{S}_s,\\
    10^\lambda, &\text{if } \eta\in\mathcal{S}_e.
    \label{single_nug}
\end{cases}
\end{equation}
This introduces a single nugget hyperparameter, \(\lambda\) which can either be learned through statistical inference or chosen as the smallest value that ensures numerical stability. In the latter case, we set \( \lambda = \xi_{ms} - \xi_R\) where \(\xi_R\) is the smallest eigenvalue of the correlation matrix and \( \xi_{ms}\) is the minimum eigenvalue required for invertibility (generally taken as the machine precision, \( \sim10^{-8}\)). A second approach is to allow output-specific variances,
\begin{equation}
\pi(\eta;\bm{\lambda})=\begin{cases}
    0, & \text{if } \eta\in\mathcal{S}_s,\\
    10^{\lambda_1}, &\text{if } \eta\in\mathcal{S}_{1,e}\\
    \vdots&\vdots\\
    10^{\lambda_{d_e}},&\text{if } \eta\in\mathcal{S}_{d_e,e},
    \label{multi_nug}
\end{cases}
\end{equation}
where \(\mathcal{S}_{i,e}\) denotes the set of observations corresponding to the \(i^{th}\) physical experiment. This vector-valued nugget \(\boldsymbol{\lambda}\) provides model flexibility to capture output dependent experimental uncertainty. Introducing multiple nugget hyperparameters can increase model complexity and pose identifiability challenges, especially with limited data per output. Regularization or hierarchical priors on \(\boldsymbol{\lambda}\) can help mitigate over fitting and improve stable estimation (see Sec.~5.2 of \cite{gelman2013}).

\subsection{Training and Prediction}
\label{train_predict}
The emulator introduced in Secs.~\ref{modeling_settings}, and \ref{output_noise} is governed by a set of hyperparameters \(\bm{\omega}\),\(\bm{\beta}\),\(\sigma^2\),\(\mathbf{A}\),\(\bm{\lambda}\), and \(\bm{\theta}_e\) which must be inferred from the observed training data \(D_n\). The basis function weights \(\bm{\beta}\) and prior variance \(\sigma^2\) are treated as nuisance parameters and profiled out via closed-form conditional maximum likelihood estimates (MLE) given \(\bm{\omega}\), \(\mathbf{A}\), \(\bm{\lambda}\) and \(\bm{\theta}_e\), thereby reducing the effective hyperparameter dimensionality.

Rather than relying solely on deterministic resampling techniques such as leave-one-out cross-validation, we adopt a Bayesian inference formulation that enables incorporation of domain-informed priors. This approach regularizes parameter estimation, helping to avoid physically unrealistic values that may otherwise arise in purely deterministic or unconstrained optimization \cite{williams2006}. The posterior distribution over hyperparameters is given by
\begin{equation}
    P\left(\bm{\omega},\mathbf{A},\bm{\lambda},\bm{\theta}_e\mid D_n\right) \propto P\left(D_n \mid \bm{\omega},\mathbf{A},\bm{\lambda},\bm{\theta}_e\right)P\left(\bm{\omega},\mathbf{A},\bm{\lambda},\bm{\theta}_e\right),
    \label{post_calparm}
\end{equation}
where \(P\left(\bm{\omega},\mathbf{A},\bm{\lambda},\bm{\theta}_e\right)\) denotes the prior distribution of hyperparameters with a likelihood given as
\begin{equation}
    P\left(D_n \mid \bm{\omega},\mathbf{A},\bm{\lambda},\bm{\theta}_e\right) = (2\pi\sigma^2)^{-\frac{n}{2}}| \mathbf{R}_{\delta} |^{-\frac{1}{2}}\exp\left( - \dfrac{\left(\mathbf{Y}-\mathbf{M}\bm{\beta}\right)^T \mathbf{R}_{\delta}^{-1}\left(\mathbf{Y}-\mathbf{M}\bm{\beta}\right)}{2\sigma^2} \right).
    \label{llh}
\end{equation}
Note that the uncertainty from observations are modeled by the nugget parameters \(\bm{\lambda}\) through the correlation matrix  \(\mathbf{R}_{\delta} = \mathbf{R} + \bm{\pi}\), consistent with the approach of Kennedy and O’Hagan \cite{kennedy2001}.

Once hyperparameters are inferred, the emulator provides a posterior predictive distribution at any new input \(\mathbf{x}_0\) as \(\hat{f}\mid\mathbf{x}_0\sim\mathcal{N}\left(\mu(\mathbf{x}_0),s^2(\mathbf{x}_0)\right)\) in which the predictive mean and variance are given by
\begin{align}
    \mu(\mathbf{x}_0) &=\mathbf{m}^T(\mathbf{x}_0)\hat{\bm{\beta}} + \mathbf{r}(\mathbf{x}_0,\mathbf{X})\mathbf{R}_{\delta}^{-1}\left(\mathbf{Y} - \mathbf{M}\hat{\bm{\beta}}\right),\\
    s^2(\mathbf{x}_0) &= \hat{\sigma}^2\left(1-\mathbf{r}^T(\mathbf{x}_0,\mathbf{X})\mathbf{R}_{\delta}^{-1}\mathbf{r}(\mathbf{x}_0,\mathbf{X})+\mathbf{W}^T\left( \mathbf{M}^T \mathbf{R}_{\delta}^{-1}\mathbf{M}\right) \mathbf{W}\right),
\end{align}
where \(\mathbf{W} = \mathbf{m}(\mathbf{x}_0)-\mathbf{M}^T\mathbf{R}^{-1}_{\delta}\mathbf{r}(\mathbf{x}_0,\mathbf{X})\) and \(\mathbf{m}(\mathbf{x}_0)=\left\{m_1(\mathbf{x}_0),\ldots,m_{\rho}(\mathbf{x}_0) \right\}^T\) is a \(\rho\) dimensional vector of basis functions. 

The posterior predictive distribution enables estimation of the predictive bias between the \(i^{th}\) simulation and the \(j^{th}\) physical experiment at \(\mathbf{x}_0\), defined as \(\delta_{i,j}(\mathbf{x}_0) = \mathbb{E}\left( \hat{f}^{(e)}_j(\mathbf{x}_0) - \hat{f}^{(s)}_i(\mathbf{x}_0) \mid D_n \right)\). Note that such bias estimation is meaningful primarily when comparing outputs that represent the same system and physical variable. Given the complexity of joint modeling and the high dimensionality of the parameter space, the next section focuses on efficient estimation of calibration parameters using the expected Fisher information matrix to improve numerical stability and tractability.

\subsection{Identifiability Constraints and Fisher Information of Calibration Parameters}
\label{parm_est}

Estimating the hyperparameters \(\left\{\bm{\omega}, \mathbf{A}, \bm{\lambda}, \bm{\theta}_e\right\}\) can be challenging due to their high dimensionality and potential identifiability issues. To address these challenges, we impose structural constraints on the latent embedding matrix \(\mathbf{A}\). Specifically, we fix the first categorical level at the origin to remove translational invariance, and enforce a lower-triangular form on \(\mathbf{A}\) to eliminate rotational and reflectional symmetries, following \cite{zhang2020}. This reduces the effective number of free parameters in \(\mathbf{A}\), enhancing identifiability while preserving model expressiveness. This anchoring and triangular constraint using the triangular number \(T_{d_z}=\sum_{i=1}^{d_z}i\) reduce the number of free parameters in \(\mathbf{A}\) from \(q \times d_z\) to \(q \times d_z - T_{d_z}\), corresponding to the removal of symmetric degrees of freedom. Consequently, the total number of hyperparameters estimated in Eqn.~\ref{post_calparm} is \((d+p) + (q\times d_z - T_{d_z}) + d_e + p\), associated with \(\bm{\omega}\), \(\mathbf{A}\), \(\bm{\lambda}\), and \(\bm{\theta}_e\), respectively.

Exact Bayesian inference over all model hyperparameters is computationally demanding for the proposed latent-variable calibration model. We therefore employ an empirical Bayes approximation in which the nuisance hyperparameters are first estimated by maximum likelihood. Conditional on these estimates, we approximate the posterior distribution of the calibration parameters using a local Gaussian approximation derived from the expected Fisher information. Under these conditions, the calibration parameters satisfy
\begin{equation}
\bm{\theta}_e \mid D_n\; \dot{\sim} \;\mathcal{N}\left(\bm{\theta}_{{\mathrm{MLE}}},\bm{\mathcal{J}}^{-1} \right)
\end{equation}
where \(\bm{\mathcal{J}}\) is the expected Fisher information matrix evaluated at \(\bm{\theta}_{{\mathrm{MLE}}}\).

The MLE is obtained by minimizing the negative twice log-likelihood (Eqn.~\ref{llh}), resulting in
\begin{equation}
    \left\{ \bm{\omega}_{{\mathrm{MLE}}}, \mathbf{A}_{{\mathrm{MLE}}}, \bm{\lambda}_{{\mathrm{MLE}}}, \bm{\theta}_{{\mathrm{MLE}}} \right\} = \argmin_{\bm{\omega}\in\Omega, \mathbf{A}\in\mathcal{A},\bm{\lambda}\in\Lambda,\bm{\theta}_e\in\Theta} n\log (\hat{\sigma}^2)+\log\left(\left|\mathbf{R}_{\delta}\right|\right),
\end{equation}
where constant terms are omitted. In practice, we normalize inputs \(\mathbf{x},\bm{\theta} \) to the unit hypercube and standardize outputs. We recommend the search domains \(\mathcal{A}=[-2,2]^{q\times d_z-T_{d_z}}\), \(\Lambda=[-8,0]\) and \(\Theta=[0,1]^p\) which balance numerical stability, expressiveness, and computational efficiency.

Focusing on \(\bm{\theta}_e = \left\{\theta_1,\ldots,\theta_p \right\}^T \), we approximate the inverse covariance of the posterior by the expected Fisher information matrix \(\bm{\mathcal{J}}\in\mathbb{R}^{p\times p}\), treating all other hyperparameters as fixed at their MLE values. Its entries are
\begin{equation}
\bm{\mathcal{J}}_{ij} = \mathbb{E}\!\left( -\dfrac{\partial^2\log P\left(D_n \mid\bm{\omega},\mathbf{A},\bm{\lambda},\bm{\theta}_e\right)}{\partial \theta_i \partial \theta_j} \right)=\dfrac{1}{2}\text{tr}\!\left( \mathbf{R}_{\delta}^{-1} \dfrac{\partial \mathbf{R}}{\partial \theta_i} \mathbf{R}_{\delta}^{-1} \frac{\partial \mathbf{R}}{\partial \theta_j} \right),
\label{fish_gen}
\end{equation}
where the expectation is taken with respect to the GP likelihood and \(\mathrm{tr}(\cdot)\) denotes the matrix trace. The expected Fisher information is computed using the full covariance \( \hat{\sigma}^2\left(\mathbf{R}+ \bm{\pi}\right) \) where the additive nugget terms down weigh noisy outputs. We use the expected Fisher information rather than the observed Fisher because it admits a closed-form expression for GP models, does not depend on a specific realization of the data, and provides a smoother local metric for uncertainty quantification.

The expected Fisher information depends on the derivatives of the correlation matrix \(\mathbf{R}\) with respect to the calibration parameters. For the Squared-exponential covariance, the derivative of an individual correlation entry with respect to the \(k^{\text{th}}\) calibration parameter is given by
\begin{equation}
\frac{\partial r_{i,j}}{\partial \theta_k}
=
\begin{cases} 
    -2\times 10^{\omega_{d+k}}\left( \theta_{i,k}-\theta_{j,k}\right)r_{i,j}, &\text{if } i\in \mathcal{S}_k,j\notin\mathcal{S}_k,\\
    2\times 10^{\omega_{d+k}}\left( \theta_{i,k}-\theta_{j,k}\right)r_{i,j}, &\text{if } i\notin \mathcal{S}_k,j\in\mathcal{S}_k,\\
    0, &\text{otherwise},
    \end{cases} 
    \label{deriv_SE}
\end{equation}
and for the Mat\'ern covariance,
\begin{equation}
\frac{\partial r_{i,j}}{\partial \theta_k}
=
\begin{cases} 
    -2\times 10^{\omega_{d+k}}\left( \theta_{i,k}-\theta_{j,k}\right)\dfrac{\nu}{\kappa_{i,j}} \dfrac{2^{1-\nu}}{\Gamma(\nu)}\kappa_{i,j}^{\nu}K_{\nu-1}(\kappa_{i,j}), &\text{if } i\in \mathcal{S}_k,j\notin\mathcal{S}_k,\\
    2\times 10^{\omega_{d+k}}\left( \theta_{i,k}-\theta_{j,k}\right)\dfrac{\nu}{\kappa_{i,j}} \dfrac{2^{1-\nu}}{\Gamma(\nu)}\kappa_{i,j}^{\nu}K_{\nu-1}(\kappa_{i,j}), &\text{if } i\notin \mathcal{S}_k,j\in\mathcal{S}_k,\\
    0, &\text{otherwise},
    \end{cases}
\label{Fish_mat}
\end{equation}
where \(\kappa_{ij}=\sqrt{2\nu\Delta(\mathbf{X}_i,\mathbf{X}_j)}\) and \(r_{i,j} =  r(\mathbf{X}_i,\mathbf{X}_j)\).

The index set \(\mathcal{S}_k\) denotes observations for which correlation entries depend on the calibration parameter \(\theta_k\), and is distinct from the set of experimental observations \(\mathcal{S}_e\). This distinction allows different calibration parameters to influence different subsets of the experimental outputs. Notably, for the Mat\'ern covariance, the derivatives retain the antisymmetric structure seen in the Squared-exponential case, differing only by a smoothness-dependent scalar factor involving modified Bessel functions. Consequently, the expected Fisher information-based Gaussian approximation yields a computationally tractable and numerically stable method to estimate calibration parameters and quantify their uncertainty within the joint GP emulator. Additional details on the derivation of the expected Fisher information for the hyperparameters can be found in the Appx.~\ref{Fish_info}. 
\section{Empirical Evaluation of Local Posterior Geometry}
\label{results}
We evaluate how output-dependent experimental uncertainty and the number of observed responses affect calibration accuracy, posterior calibration, and local posterior geometry under a squared-exponential covariance function. A pedagogical illustration is provided in Appx.~\ref{peda_exam}.

\subsection{Experimental Uncertainties and Accuracy}
\label{accuracy_and_noise_sec}
A central question in calibration and bias-correction models is how heterogeneous experimental uncertainty across outputs affects parameter recovery, posterior uncertainty, and predictive accuracy. To study this, we consider numerical test Problems 2 and 3 (Tbl.~\ref{sample_fun_tbl}, Appx.~\ref{example_form}) and vary the standard deviation of the experimental uncertainty to \(\bar{\mathbf{\varepsilon}}_i \in \{0, 0.05, 0.1, 0.25\}^T\) times the range of the experimental outputs multiplied by \( \left\{ 0.8, 0.2, 1.5\right\}^T\) (i.e., different experimental variance for each output). The experiments use \(10\times (d+p)\) simulation samples and \(5\times(d+p)\) physical samples.

\begin{figure}[t]
\centering
\begin{tikzpicture}
\node[inner sep=0pt] (F) at (16mm,28mm){\includegraphics[width=35mm]{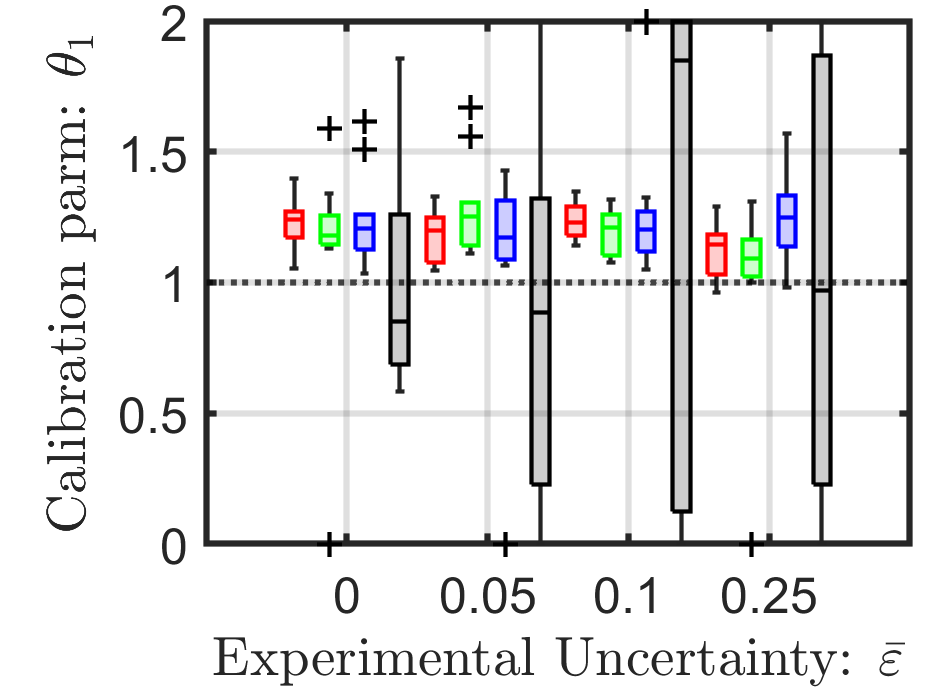}};
\node[inner sep=0pt] (F) at (50mm,28mm){\includegraphics[width=35mm]{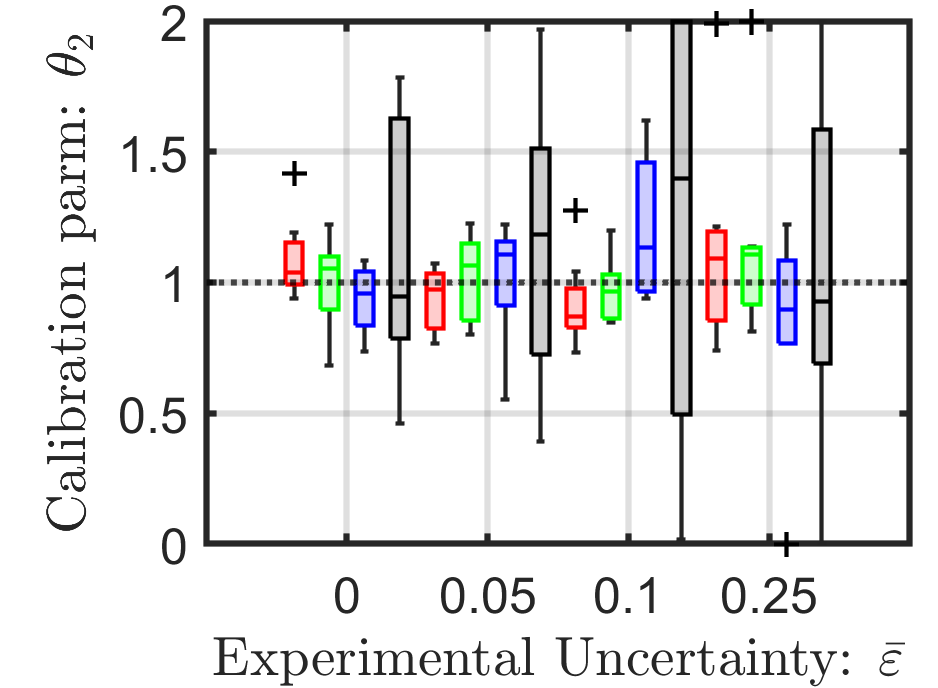}};
\node[inner sep=0pt] (F) at (84mm,28mm){\includegraphics[width=35mm]{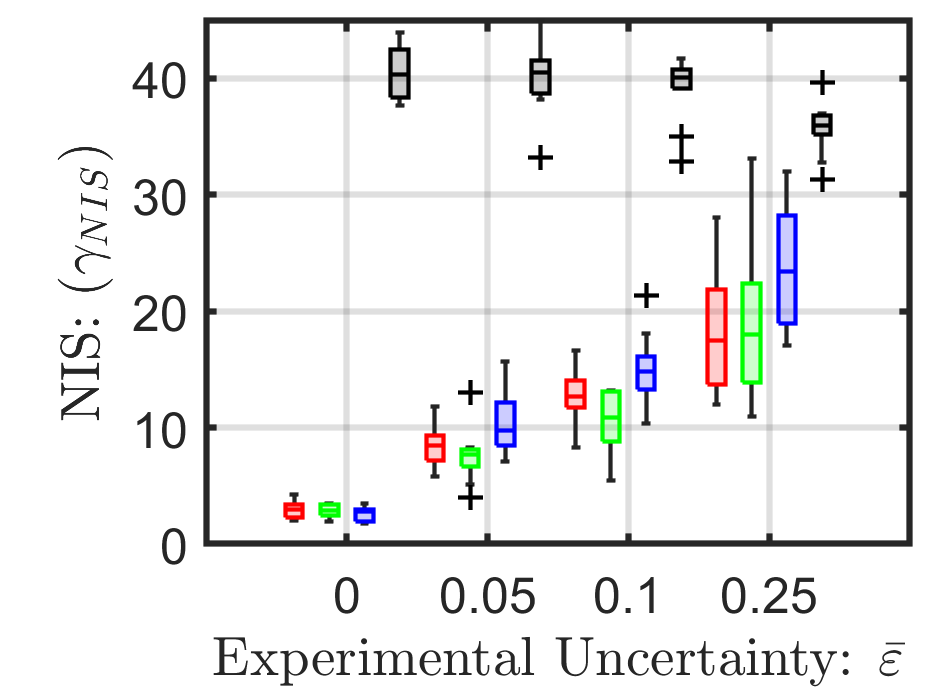}};
\node[inner sep=0pt] (F) at (121mm,30mm){\includegraphics[width=25mm]{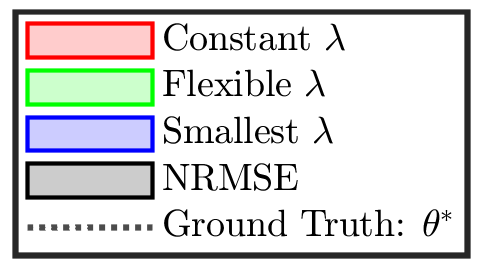}};

\node[inner sep=0pt] (F) at (16mm,0mm){\includegraphics[width=35mm]{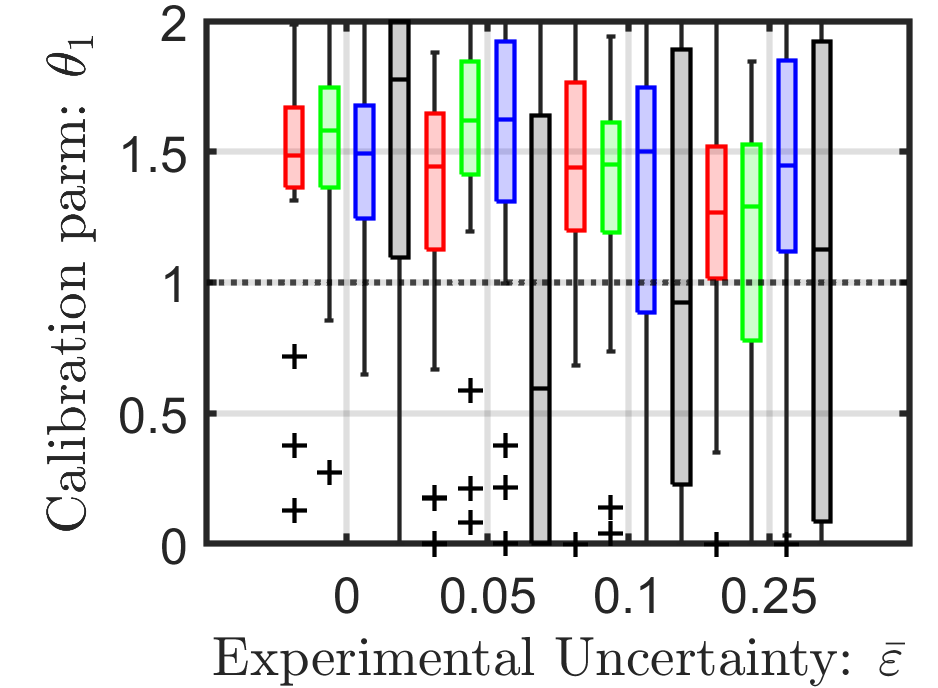}};
\node[inner sep=0pt] (F) at (50mm,0mm){\includegraphics[width=35mm]{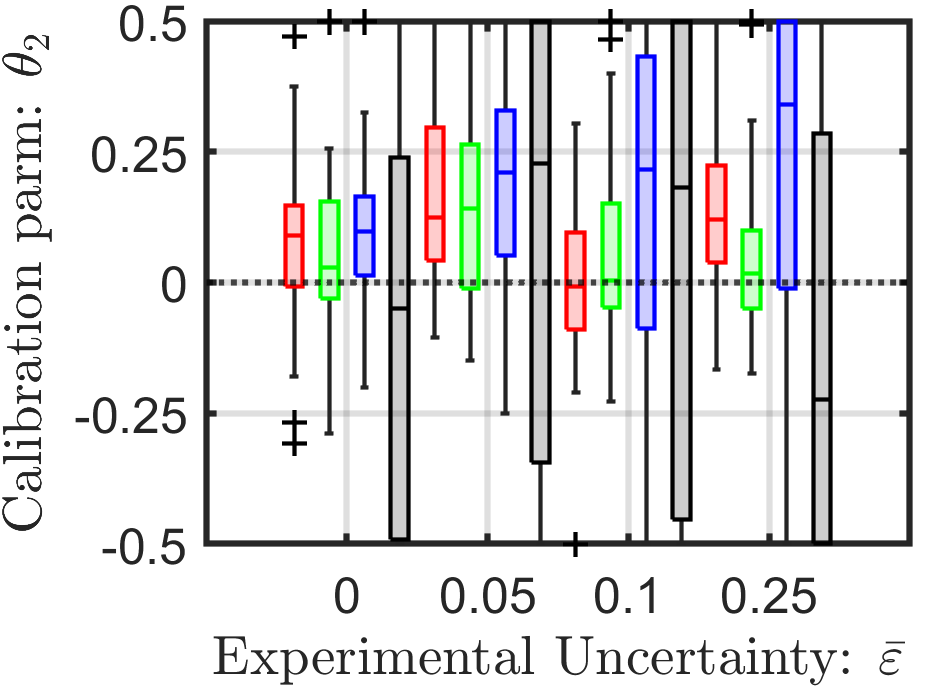}};
\node[inner sep=0pt] (F) at (84mm,0mm){\includegraphics[width=35mm]{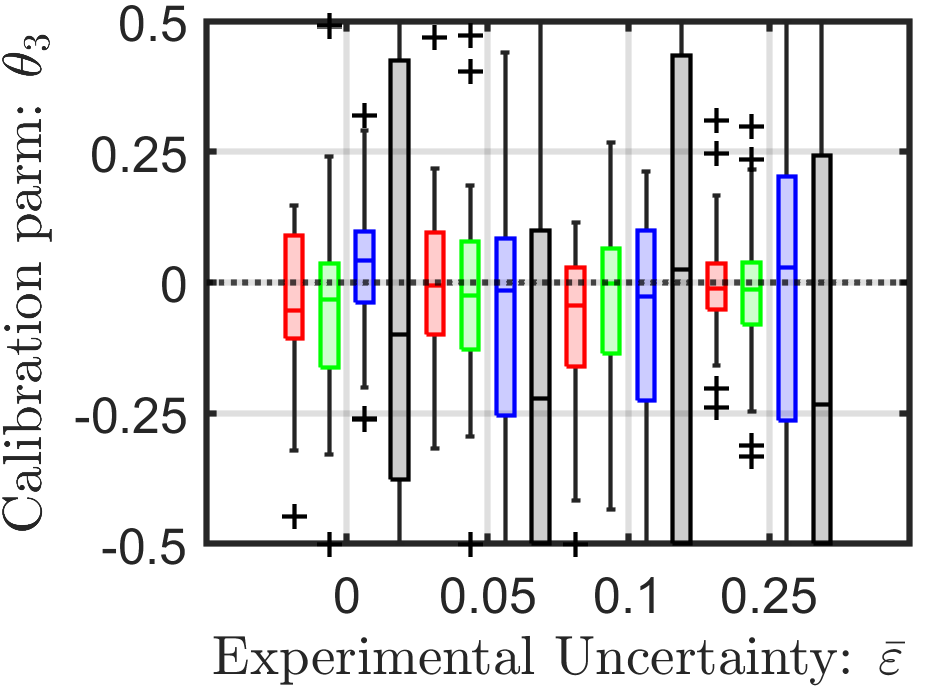}};
\node[inner sep=0pt] (F) at (118mm,0mm){\includegraphics[width=35mm]{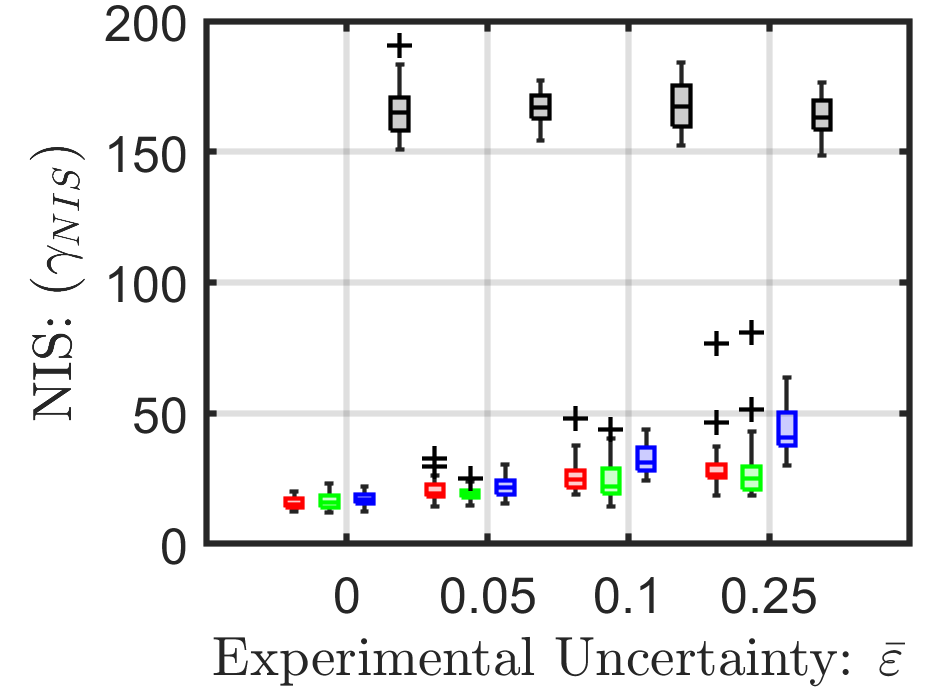}};

\draw[ line width=0.2mm, fill = white] (6mm,38mm) --++ (0,4mm) --++ (4mm,0mm) --++ (0mm,-4mm) -- cycle;
\node[mynode, anchor=center, align=center, text width = 4mm] at (6mm+2mm,38mm+2mm) {A};

\draw[ line width=0.2mm, fill = white] (40mm,38mm) --++ (0,4mm) --++ (4mm,0mm) --++ (0mm,-4mm) -- cycle;
\node[mynode, anchor=center, align=center, text width = 4mm] at (40mm+2mm,38mm+2mm) {B};

\draw[ line width=0.2mm, fill = white] (74mm,38mm) --++ (0,4mm) --++ (4mm,0mm) --++ (0mm,-4mm) -- cycle;
\node[mynode, anchor=center, align=center, text width = 4mm] at (74mm+2mm,38mm+2mm) {C};

\draw[ line width=0.2mm, fill = white] (6mm,10mm) --++ (0,4mm) --++ (4mm,0mm) --++ (0mm,-4mm) -- cycle;
\node[mynode, anchor=center, align=center, text width = 4mm] at (6mm+2mm,10mm+2mm) {D};

\draw[ line width=0.2mm, fill = white] (40mm,10mm) --++ (0,4mm) --++ (4mm,0mm) --++ (0mm,-4mm) -- cycle;
\node[mynode, anchor=center, align=center, text width = 4mm] at (40mm+2mm,10mm+2mm) {E};

\draw[ line width=0.2mm, fill = white] (74mm,10mm) --++ (0,4mm) --++ (4mm,0mm) --++ (0mm,-4mm) -- cycle;
\node[mynode, anchor=center, align=center, text width = 4mm] at (74mm+2mm,10mm+2mm) {F};

\draw[ line width=0.2mm, fill = white] (108mm,10mm) --++ (0,4mm) --++ (4mm,0mm) --++ (0mm,-4mm) -- cycle;
\node[mynode, anchor=center, align=center, text width = 4mm] at (108mm+2mm,10mm+2mm) {G};

\end{tikzpicture}
\caption{Calibration results across varying levels of output-dependent experimental uncertainty (25 runs; ‘+’ denotes outliers). A,B) parameter estimates and C) NIS for problem 2; D–F) parameter estimates and G) NIS for problem 3. Noise magnitude and treatment influence calibration accuracy.}
\label{exp_noise_results}
\end{figure}
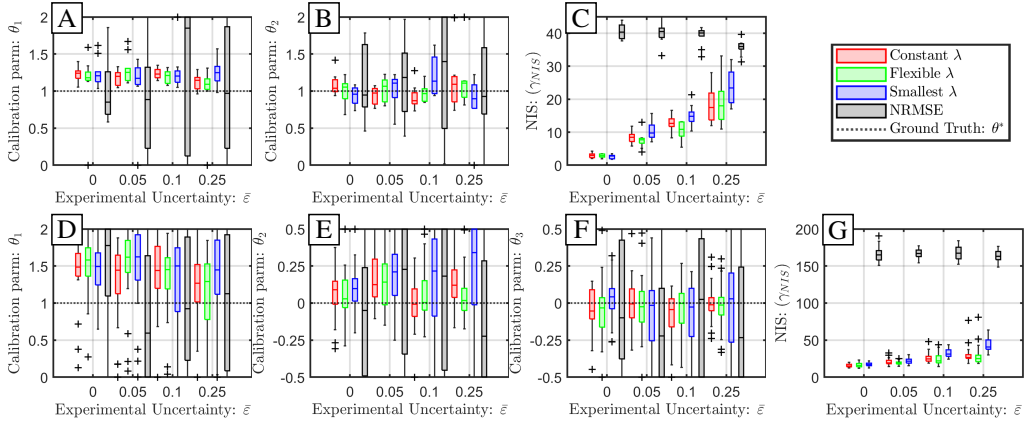

Panels A, B, D, E, and F in Fig.~\ref{exp_noise_results} show whisker plots of the calibration-parameter estimates across 25 repetitions for three treatments of experimental uncertainty: a homogeneous nugget (Const), an output-dependent nugget (Flex), and the smallest nugget ensuring positive definiteness (Min). Compared with calibration based on minimizing the normalized root mean squared error (NRMSE), the probabilistic calibration treatments generally produce estimates that are closer to the ground truth and less variable across repetitions. The Const and Flex treatments also remain comparatively insensitive to increasing experimental uncertainty. Conversely, the Min treatment becomes increasingly unstable as experimental uncertainty grows, behavior that is consistent with insufficient regularization of noise-induced variation in the local likelihood geometry.

Point estimates alone do not assess the calibration of the local Gaussian approximation. We therefore evaluate posterior calibration using the Wasserstein-1 distance \(\gamma_d\) between the empirical distribution of standardized calibration residuals and the standard normal distribution (Appx.~\ref{example_form}). Table~\ref{area_per_noise} shows that no nugget treatment provides uniformly superior calibration across parameters and noise levels. Increasing the experimental uncertainty sometimes reduces \(\gamma_d\), indicating improved agreement with the Gaussian approximation. This should not be interpreted as increased information or stronger parameter localization; rather, a larger nugget broadens overconfident posterior approximations and improves repeated-run uncertainty calibration. The effect is parameter dependent, suggesting anisotropic changes in local posterior geometry. NRMSE is excluded because it provides no posterior uncertainty distribution.

\begin{table}[t]

\centering

\setlength{\tabcolsep}{3pt}
\renewcommand{\arraystretch}{1}

\caption{Calibration metric \(\gamma_d\) for Problems 2 and 3 across experimental-noise levels and nugget treatments (lower is better). Bold indicates the lowest value for each parameter and noise level.}
\label{area_per_noise}

{\scriptsize
\begin{tabular}{ccccc ccccc ccccc ccccc}
\hline
& \multicolumn{7}{c}{\textbf{Problem 2}} && \multicolumn{11}{c}{\textbf{Problem 3}} \\

\cline{2-8} \cline{10-20}

& \multicolumn{3}{c}{\(\gamma_d(\cdot)\) for \(\theta_1\)} && \multicolumn{3}{c}{\(\gamma_d(\cdot)\) for \(\theta_2\)} && \multicolumn{3}{c}{\(\gamma_d(\cdot)\) for \(\theta_1\)} && \multicolumn{3}{c}{\(\gamma_d(\cdot)\) for \(\theta_2\)} && \multicolumn{3}{c}{\(\gamma_d(\cdot)\) for \(\theta_3\)} \\

\cline{2-4} \cline{6-8} \cline{10-12} \cline{14-16} \cline{18-20} 

\( \bar{\bm{\varepsilon}}\) & Const & Flex & Min && Const & Flex & Min && Const & Flex & Min && Const & Flex & Min && Const & Flex & Min \\
\hline
0.00 & \textbf{3.859} & 5.409 & 7.033 && \textbf{1.695} & 3.564 & 5.566 && 5.556 & \textbf{5.000} & 5.698 && 2.138 & 1.806 & \textbf{1.756} && 0.947 & 1.189 & \textbf{0.584} \\
0.05 & \textbf{2.125} & 4.188 & 2.620 && \textbf{2.126} & 3.541 & 2.914 && 6.139 & 6.129 & \textbf{4.506} && \textbf{1.406} & 2.127 & 2.774 && \textbf{0.716} & 1.190 & 1.514 \\
0.10 & \textbf{1.729} & 1.970 & 2.586 && 1.790 & \textbf{0.941} & 2.480 && 5.615 & 5.464 & \textbf{2.561} && 2.190 & \textbf{1.896} & 2.251 && 1.177 & 1.097 & \textbf{1.011} \\
0.25 & \textbf{0.734} & 2.916 & 1.683 && \textbf{1.031} & 2.791 & 2.190 && 3.703 & 3.937 & \textbf{1.702} && 1.684 & \textbf{1.272} & 2.159 && \textbf{0.502} & 0.795 & 0.745 \\
\hline
\end{tabular}
}
\end{table}

We next assess predictive uncertainty using the normalized interval score (NIS) for 95\% intervals (panels C and G in Fig.~\ref{exp_noise_results}). The Bayesian approaches produce the most accurate predictive intervals when experimental uncertainty is small, with the Flex treatment providing a moderate advantage. These results show that modeling output-dependent uncertainty alone does not uniformly improve parameter recovery, posterior calibration, and predictive accuracy. We therefore next examine whether additional experimental responses provide complementary constraints on the local posterior geometry.

\subsection{Local Posterior Geometry with Single and Multiple Responses}
\label{identifiability_insights}
Previous work has shown that multiple experimental responses can reduce confounding in Kennedy–O’Hagan models improves with multiple experimental responses \cite{arendt2012}; We examine how this effect appears in parameter recovery and local posterior geometry when simulation and experimental outputs are coupled through latent embeddings. Using the same benchmark problems as in Sec.~\ref{accuracy_and_noise_sec}, we vary the number of experimental outputs used during calibration and keep the number of simulation experiments constant at \(d_s=3 \).

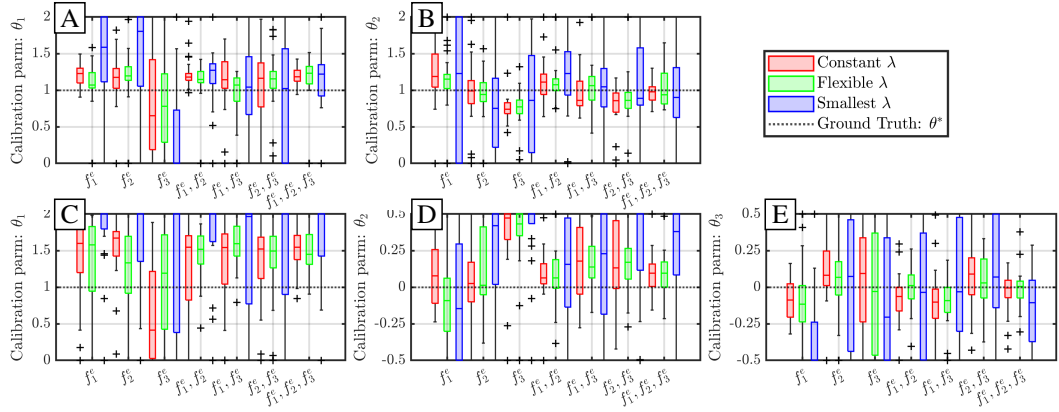
\begin{figure}[t]
\centering
\begin{tikzpicture}

\node[inner sep=0pt] (F) at (20mm,0mm){\includegraphics[width=51mm]{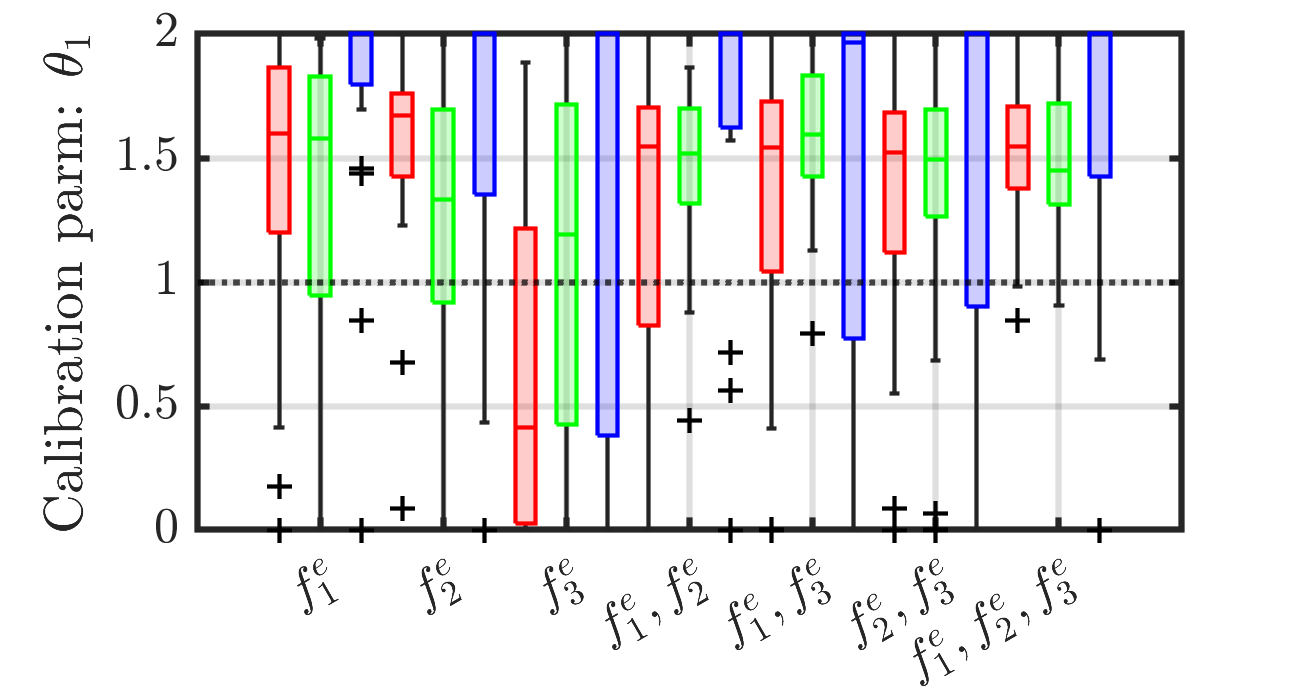}};
\node[inner sep=0pt] (F) at (67mm,0mm){\includegraphics[width=51mm]{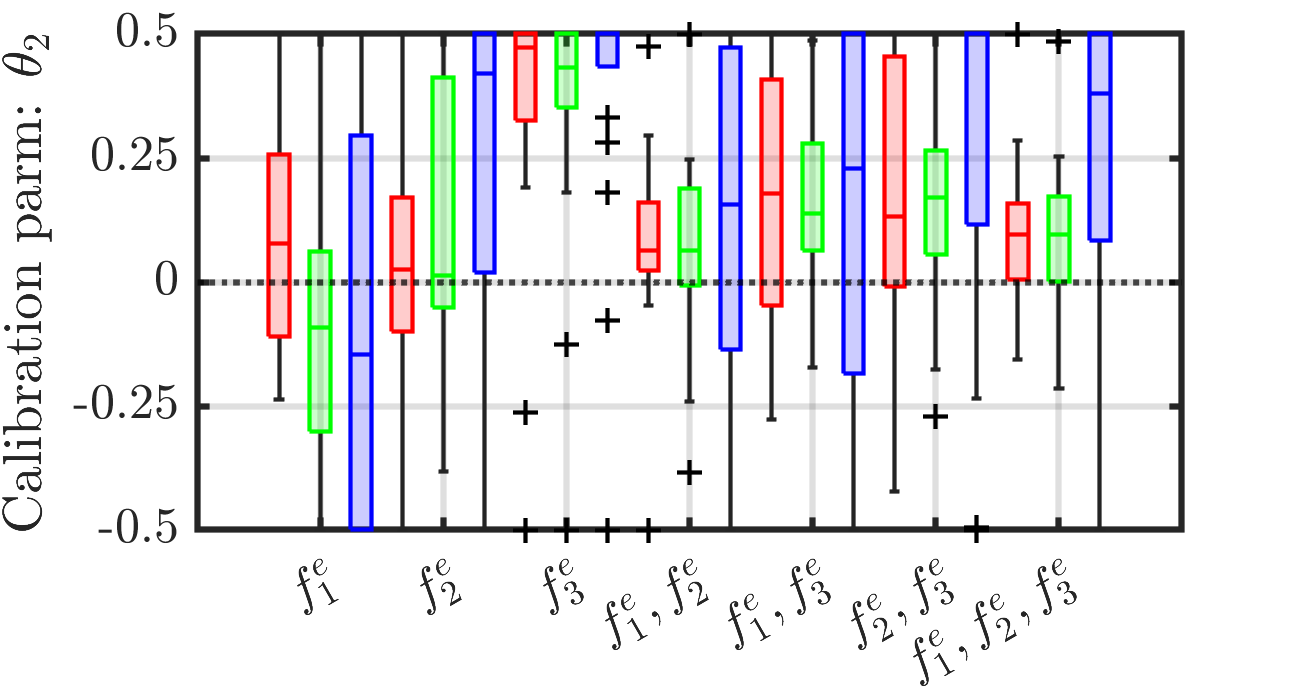}};
\node[inner sep=0pt] (F) at (114mm,0mm){\includegraphics[width=51mm]{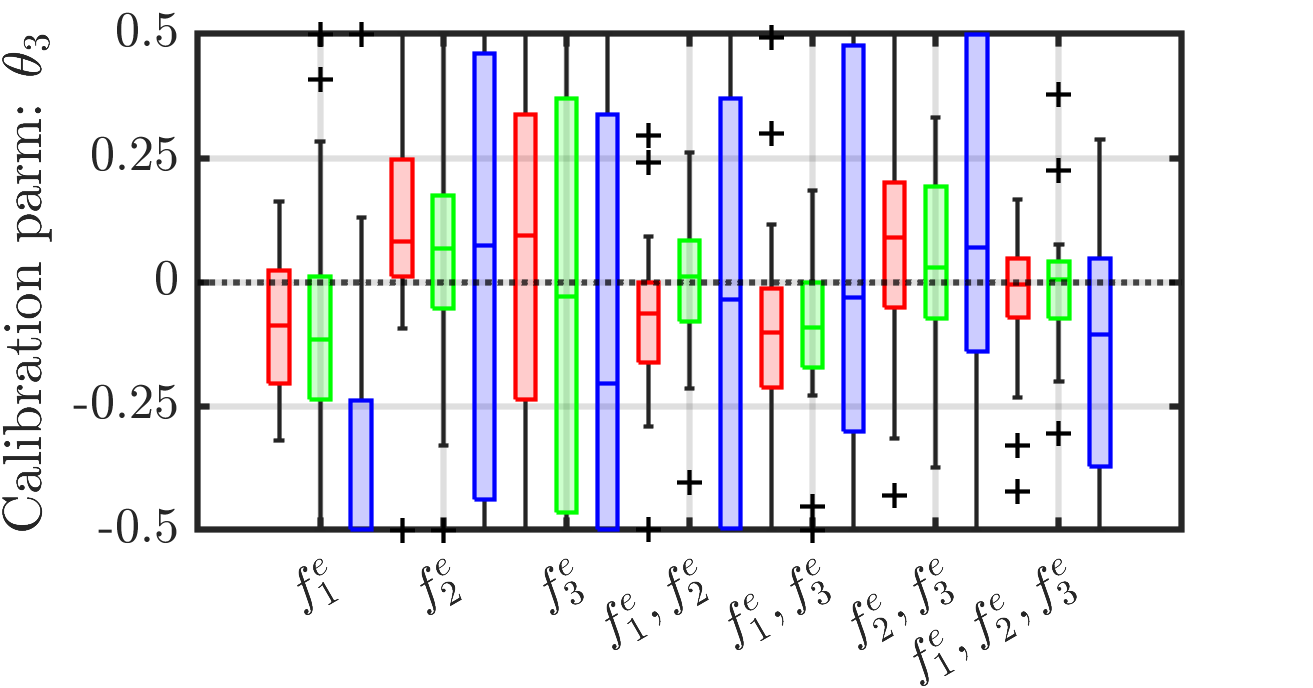}};

\node[inner sep=0pt] (F) at (20mm,26mm){\includegraphics[width=51mm]{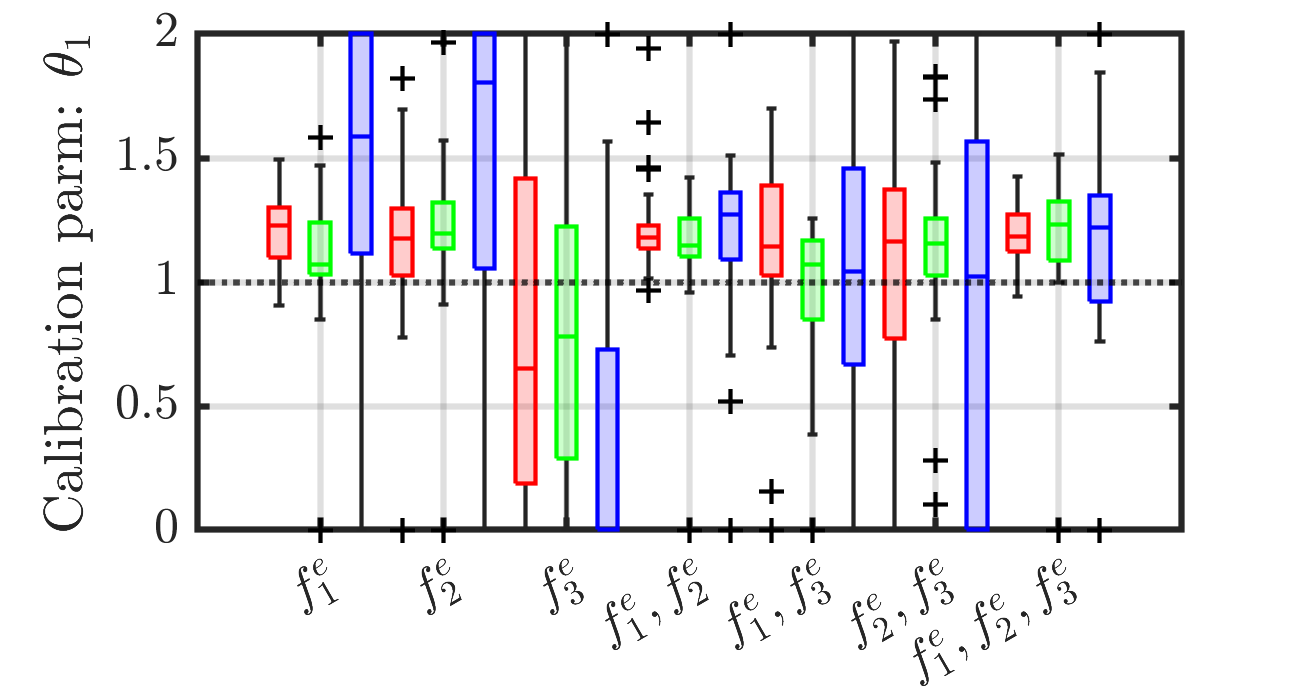}};
\node[inner sep=0pt] (F) at (67mm,26mm){\includegraphics[width=51mm]{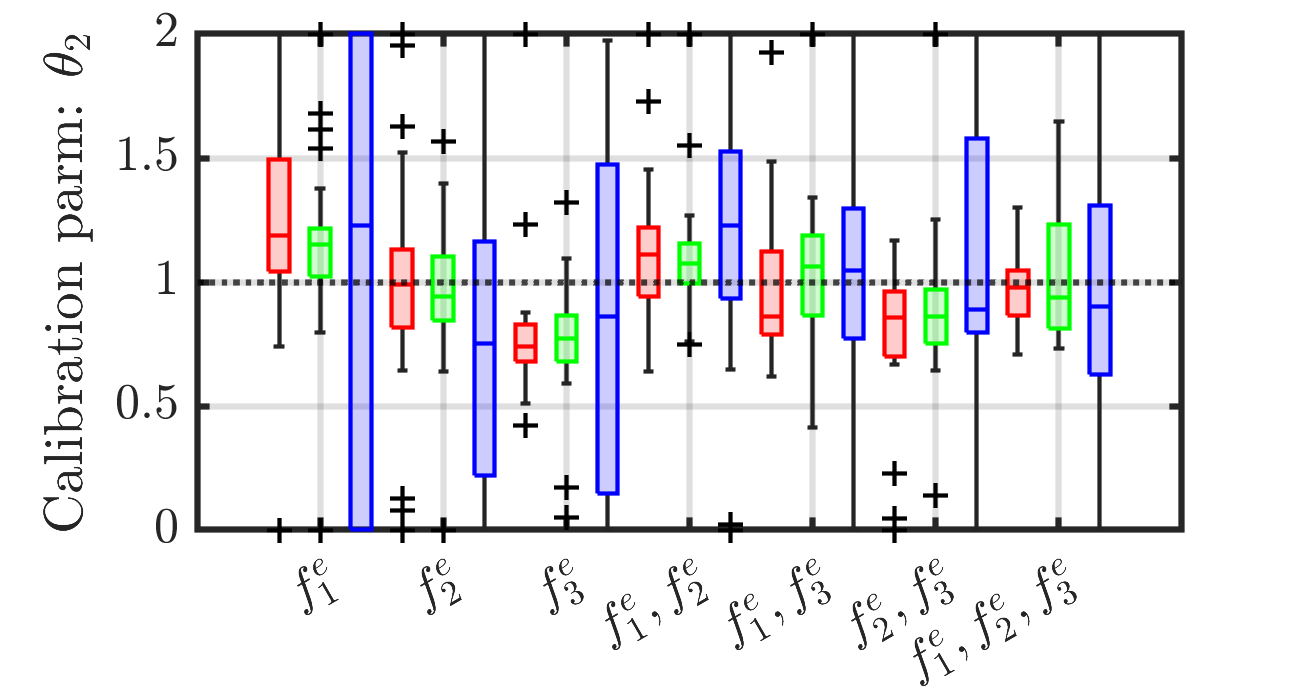}};
\node[inner sep=0pt] (F) at (108mm,28mm){\includegraphics[width=25mm]{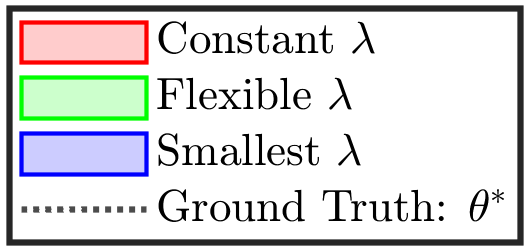}};

\draw[ line width=0.2mm, fill = white] (2mm,36mm) --++ (0,4mm) --++ (4mm,0mm) --++ (0mm,-4mm) -- cycle;
\node[mynode, anchor=center, align=center, text width = 4mm] at (2mm+2mm,38mm) {A};

\draw[ line width=0.2mm, fill = white] (49.25mm,36mm) --++ (0,4mm) --++ (4mm,0mm) --++ (0mm,-4mm) -- cycle;
\node[mynode, anchor=center, align=center, text width = 4mm] at (49.25mm+2mm,38mm) {B};

\draw[ line width=0.2mm, fill = white] (2mm,10mm) --++ (0,4mm) --++ (4mm,0mm) --++ (0mm,-4mm) -- cycle;
\node[mynode, anchor=center, align=center, text width = 4mm] at (2mm+2mm,10mm+2mm) {C};

\draw[ line width=0.2mm, fill = white] (49.25mm,10mm) --++ (0,4mm) --++ (4mm,0mm) --++ (0mm,-4mm) -- cycle;
\node[mynode, anchor=center, align=center, text width = 4mm] at (49.25mm+2mm,10mm+2mm) {D};

\draw[ line width=0.2mm, fill = white] (96mm,10mm) --++ (0,4mm) --++ (4mm,0mm) --++ (0mm,-4mm) -- cycle;
\node[mynode, anchor=center, align=center, text width = 4mm] at (96mm+2mm,10mm+2mm) {E};

\end{tikzpicture}
\caption{Calibration results for varying numbers of outputs (\(\bar{\varepsilon}=0.1\); 25 runs; ‘+’ denotes outliers). A,B) parameter estimates for problem 2; C–E) parameter estimates for problem 3. Increasing the number of outputs selectively strengthens local posterior constraints.}
\label{exp_num_out_results}
\end{figure}

Figure~\ref{exp_num_out_results} shows calibration-parameter estimates obtained from different subsets of experimental outputs. The Min treatment consistently exhibits greater variability and larger deviations from the ground truth. In contrast, the Const and Flex treatments produce more accurate estimates, with performance generally improving as additional outputs are included. Although one or two responses sufficiently constrain some parameters, others require all three responses to achieve comparable localization. This parameter-dependent behavior is consistent with different outputs contributing complementary directions of local posterior curvature rather than redundant information.

In Tbl.~\ref{diff_out_res_tbl}, we report performance at a fixed experimental-noise level \(\bar{\sigma}=0.10\) using predictive MSE, NIS, and the repeated-run posterior-calibration metric \(\gamma_d\) Const and Flex provide the most accurate posterior predictions across most output configurations, with Flex showing a modest advantage when only a few responses are available. In contrast, Min often yields the lowest \(\gamma_d\) values despite less accurate point estimates. This reflects wider posterior approximations that reduce overconfidence and produce standardized residuals closer to the standard normal reference, rather than stronger parameter localization. Overall, the results reveal a trade-off between parameter localization and uncertainty calibration: Const and Flex provide sharper estimates, whereas Min often yields better-calibrated uncertainty at the cost of weaker localization. Similar trends are observed across other experimental-noise levels (Appx.~\ref{additiona_results_diff_outputs_appx}

\begin{table}[t]
\centering

\setlength{\tabcolsep}{1.8pt}
\renewcommand{\arraystretch}{1}

\caption{Calibration performance for Problems 2 and 3 at fixed experimental variance (\(\bar{\varepsilon}=0.1\)) across different output sets. Const and Flex generally yield the best predictive accuracy, whereas Min often yields the lowest posterior-calibration discrepancy \(\gamma_d\). Lower \(\gamma_d\) indicates better repeated-run uncertainty calibration, not necessarily sharper parameter localization.} Bold indicates the best value per metric and output set.
\label{diff_out_res_tbl}

{\scriptsize
\begin{tabular}{cc cccc cccc cccc cccc cccc ccc}
\hline
&& \multicolumn{11}{c}{\textbf{Problem 2}} && \multicolumn{11}{c}{\textbf{Problem 3}} \\

\cline{3-13} \cline{15-25}

&& \multicolumn{3}{c}{\(\gamma_{MSE}\)} && \multicolumn{3}{c}{\(\gamma_{NIS}\)} && \multicolumn{3}{c}{\(\mathbb{E}(\gamma_{d})\)} && \multicolumn{3}{c}{\(\gamma_{MSE}\)} && \multicolumn{3}{c}{\(\gamma_{NIS}\)} && \multicolumn{3}{c}{\(\mathbb{E}(\gamma_{d})\)}\\

\cline{3-5} \cline{7-9} \cline{11-13} \cline{15-17} \cline{19-21} \cline{23-25}

\( \bar{\varepsilon}\) && Const & Flex & Min && Const & Flex & Min && Const & Flex & Min && Const & Flex & Min && Const & Flex & Min && Const & Flex & Min \\
\hline
\( f_1^{(e)}\)                        && 0.424 & \textbf{0.434} & 0.533 && 18.15 & \textbf{17.47} & 21.62 && 1.020 & 0.882 & \textbf{0.642} && 0.643 & \textbf{0.594} & 0.688 && 30.82 & \textbf{27.32} & 33.69 && 2.689 & 2.781 & \textbf{0.940} \\
\( f_2^{(e)}\)                        && 0.465 & \textbf{0.437} & 0.528 && 19.27 & \textbf{18.00} & 21.28 && 1.680 & 2.168 & \textbf{0.734} && 0.615 & \textbf{0.586} & 0.699 && 31.18 & \textbf{28.19} & 36.30 && 2.612 & 1.746 & \textbf{0.893} \\
\( f_3^{(e)}\)                        && \textbf{0.395} & 0.397 & 0.521 && \textbf{16.04} & 16.63 & 22.63 && 5.498 & 4.120 & \textbf{2.983} && \textbf{0.589} & 0.655 & 0.701 && \textbf{30.77} & 32.62 & 37.38 && 2.436 & 3.149 & \textbf{0.739} \\
\( f_{1,2}^{(e)}\)            && \textbf{0.384} & 0.394 & 0.495 && \textbf{16.24} & 17.01 & 21.66 && \textbf{1.301} & 1.504 & 1.340 && \textbf{0.520} & 0.526 & 0.684 && \textbf{27.06} & 27.33 & 35.24 && 3.068 & 1.847 & \textbf{1.368} \\
\( f_{1,2}^{(e)}\)            && \textbf{0.376} & 0.387 & 0.473 && \textbf{16.34} & 16.54 & 20.27 && 2.090 & \textbf{1.530} & 2.350 && \textbf{0.560} & 0.566 & 0.681 && 30.39 & \textbf{29.97} & 34.75 && 2.697 & 2.730 & \textbf{0.868} \\
\( f_{2,3}^{(e)}\)            && 0.402 & \textbf{0.400} & 0.491 && \textbf{17.01} & 17.49 & 20.89 && 3.700 & 3.303 & \textbf{2.922} && \textbf{0.558} & 0.610 & 0.696 && \textbf{28.80} & 32.19 & 36.60 && 2.529 & 3.666 & \textbf{1.208} \\
\( f_{1,2,3}^{(e)}\) && \textbf{0.215} & 0.240 & 0.372 && \textbf{12.29} & 12.39 & 19.28 && \textbf{1.579} & 2.343 & 2.702 && \textbf{0.399} & 0.424 & 0.508 && \textbf{22.11} & 24.55 & 29.54 && 2.571 & 2.520 & \textbf{1.645} \\
\hline
\end{tabular}
}
\end{table}

\subsection{Practical Example for Battery Modeling and Calibration}
\label{battery}
To demonstrate practical utility, we apply the calibration and bias-correction model to a simulation model of battery electrodes. The problem involves one design variable and six calibration parameters with three experimental outputs. Physical experiments consist of five measurements per output (each repeated three times), while 383 simulation runs are generated using a Sobol sequence (Appx.~\ref{battery_experiments}).

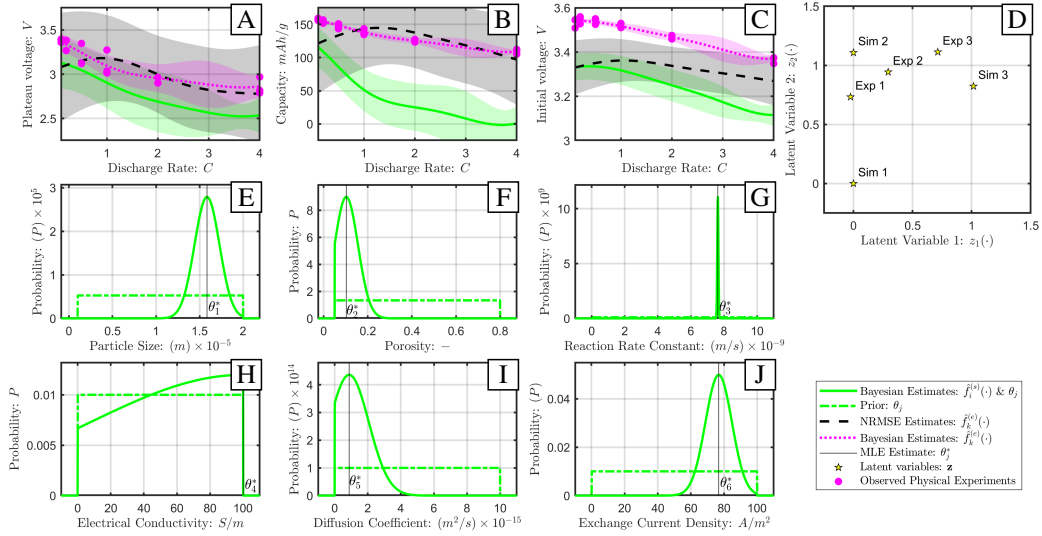
\begin{figure}[t]
\centering
\begin{tikzpicture}
\node[inner sep=0pt] (F) at (16mm,28mm){\includegraphics[width=35mm]{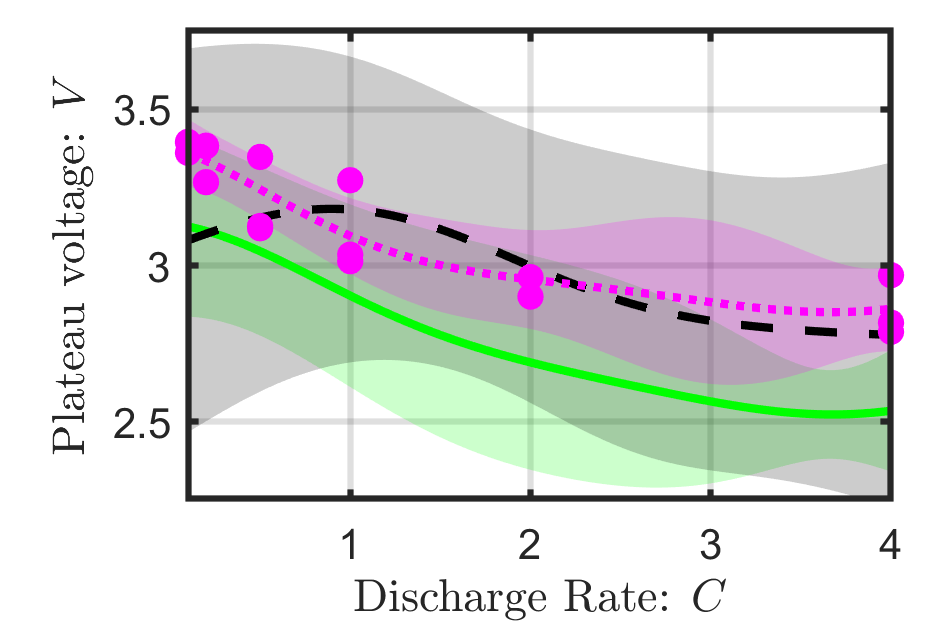}};
\node[inner sep=0pt] (F) at (50mm,28mm){\includegraphics[width=35mm]{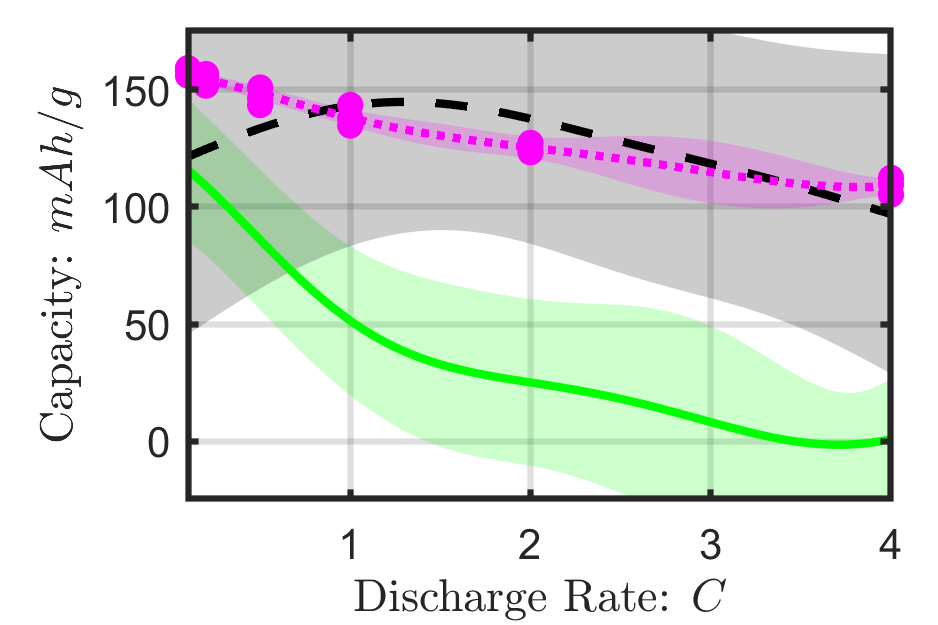}};
\node[inner sep=0pt] (F) at (84mm,28mm){\includegraphics[width=35mm]{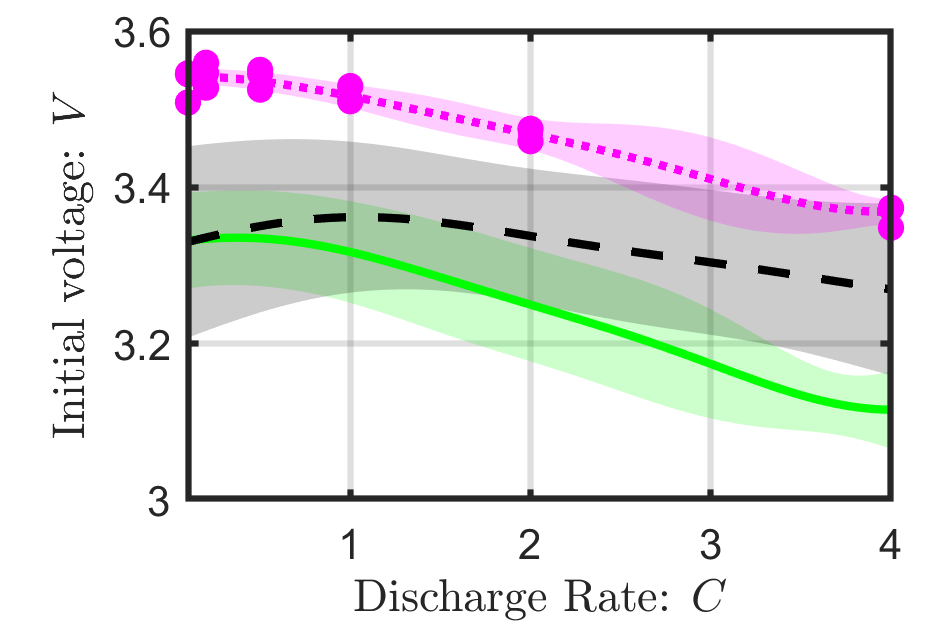}};
\node[inner sep=0pt] (F) at (118mm,23mm){\includegraphics[width=35mm]{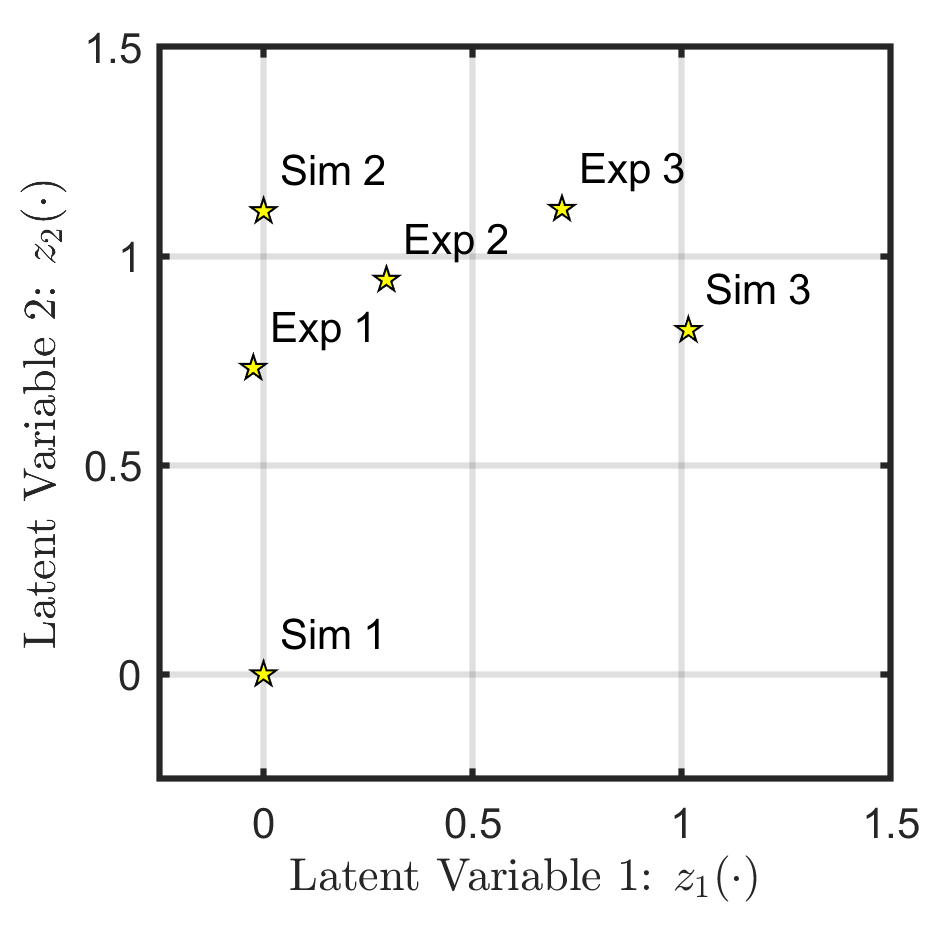}};

\node[inner sep=0pt] (F) at (16mm,4.5mm){\includegraphics[width=35mm]{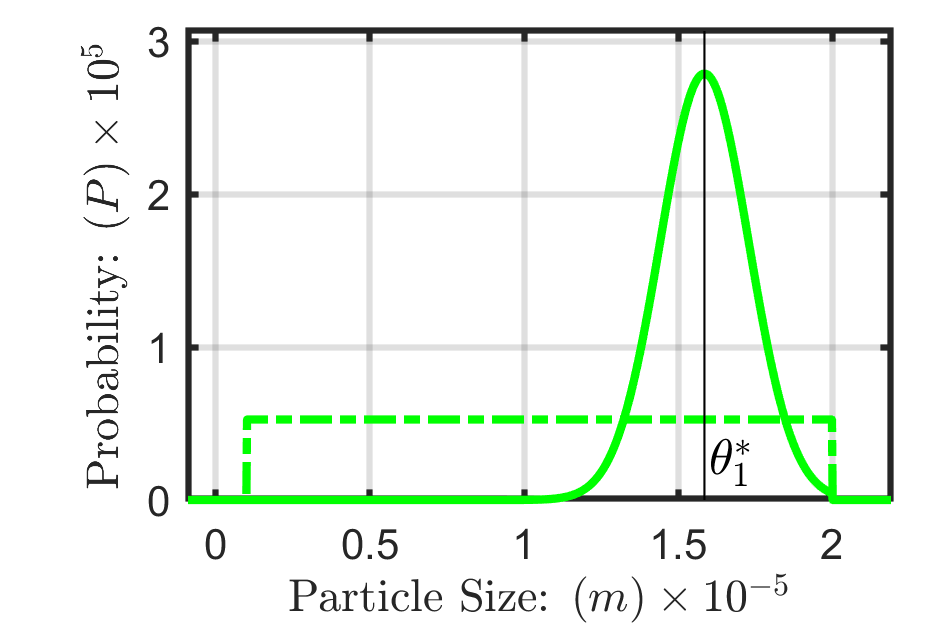}};
\node[inner sep=0pt] (F) at (50mm,4.5mm){\includegraphics[width=35mm]{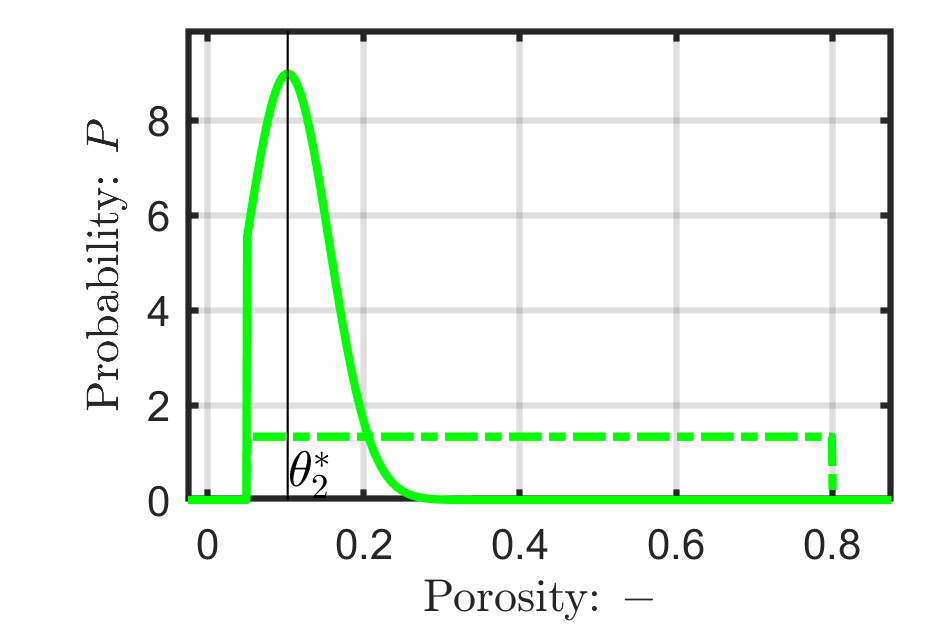}};
\node[inner sep=0pt] (F) at (84mm,4.5mm){\includegraphics[width=35mm]{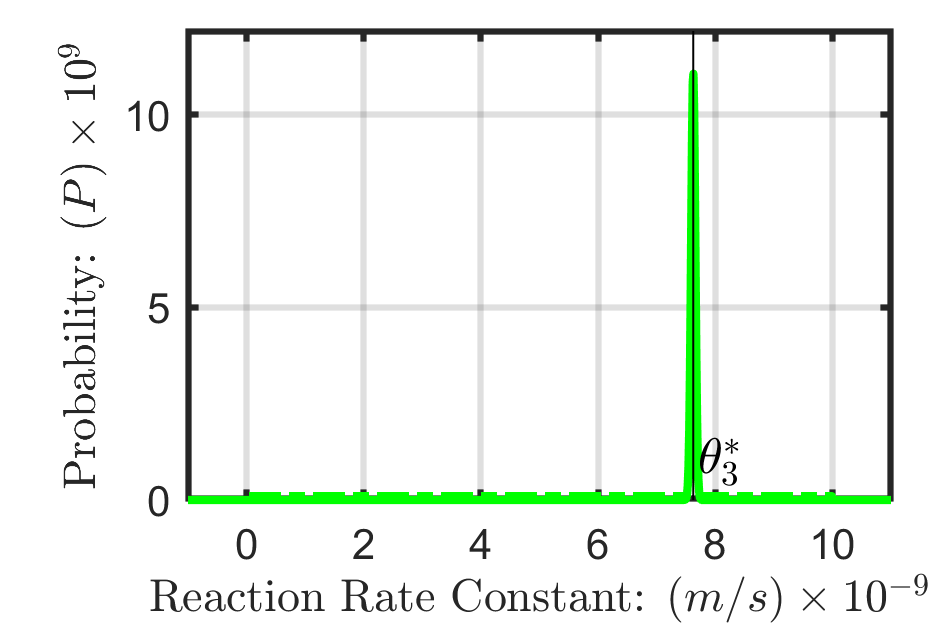}};
\node[inner sep=0pt] (F) at (119mm,-18mm){\includegraphics[width=28mm]{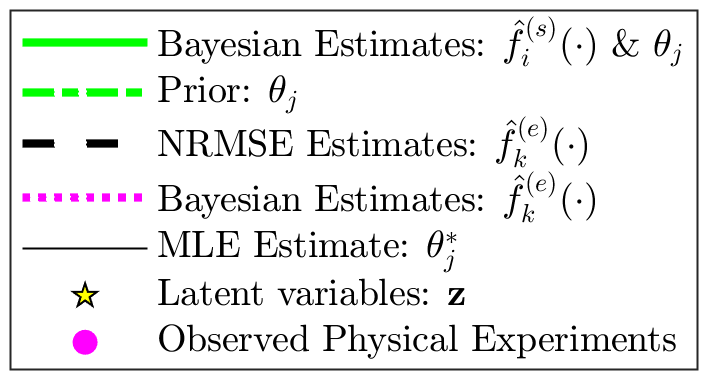}};

\node[inner sep=0pt] (F) at (16mm,-19mm){\includegraphics[width=35mm]{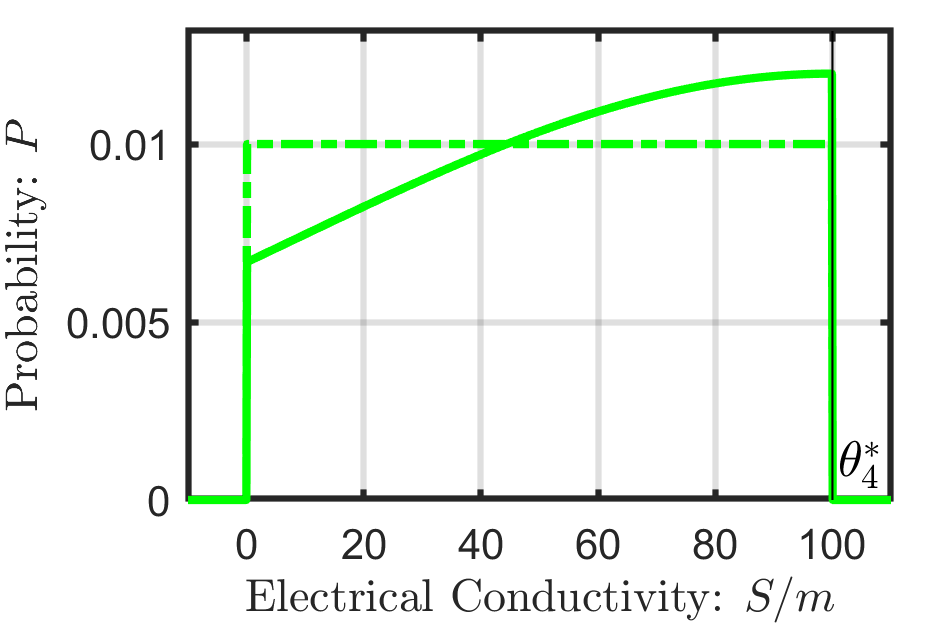}};
\node[inner sep=0pt] (F) at (50mm,-19mm){\includegraphics[width=35mm]{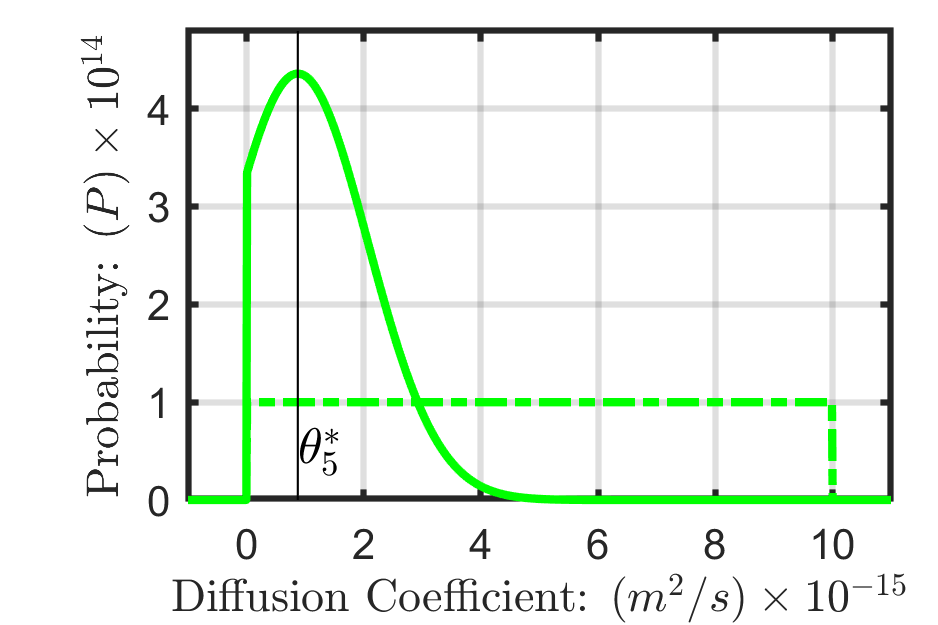}};
\node[inner sep=0pt] (F) at (84mm,-19mm){\includegraphics[width=35mm]{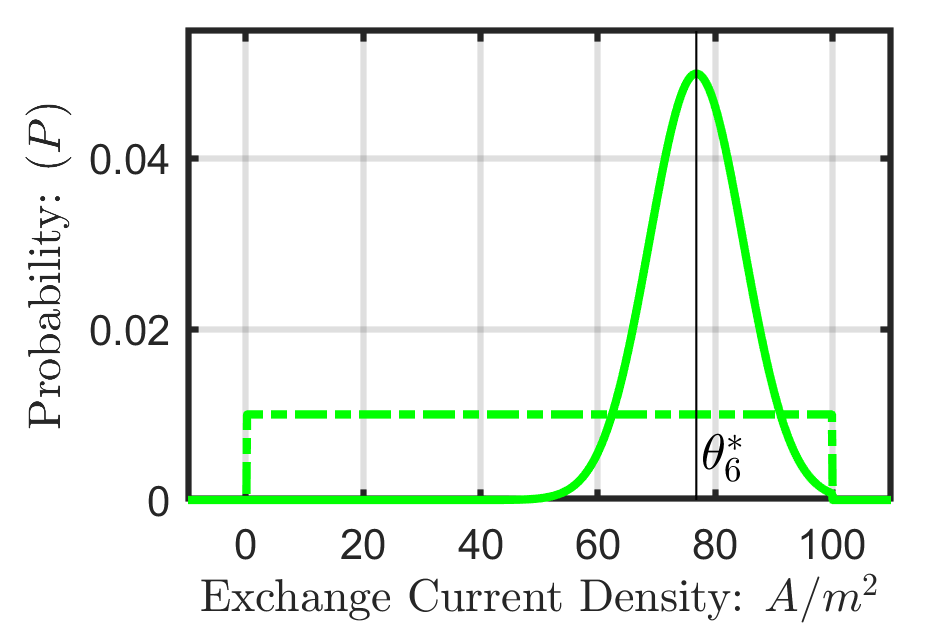}};

\draw[ line width=0.2mm, fill = white] (28mm,35mm) --++ (0,4mm) --++ (4mm,0mm) --++ (0mm,-4mm) -- cycle;
\node[mynode, anchor=center, align=center, text width = 4mm] at (30mm,37mm) {A};

\draw[ line width=0.2mm, fill = white] (62mm,35mm) --++ (0,4mm) --++ (4mm,0mm) --++ (0mm,-4mm) -- cycle;
\node[mynode, anchor=center, align=center, text width = 4mm] at (64mm,37mm) {B};

\draw[ line width=0.2mm, fill = white] (96mm,35mm) --++ (0,4mm) --++ (4mm,0mm) --++ (0mm,-4mm) -- cycle;
\node[mynode, anchor=center, align=center, text width = 4mm] at (98mm,37mm) {C};

\draw[ line width=0.2mm, fill = white] (130mm,35mm) --++ (0,4mm) --++ (4mm,0mm) --++ (0mm,-4mm) -- cycle;
\node[mynode, anchor=center, align=center, text width = 4mm] at (132mm,37mm) {D};

\draw[ line width=0.2mm, fill = white] (28mm,11.5mm) --++ (0,4mm) --++ (4mm,0mm) --++ (0mm,-4mm) -- cycle;
\node[mynode, anchor=center, align=center, text width = 4mm] at (30mm,13.5mm) {E};

\draw[ line width=0.2mm, fill = white] (62mm,11.5mm) --++ (0,4mm) --++ (4mm,0mm) --++ (0mm,-4mm) -- cycle;
\node[mynode, anchor=center, align=center, text width = 4mm] at (64mm,13.5mm) {F};

\draw[ line width=0.2mm, fill = white] (96mm,11.5mm) --++ (0,4mm) --++ (4mm,0mm) --++ (0mm,-4mm) -- cycle;
\node[mynode, anchor=center, align=center, text width = 4mm] at (98mm,13.5mm) {G};

\draw[ line width=0.2mm, fill = white] (28mm,-12mm) --++ (0,4mm) --++ (4mm,0mm) --++ (0mm,-4mm) -- cycle;
\node[mynode, anchor=center, align=center, text width = 4mm] at (30mm,-10mm) {H};

\draw[ line width=0.2mm, fill = white] (62mm,-12mm) --++ (0,4mm) --++ (4mm,0mm) --++ (0mm,-4mm) -- cycle;
\node[mynode, anchor=center, align=center, text width = 4mm] at (64mm,-10mm) {I};

\draw[ line width=0.2mm, fill = white] (96mm,-12mm) --++ (0,4mm) --++ (4mm,0mm) --++ (0mm,-4mm) -- cycle;
\node[mynode, anchor=center, align=center, text width = 4mm] at (98mm,-10mm) {J};

\end{tikzpicture}
\caption{Calibration of a lithium-ion battery model. A–C) Posterior predictive distributions, D) latent space, and E–J) calibration parameter posteriors. The probabilistic calibration produces smooth predictive trends, latent alignment across data sources, and parameter-specific uncertainty estimates that are unavailable from NRMSE.}
\label{battery_results}
\end{figure}

Panels A–C in Fig.~\ref{battery_results} show posterior predictive distributions across discharge rates \(x\) using output dependent noise (i.e., Flex) and compare it with an NRMSE fit. Although the NRMSE approach achieves a closer point-wise fit, the Bayesian calibration produces smoother predictive trends in panels A and C, consistent with the model allocating systematic differences between simulation and experiment to the inferred discrepancy structure rather than forcing the calibration parameters alone to account for them. In panel B, the Bayesian predictive distribution appears less accurate but remains convex across discharge rates, whereas the NRMSE fit exhibits concave behavior inconsistent with the experimental trend. Specifically, the convex predictive structure produced by the Bayesian model reflects consistency with the underlying discharge-rate dependence, whereas the concave NRMSE fit may reflect local over fitting to measurement noise.

Panels E–J show posterior distributions of the calibration parameters. Under the fitted model, several parameters, including the reaction-rate constant, have concentrated local posterior approximations, whereas others, such as the electrical conductivity, remain weakly constrained. This illustrates how the proposed local posterior characterization distinguishes parameters that are strongly constrained by the available responses from those for which substantial uncertainty remains. This separation reflects parameter dependent identifiability and provides a model-dependent indication of which physical mechanisms are more strongly informed by the available data.

Panel D shows the two-dimensional latent embedding of simulation (Sim) and experimental (ExP) data sources. Physical experiments cluster closely with their corresponding simulation outputs, showing that the fitted latent representation places each experiment near its corresponding simulation output and thereby induces strong cross-source correlations within the GP model. This fitted alignment provides a compact representation of the cross-source relationships used by the calibration model.

Collectively, these results illustrate the practical role of the proposed posterior-geometry framework in a multi-response calibration problem with model discrepancy. The fitted latent embedding provides a parsimonious alignment between simulation and experimental outputs, while the local Gaussian approximations distinguish parameters that are strongly constrained by the available responses from those that remain uncertain. The example therefore demonstrates the interpretive utility of the framework without requiring that the physical calibration parameters be known.

\section{Concluding Remarks}
\label{conclusion}
We presented a computationally tractable characterization of local posterior geometry for latent-variable multi-response Bayesian calibration. Inference is performed using an empirical Bayes approximation, in which nuisance hyperparameters are estimated by maximum likelihood and the conditional posterior over the calibration parameters is approximated using the expected Fisher information. Through controlled numerical experiments and a lithium-ion battery calibration study, we demonstrated how multiple responses, latent alignment, and output-specific uncertainty shape local posterior geometry and the resulting practical identifiability of calibration parameters.

The proposed characterization is local and model dependent and therefore does not guarantee statistical identifiability or recovery of the true calibration parameters. Future work should establish stronger conditions for parameter separation by incorporating structural information from governing physics, accounting for uncertainty in nuisance hyperparameters, and extending the framework beyond Euclidean parameter spaces to structured or manifold-valued hypothesis spaces. The proposed framework provides a tractable foundation for studying posterior geometry in more general multi-response and cross-source calibration problems.

\begin{ack}
WJD and JRB acknowledge the Faraday Institution through the Characterisation and Manufacturing of Advanced LFP Batteries grant (FIRG081).
\end{ack}

\bibliographystyle{unsrt}
\bibliography{Bibliography}
\appendix
\clearpage
\section{Pedagogical Illustration of Calibration and Model Misspecification}
\label{peda_exam}
A key distinction between Bayesian calibration and norm-minimization is that the latter implicitly attributes all model misspecification/bias to the calibration parameters. In contrast, Bayesian calibration enables separation of structural discrepancy from parameter uncertainty and yields a joint posterior distribution over calibration parameters and the bias function rather than a single point estimate. To illustrate this distinction, we apply the model introduced in Sec.~\ref{problem_form} to a pedagogical example described as Problem 1 in Tbl.~\ref{sample_fun_tbl} in Appx~\ref{example_form}. In this example, the simulation model is structurally misspecified relative to the physical process. We observe six noisy ``physical measurements'', where the observational noise has a standard deviation equal to 10\% of the response range (i.e., \( \bar{\varepsilon} = 0.1 \)), together with twelve simulation outputs.

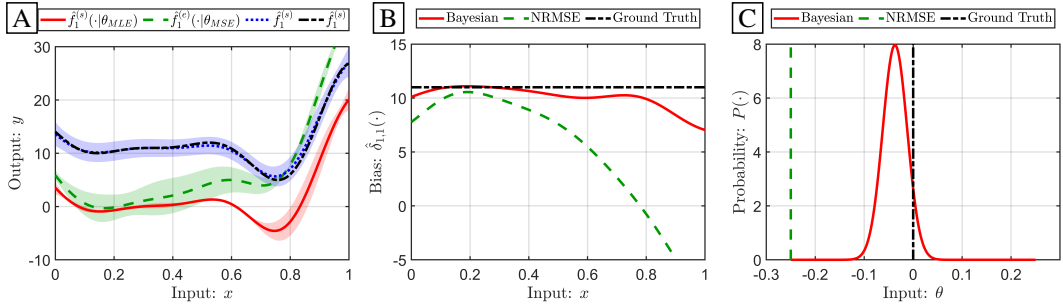
\begin{figure}[t]
\centering
\begin{tikzpicture}
\node[inner sep=0pt] (F) at (25mm,15mm){\includegraphics[width=50mm]{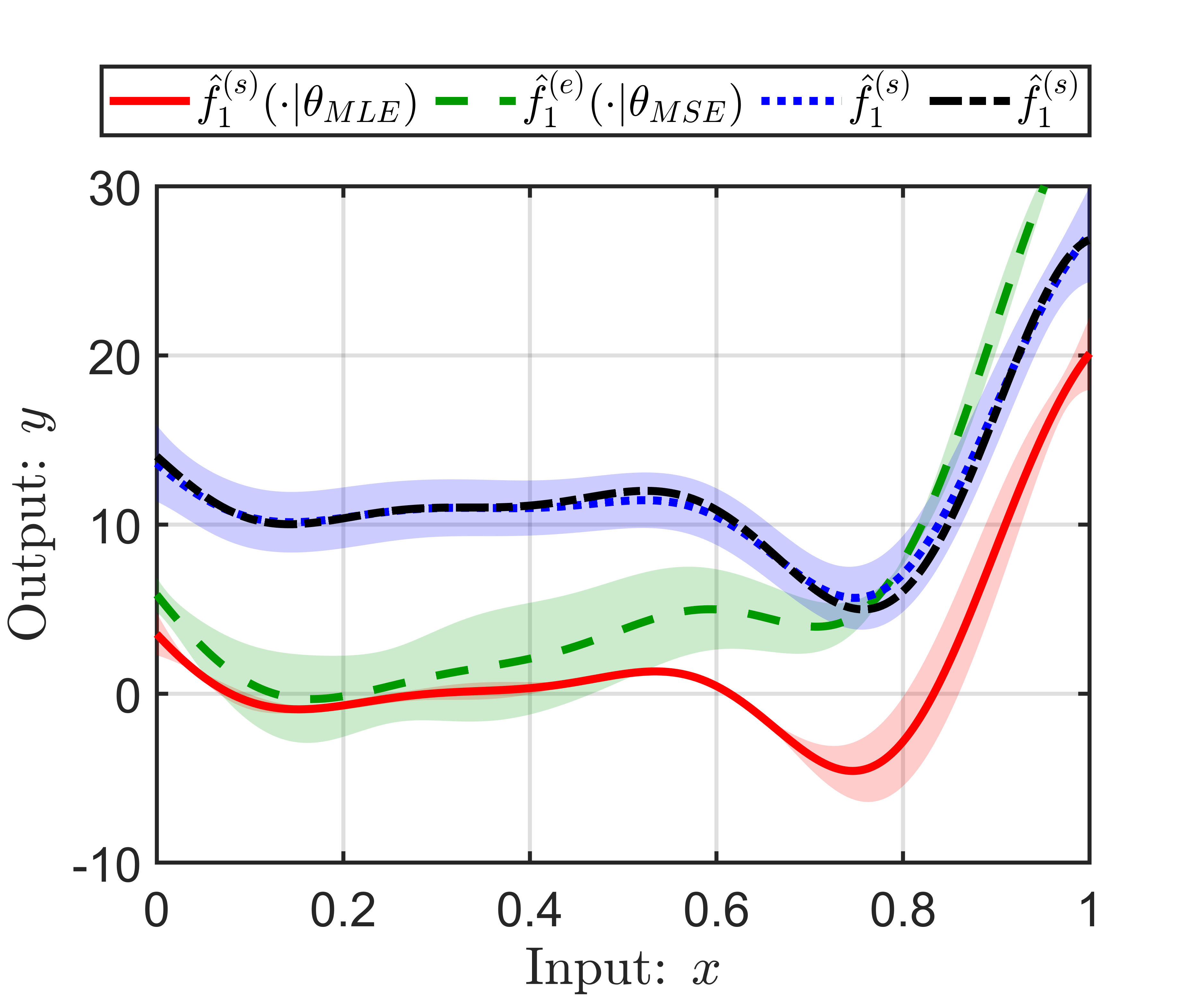}};
\node[inner sep=0pt] (F) at (72mm,15mm){\includegraphics[width=50mm]{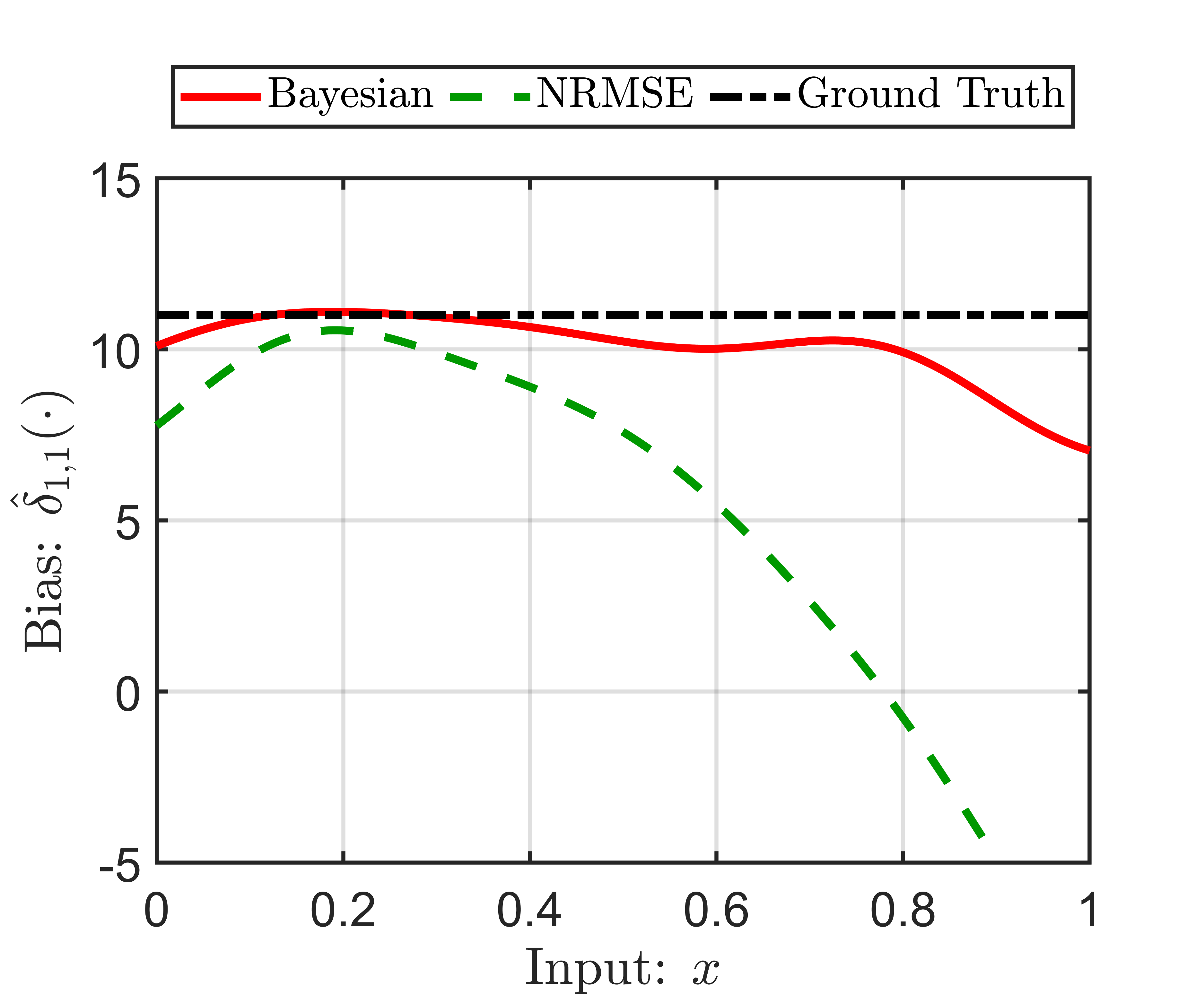}};
\node[inner sep=0pt] (F) at (119mm,15mm){\includegraphics[width=50mm]{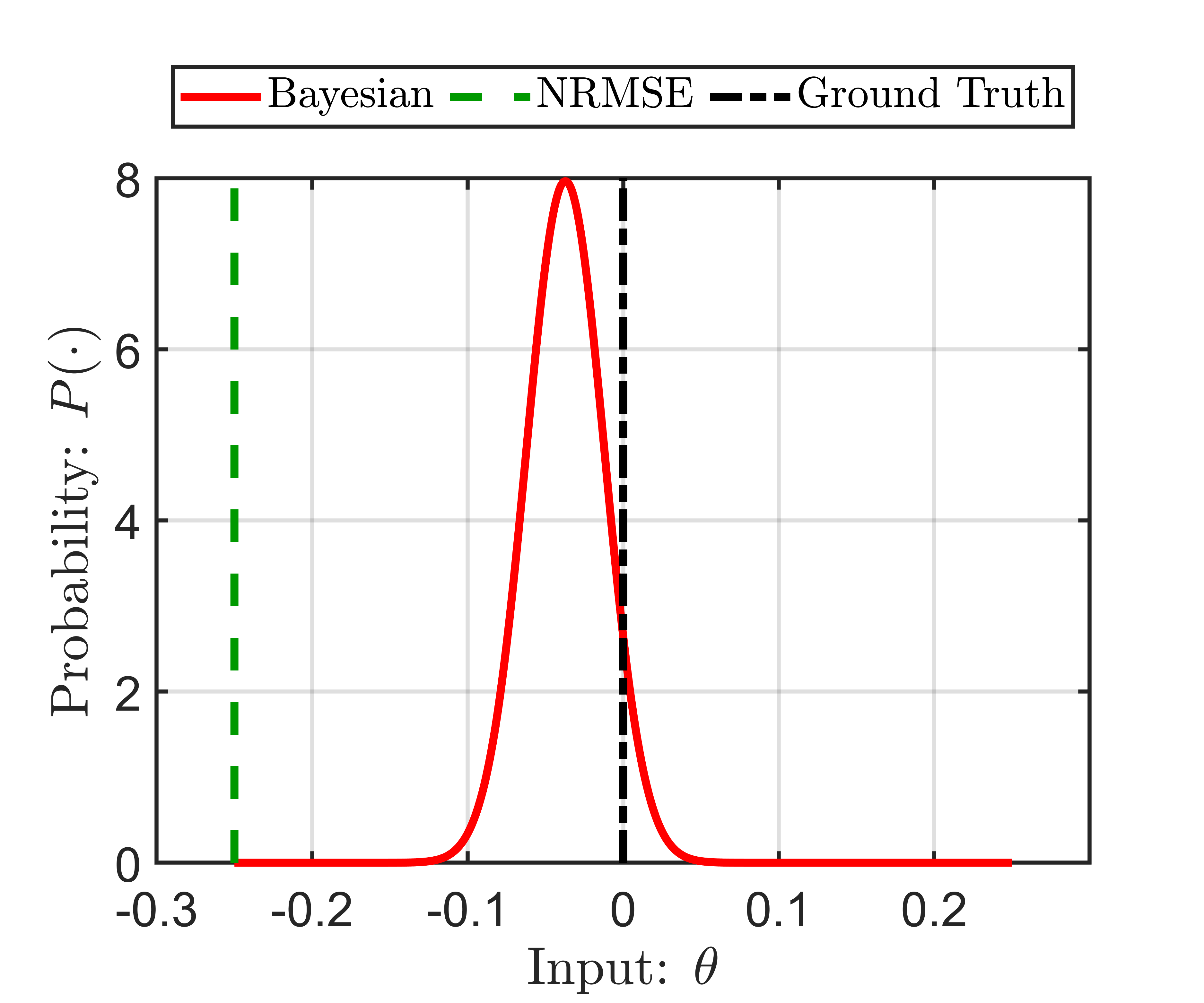}};

\draw[ line width=0.2mm] (0mm,30mm) --++ (0,4mm) --++ (4mm,0mm) --++ (0mm,-4mm) -- cycle;
\node[mynode, anchor=center, align=center, text width = 4mm] at (2mm,32mm) {A};

\draw[ line width=0.2mm] (48.5mm,30mm) --++ (0,4mm) --++ (4mm,0mm) --++ (0mm,-4mm) -- cycle;
\node[mynode, anchor=center, align=center, text width = 4mm] at (50.5mm,32mm) {B};

\draw[ line width=0.2mm] (96mm,30mm) --++ (0,4mm) --++ (4mm,0mm) --++ (0mm,-4mm) -- cycle;
\node[mynode, anchor=center, align=center, text width = 4mm] at (98mm,32mm) {C};

\end{tikzpicture}
\caption{Pedagogical example illustrating calibration under structural discrepancy and its influence on calibration parameters. (A) Posterior predictive distribution for the physical response under Bayesian calibration (blue dotted) and MSE (green dashed line) each plotted with their 95\% prediction intervals compared with ground truth given by the (black dash-dotted). (B) Estimated bias function under Bayesian calibration (red solid) and NRMSE (green dashed), with true discrepancy (black dash-dotted). (C) Posterior distribution of the calibration parameter; the true value is \(\theta^*=0\)}
\label{peda_fig}
\end{figure}

Panel A of Fig.~\ref{peda_fig} shows the posterior predictive distribution for the physical response. Despite the limited number of physical observations, the Bayesian prediction (blue dotted) closely follows the ground truth (black dash-dotted). This improvement is enabled by leveraging the additional simulation data while accounting for structural discrepancy. The corresponding calibration estimates are \( \theta_{MLE}=-0.0372\) and \( \theta_{MSE}=0.25\). 

The NRMSE approach selects \(\theta_{\text{MSE}}\) by directly minimizing the discrepancy between simulation and physical outputs, thereby forcing calibration parameters to compensate for structural bias. The effect becomes more apparent in Panel B, which shows the inferred bias functions. The Bayesian estimate (red solid) closely matches the ground-truth discrepancy (black dash-dotted), whereas the NRMSE approach (green dashed) fails to recover the correct bias structure. Instead, it shifts the calibration parameter so that structural discrepancy is absorbed into parameter mismatch, resulting in systematic deviation that trends to zero and is inconsistent with the ground truth (\(\delta_{1,1}(\cdot)=11\)). 

Panel C presents the posterior distribution of the calibration parameter under the Bayesian approach. The true value is \(\theta^\star = 0\). Using a  approximation based on the expected Fisher information, we obtain \(\theta\mid D_n\;\dot{\sim}\;\mathcal{N}(-0.0372,0.0501).\) Although slightly biased due to limited data, the posterior mass concentrates near the true value. In contrast, the NRMSE estimate \(\theta_{\text{MSE}}=-0.25\) lies at the edge of the admissible calibration parameter range and is substantially farther from \(\theta^\star\).

The relatively tight posterior uncertainty arises from the strong latent-space correlation between simulation and physical observations, \(0.9205=\exp\left(-\left\|\mathbf{z}(\eta)-\mathbf{z}(\eta^{\prime})\right\|_2^2\right)=\exp\left(-\left\|0-0.0829\right\|_2^2\right) \) (note that these numbers where obtained through MLE of the hyperparameters in \(\mathbf{A}\). While such a high correlation may be optimistic in practice, this example highlights how explicitly modeling the bias can prevent structural discrepancy from being incorrectly absorbed into calibration parameters.

\section{Fisher Information for Gaussian Process Calibration Models}
\label{Fish_info}
Here we provide a detailed discussion of the derivation of the expected Fisher information of the calibration parameters and other hyperparameters. 

\subsection{General Expression for Expected Fisher Information}
\label{fish_gen_sec}
For the prior distribution defined in Eqn.~\ref{priorGP}, the log likelihood is given as
\begin{equation}
    \log \left(P(D_n \mid \bm{\omega},\mathbf{A},\bm{\lambda},\bm{\theta}_e)\right)= -\dfrac{1}{2}\left( n\log (\sigma^2)+\log\left(\left|\mathbf{R}_{\delta}\right|\right)+\frac{1}{\sigma^2}\mathbf{Z}^T \mathbf{R}_{\delta}^{-1}\mathbf{Z}+n\log(2\pi) \right),
    \label{full_llh}
\end{equation}
where \(\mathbf{Z}=\mathbf{Y}-\mathbf{M}\bm{\beta}\). We are interested in efficiently approximating the confidence that one can have in the hyperparameters \(\bm{\omega},\mathbf{A},\bm{\lambda},\bm{\theta}_e\), especially the calibration parameters \(\boldsymbol{\theta}_e \). This can be achieved through the expected Fisher information \cite{frieden2004} for which the general expression is given as
\begin{equation}
    \mathcal{J}_{i,j}=-\mathbb{E}\left( \dfrac{\partial^2 \log (P\mid\bm{\omega},\mathbf{A},\bm{\lambda},\bm{\theta}_e )}{\partial \theta_i\partial\theta_j} \right),
    \end{equation}
where the expectation is taken with respect to the training data \(\mathbf{Y}\sim\mathcal{N}\left( \mathbf{0},\sigma^2\left(\mathbf{R}+\bm{\pi}\right) \right)\), comprising simulation and physical observations. Consequently, taking the first derivative of the log likelihood in Eqn.~\ref{full_llh} gives us 
\begin{equation}
    \dfrac{\partial \log (P\mid\bm{\omega},\mathbf{A},\bm{\lambda},\bm{\theta}_e )}{\partial \theta_i} = - \dfrac{1}{2}\text{tr}\left( \mathbf{R}_{\delta}^{-1} \dfrac{\partial\mathbf{R}}{\partial\theta_i}  \right) +\dfrac{1}{2\sigma^2}\mathbf{Z}^T\mathbf{R}_\delta^{-1}\dfrac{\partial\mathbf{R}}{\partial\theta_i}\mathbf{R}_\delta^{-1}\mathbf{Z} ,
\end{equation}
where we used Jacobi's formula that provided the trace \( \text{tr}(\cdot)\) expression and the chain rule to take the derivative of the inverse correlation matrix. Moreover, it should be noted that we used \(\partial\mathbf{R}_{\delta}/\partial\theta_i = \partial\mathbf{R}/\partial\theta_i\). Taking the derivative again of the second term on the right-hand side gives us 
\begin{align}
    \dfrac{\partial}{\partial \theta_j}\left( \dfrac{1}{2\sigma^2}\mathbf{Z}^T\mathbf{R}_\delta^{-1}\dfrac{\partial\mathbf{R}}{\partial\theta_i}\mathbf{R}_\delta^{-1}\mathbf{Z} \right) = &\dfrac{1}{2\sigma^2}\Big( - \mathbf{Z}^T\mathbf{R}_\delta^{-1}\dfrac{\partial\mathbf{R}}{\partial\theta_j}\mathbf{R}_\delta^{-1}\dfrac{\partial\mathbf{R}}{\partial\theta_i}\mathbf{R}_\delta^{-1}\mathbf{Z} \nonumber\\ &+\mathbf{Z}^T\mathbf{R}_\delta^{-1}\dfrac{\partial\mathbf{R}}{\partial\theta_i\partial\theta_j}\mathbf{R}_\delta^{-1}\mathbf{Z} \nonumber\\&- \mathbf{Z}^T\mathbf{R}_\delta^{-1}\dfrac{\partial\mathbf{R}}{\partial\theta_i}\mathbf{R}_\delta^{-1}\dfrac{\partial\mathbf{R}}{\partial\theta_j}\mathbf{R}_\delta^{-1}\mathbf{Z}\Big). 
\end{align}
Conversely, the second derivative on the right-hand side gives
\begin{equation}
    \dfrac{\partial}{\partial \theta_j} \dfrac{1}{2}\text{tr}\left( \mathbf{R}_{\delta}^{-1} \dfrac{\partial\mathbf{R}}{\partial\theta_i}  \right) =\dfrac{1}{2}
    \text{tr}\left( -\mathbf{R}_{\delta}^{-1}\dfrac{\partial\mathbf{R}}{\partial\theta_j}\mathbf{R}_{\delta}^{-1}\dfrac{\partial\mathbf{R}}{\partial\theta_i} + \mathbf{R}_{\delta}^{-1}\dfrac{\partial^2\mathbf{R}}{\partial\theta_i\partial\theta_j} \right).
\end{equation}
By rewriting our prior definition in Eqn.~\ref{priorGP} as \(\mathbf{Y}-\mathbf{M}\bm{\beta}=\mathbf{Z}\sim\mathcal{N}\left( \mathbf{0},\sigma^2\left(\mathbf{R}+\bm{\pi}\right) \right)\) we can conclude that \(\mathbb{E}\left( \mathbf{Z}\mathbf{Z}^T\right)=\sigma^2\left(\mathbf{R}+\bm{\pi} \right)\) as the covariance is defined as \(\text{Cov}\left( \mathbf{Z} \right) = \mathbb{E}\left( \mathbf{Z}\mathbf{Z}^T \right) - \mathbb{E}\left( \mathbf{Z} \right) \mathbb{E}\left( \mathbf{Z} \right)^T\) and \( \mathbb{E}\left( \mathbf{Z}\right) = \mathbf{0}\). Subsequently, by taking the negative expectation we retrieve the general expression for the expected Fisher information as given in Eqn.~\ref{fish_gen}. 

It should be noted that the addition of a nugget reduces the magnitude of the Fisher information by inflating the effective covariance. Under this parameterization, calibration and nugget parameters exhibit weak, but generally non-zero, coupling in expectation. 

To see this, consider the cross Fisher information between \( \lambda_i \) and \( \theta_j \),
\begin{equation}
    \mathcal{J}_{\lambda_i,\theta_j} = \dfrac{1}{2}\text{tr}\left( \mathbf{R}_{\delta}^{-1} \dfrac{\partial \mathbf{R}}{\partial \theta_j} \mathbf{R}_{\delta}^{-1} \mathbf{e}_i\mathbf{e}_i^T \right) = \dfrac{1}{2} \mathbf{e}_i^T \mathbf{R}_{\delta}^{-1} \dfrac{\partial \mathbf{R}}{\partial \theta_j} \mathbf{R}_{\delta}^{-1} \mathbf{e}_i,
\end{equation}
where we used \( \dfrac{\partial \mathbf{R}_{\delta}}{\partial \lambda_i}=\mathbf{e}_i\mathbf{e}_i^T\). 

Although \( \dfrac{\partial \mathbf{R}}{\partial \theta_j} \) has zero diagonal entries, multiplication by \( \mathbf{R}_\delta^{-1} \) on both sides introduces non-zero diagonal contributions through interactions across observations. Consequently, the cross Fisher information does not vanish in general. However, these terms are often small in practice due to the localized effect of nugget parameters and the structured sparsity of calibration derivatives.

Under the correlation parameterization \(R_{ii}=1\), calibration parameters influence the kernel only through pairwise distances, implying that \(\partial \mathbf{R}/\partial \theta_j\) has diagonals of zero. Consequently, the matrix \( \partial \mathbf{R}/\partial \theta_j \) is dominated by off-diagonal structure, while the nugget derivative is purely diagonal. Although the multiplication by \( \mathbf{R}_{\delta}^{-1} \) introduces mixing between rows and columns, the resulting diagonal entries in Eqn.~\ref{noise_theta_deriv} remain relatively small in magnitude. This implies weak coupling between calibration and nugget parameters in the expected Fisher information.

From a practical perspective, the nugget \(\bm{\lambda}\) primarily rescales marginal variances, whereas calibration parameters \(\bm{\theta}\) govern correlation structure. As a result, observations with lower experimental uncertainty contribute more strongly to the likelihood, and calibration is therefore driven primarily by low-noise experimental outputs.

\subsection{Fisher Information of Calibration Parameters for Squared Exponential and Mat\'ern Covariance Functions}
Given the general expression for the expected Fisher information in Eqn.~\ref{fish_gen}, we need to take the derivative of the matrix \(\mathbf{R}\) with respect to \(\bm{\theta}\) given the different correlation structures (i.e., Squared exponential and Mat\'ern).  Given the \((i,j)^{th}\) term of the covariance matrix is given as \( r_{i,j}=\exp\left(-\Delta(\mathbf{X}_i,\mathbf{X}_j) \right) \) the derivative with respect to the \( k^{th}\) calibration parameter is given as
\begin{equation}
    \dfrac{\partial r_{i,j}}{\partial \theta_k} = -r_{i,j}\dfrac{\partial\Delta_{i,j}}{\partial \theta_k},
\end{equation}
where \(\Delta_{ij}=(\mathbf{x}_i-\mathbf{x}_j)^T\mathbf{\Omega}_{\chi}(\mathbf{x}_i-\mathbf{x}_j) +(\bm{\theta}_i-\bm{\theta}_j)^T\bm{\Omega}_{\Theta}(\bm{\theta}_i-\bm{\theta}_j)+\left\|\mathbf{z}(\eta_i)-\mathbf{z}(\eta_j)\right\|_2^2\) and \(r_{i,j} =  r(\mathbf{X}_i,\mathbf{X}_j)\). As \(\theta_k\) only occurs in the second element on the right hand side of \(\Delta_{i,j}\) we can obtain the expression in Eqn.~\ref{deriv_SE}. 

Derivatives are non-zero only for pairs of observations whose correlation depends on the calibration parameters. In particular, for simulation outputs in \(\mathcal{S}_s\), calibration parameters are fixed and \(\partial \Delta_{i,j}/\partial \theta_k = 0\). Consequently, non-zero contributions arise primarily from pairs involving at least one calibrated experimental observation. This structure implies that the expected Fisher information is dominated by cross-correlations between calibrated and uncalibrated observations, rather than by within-simulation interactions.

For the Mat\'ern covariance the correlation is given as \( r_{i,j}=\dfrac{2^{1-\nu}}{\Gamma(\nu)}\kappa_{i,j}^{\nu}K_{\nu}(\kappa_{i,j}) \) where \(\kappa_{ij}=\sqrt{2\nu\Delta(\mathbf{X}_i,\mathbf{X}_j)}\). Differentiation with respect to calibration parameters proceeds via the chain rule, \(\partial r/\partial\theta_k = (\partial r/\partial \kappa_{i,j})(\partial \kappa_{ij} / \partial \Delta_{i,j})(\partial \Delta_{i,j}/\partial\theta_k)\), using the identity \( \dfrac{d \kappa^{\nu}K_{\nu}(\kappa)}{d\kappa} = \kappa^{\nu}K_{\nu-1}(\kappa) \) for the modified Bessel functions we can get the derivative of the \( (i,j)^{th} \) term of a Mat\'ern correlation matrix with respect to \(\theta_k\) as
\begin{equation}
    \dfrac{\partial r_{i,j}}{\partial \theta_k} =-\dfrac{\nu}{\kappa_{i,j}}\dfrac{2^{1-\nu}}{\Gamma(\nu)}\kappa_{i,j}^{\nu}K_{\nu-1}(\kappa_{i,j})\dfrac{\partial\Delta_{i,j}}{\partial \theta_k}.
    \label{matern_deriv_interim}
\end{equation}
The last term on the right-hand side of Eqn.~\ref{matern_deriv_interim} can be treated in the same way as the squared exponential and thus provides the expression in Eqn.~\ref{Fish_mat}. Although the derivative is taken with respect to a calibration parameter \(\bm{\theta}\), the Mat\'ern covariance depends on \(\bm{\theta}\) only through the induced distance \(\kappa \); hence differentiation proceeds via the chain rule using standard identities for modified Bessel functions \cite{abramowitz1948}.

Although Eq.~\ref{Fish_mat} appears to contain a singularity because of the factor \(1/\kappa_{i,j}\). his singularity is removable for all Mat\'ern kernels with smoothness parameter \( \nu > 1\). To show this, we can consider the following limit
\begin{equation}
L=\lim_{\kappa\rightarrow 0} \frac{\nu}{\kappa_{i,j}}\frac{2^{1-\nu}}{\Gamma(\nu)}\kappa_{i,j}^{\nu}K_{\nu-1}(\kappa),
\label{gen_limit}
\end{equation}
where we can use the small-argument asymptotic expansion of the modified Bessel function of the second kind (Eq. 10.30.2 in \cite{Milton2026}) that gives 
\begin{equation}
    K_{\nu-1}(\kappa) \sim \frac{1}{2}\Gamma(\nu-1)\left(\frac{\kappa}{2}\right)^{-(\nu-1)}, \quad \kappa\rightarrow 0,
    \label{DLMF}
\end{equation}
where \(\sim\) denotes asymptotic equivalence )i.e., \( f(x)\sim g(x) \text{ if }\lim_{x\rightarrow a}f(x)/g(x)=1\)). Rewriting Eq.\ref{DLMF} as \(\sim 2^{\nu-2}\Gamma(\nu-1)\kappa^{-(\nu-1)}\) and substituting into Eq.~\ref{gen_limit} yields
\begin{align}
L&= \lim_{\kappa\rightarrow 0} \nu\frac{2^{1-\nu}}{\Gamma(\nu)}\kappa_{i,j}^{\nu-1}K_{\nu-1}(\kappa_{i,j}),\nonumber \\
& = \nu\frac{2^{1-\nu}}{\Gamma(\nu)}2^{\nu-2}\Gamma(\nu-1),\nonumber \\
& =\frac{\nu}{2(\nu-1)}.
\end{align}
Consequently, the apparent singularity is removable and the derivative remains finite for \(\nu>1\). Furthermore, since \(\kappa_{i,j}=0\) if and only if \(\Delta(\mathbf{X}_i,\mathbf{X}_j)=0\), coincident inputs satisfy \(\theta_i = \theta_j\). Consequently, the multiplicative factor \(\left(\theta_i - \theta_j\right)\) in Eq.~\ref{Fish_mat} causes the derivative itself to vanish as \(\kappa \rightarrow 0\). Therefore the expected Fisher information remains well-defined at coincident inputs for all Mat\'ern kernels with \( \nu > 1\). For the battery experiments reported in this work, repeated inputs occur only when using the squared-exponential covariance, so this limiting case does not affect the reported numerical results. Moreover, the derivative is well-defined at coincident inputs, and numerical instability arises only from finite-precision arithmetic rather than from the analytical form of the derivative.

\subsection{Expected Fisher Information for all Hyperparameters}
\label{noise_theta_deriv}

Similar to the calibration parameters \(\bm{\theta}\), the derivatives of the covariance function with respect to the remaining hyperparameters \(\Xi = \{\bm{\omega}, \mathbf{A}, \bm{\lambda}\}\) can be expressed using the chain rule as
\begin{equation}
\frac{\partial r_{i,j}}{\partial \Psi}
=
\frac{\partial r_{i,j}}{\partial \Delta_{i,j}}
\cdot
\frac{\partial \Delta_{i,j}}{\partial \Xi},
\label{fish_deriiv_chain_rule}
\end{equation}
for parameters that enter through the distance metric \(\Delta_{ij}\). The first term depends on the covariance choice, while the second depends on the specific hyperparameters.

We proceed with the squared exponential covariance function, for which
\begin{equation}
\frac{\partial r_{i,j}}{\partial \Delta_{ij}} = -r_{i,j}.
\end{equation}

The remaining derivatives are given as follows:

\begin{enumerate}
\item{\textbf{Length-scale parameters:}} The derivative with respect to the length-scale parameters \(\omega_k\) is given by
\begin{equation}
\frac{\partial \Delta_{ij}}{\partial \omega_k}
=
\ln(10)\,10^{\omega_k}\,\xi_{ij}^{(k)},
\end{equation}
where
\begin{equation}
\xi_{ij}^{(k)} =
\begin{cases}
(x_{i,k}-x_{j,k})^2, & k = 1,\ldots,d, \\[4pt]
(\theta_{i,k-d}-\theta_{j,k-d})^2, & k = d+1,\ldots,d+p.
\end{cases}
\end{equation}

\item{\textbf{Embedding matrix parameters:}} The latent embedding term can be written as
\begin{equation} 
\left\|\mathbf{A} g(\eta_i)-\mathbf{A} g(\eta_j)\right\|_2^2
=
\left(g(\eta_i)-g(\eta_j)\right)^T
\mathbf{A}^T \mathbf{A}
\left(g(\eta_i)-g(\eta_j)\right),
\end{equation}
which induces a Mahalanobis-type metric in latent space. For the \((m,n)^{\text{th}}\) element of \(\mathbf A \in \mathbb{R}^{d_z \times q}\), we obtain
\begin{equation}
\frac{\partial \Delta_{i,j}}{\partial A_{mn}}
=
2\left(\mathbf A\left(g(\eta_i)-g(\eta_j)\right)\right)_m
\left(g_n(\eta_i)-g_n(\eta_j)\right).
\end{equation}

\item{\textbf{Nugget parameters:}} For nugget parameters \(\lambda_i\), the covariance matrix derivative is given by
\begin{equation}
\frac{\partial \mathbf{R}_\delta}{\partial \lambda_i}
=
\mathbf{e}_i \mathbf{e}_i^T,
\end{equation}
which affects only diagonal entries. In contrast, correlation, calibration, and embedding parameters primarily influence off-diagonal structure. Consequently, cross Fisher information terms involving nugget parameters are expected to be weaker than the corresponding within-group interactions because nugget parameters perturb only the diagonal of the covariance matrix, whereas the remaining hyperparameters primarily alter the correlation structure.
\end{enumerate}
Although the Fisher information does not strictly decouple across parameter groups, it often exhibits an approximately block-structured form in practice. This arises because correlation and embedding parameters act through the global distance metric, calibration parameters affect structured subsets of observations, and nugget parameters primarily influence marginal variances. As a result, cross-parameter interactions are typically weak, though not identically zero. A Matlab implementation of these functions can also be found in the online repository \footnote{URL link will be added after double blind review}.

To assess the validity of the proposed local Gaussian approximation, we performed two additional validation studies. The first investigates whether the Fisher information accurately captures the local posterior geometry of the calibration parameters. The second examines the approximation introduced by conditioning on the remaining hyperparameters at their maximum likelihood estimates. Together, these experiments provide empirical evidence for the assumptions underlying the proposed uncertainty quantification procedure.

We first consider Problem 2 from Tbl.~\ref{sample_fun_tbl}, which contains two calibration parameters and therefore permits direct visualization of the joint posterior distribution including a correlation term. All remaining hyperparameters were fixed at their maximum likelihood estimates, consistent with the approximation adopted in the main manuscript. Using 20 physical observations, 50 simulator evaluations, and the squared-exponential covariance function, we approximated the posterior distribution by Monte Carlo sampling and compared it with the Gaussian approximation obtained from the expected Fisher information.

\begin{figure}[ht]
\centering
\begin{tikzpicture}
\node[inner sep=0pt] (F) at (25mm,15mm){\includegraphics[width=46mm]{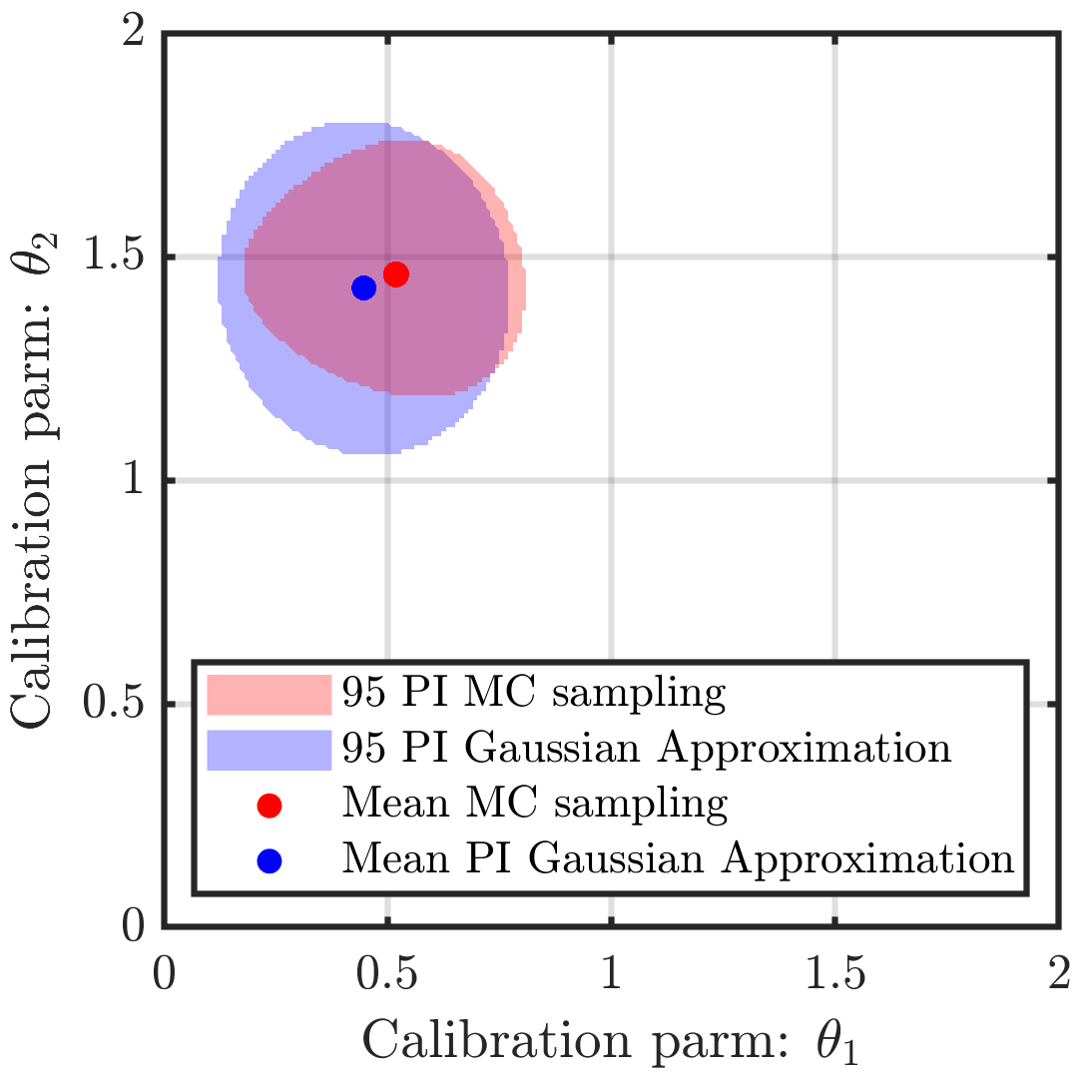}};

\end{tikzpicture}
\caption{Visualization of posterior distributions of calibration parameters for Problem 2 in Tbl.~\ref{sample_fun_tbl} approximate through Monte Carlo sampling and the Fisher approximation. }
\label{u_post_parm_comp}
\end{figure}

Fig.~\ref{u_post_parm_comp} compares the resulting 95\% credible regions, while Tbl.~\ref{posterior_comp} summarizes the corresponding posterior means, standard deviations, and parameter correlation. Although the two distributions are not identical, the Fisher approximation captures the dominant features of the posterior geometry, including its location, spread, and near-zero correlation between the calibration parameters. These results suggest that the local Gaussian approximation provides a reasonable characterization of the posterior uncertainty for this problem despite its computational simplicity.

\begin{table}
  \caption{Comparison of the first two statistical moments of the calibration parameters through Monte Carlo sampling and the Fisher Approximation for Problem 2 in Tbl.~\ref{sample_fun_tbl}.}
  \label{posterior_comp}
  \centering
  \begin{tabular}{lll}
    \toprule
   Quantity & Monte Carlo Posterior & Fisher Approximation \\
    \midrule
    Mean & \( \{0.5191,1.4607\}\) & \( \{0.4474, 1.4303\}\)\\
    Covariance & \(\{ 0.0167,    0.0139\}\) & \(\{ 0.0123, 0.0094 \}\) \\ 
    Correlation & \(-0.0011\) & \(-0.0006\) \\
    \bottomrule
  \end{tabular}
\end{table}

While the previous experiment validates the local Gaussian approximation, it does not address the approximation introduced by conditioning on the remaining hyperparameters. To investigate this effect, we consider Problem 1 from Tbl.~\ref{sample_fun_tbl}, which represents the lowest-dimensional calibration problem while still containing all classes of hyperparameters. This model contains two length-scale parameters, one nugget parameter, one calibration parameter, and one latent embedding parameter, yielding a total of five unknown hyperparameters.

The joint posterior distribution over all five hyperparameters was approximated using importance sampling. The proposal distribution was centered at the maximum likelihood estimate, with covariance determined from the local Fisher approximation. The resulting importance sampler achieved an effective sample size of approximately 1100, providing sufficient accuracy to estimate the posterior correlation matrix.

The estimated posterior correlation matrix is reported in Tbl.~\ref{posterior_comp_full_hyp}. Although moderate correlations exist among some nuisance hyperparameters, the calibration parameter exhibits only weak posterior dependence on the remaining hyperparameters, with a maximum absolute correlation of 0.11. These results suggest that conditioning on the nuisance hyperparameters introduces only a modest approximation error for this representative calibration problem.

\begin{table}
  \caption{Posterior correlation matrix estimated by importance sampling for all hyperparameters in Problem 1 from Tbl.~\ref{sample_fun_tbl}. Although the posterior is not perfectly block diagonal, the calibration parameter exhibits only weak correlation with the remaining hyperparameters (largest absolute correlation 0.11), supporting the conditional Fisher approximation adopted in this work.}
  \label{posterior_comp_full_hyp}
  \centering
  \begin{tabular}{r | rrrrr}
    \toprule
    & \(\omega_1\) & \(\omega_2\) & \( \lambda\) & \( \theta_1\) & \( A_{22}\) \\
    \midrule 
    \(\omega_1\) & 1.00 & 0.08 & 0.06 & 0.11 & -0.13 \\
    \(\omega_2\) & 0.08 & 1.00 & 0.04 &-0.05 & 0.23\\
    \( \lambda\) & 0.06 & 0.04 & 1.00 & 0.05 & 0.06 \\
    \(\theta_1\) & 0.11 &-0.05 & 0.05 & 1.00 &-0.09\\
    \( A_{22}\)  &-0.13 & 0.23 & 0.06 &-0.09 & 1.00\\
    \bottomrule
  \end{tabular}
\end{table}


\section{Test Problems and Performance Metrics}
\label{example_form}
The test problems used in Sec.~\ref{results} are presented in Tbl.~\ref{sample_fun_tbl}. These problems have been established to systematically test the effects of experimental uncertainty, the number of physical data sources, and the treatment of experimental variance in the Bayesian calibration and bias correction model. 

\begin{table}
  \caption{Test problems to validate the Bayesian calibration and bias correction method.}
  \label{sample_fun_tbl}
  \centering
  \begin{tabular}{cl}
    \toprule
    Problem Nr. & Function and modeling parameters \\
    \midrule
    1& \(f_1^{(s)}(x,\theta)=(6x-2)^2\sin\left(12x-4\right)+22\sin\left(\theta\pi\left(2x-\frac{1}{2}\right)\right)^2\), \\
     & \(f_1^{(e)}(x)=(6x-2)^2\sin\left(12x-4\right)+11 +\mathcal{N}(0,\bar{\varepsilon})\), \\
     & \(x_{min}=0,\quad x_{max}=1,\quad \theta_{min}=-0.25,\quad\theta_{max}=0.25,\quad\theta_{true}=0\) \\
     2 & \(f_1^{(s)}(\mathbf{x},\bm{\theta})=\cos\left(\frac{1}{2}\pi x_1\theta_1\right)\cos\left(\frac{1}{2}\pi x_2\right)\theta_2 \), \\
       & \(f_2^{(s)}(\mathbf{x},\bm{\theta})=\cos\left(\frac{1}{2}\pi x_1\theta_1\right)\sin\left(\frac{1}{2}\pi x_2\right)\theta_2 \), \\
       & \(f_3^{(s)}(\mathbf{x},\bm{\theta})=x_1\sin\left(\frac{1}{2}\pi x_2\right)\theta_2 \), \\
       & \(f_1^{(e)}(\mathbf{x})=\cos\left(\frac{1}{2}\pi x_1\right)\cos\left(\frac{1}{2}\pi x_2\right)+x_1x_2+\mathcal{N}(0,\bar{\varepsilon}_1) \), \\
       & \(f_2^{(e)}(\mathbf{x})=\cos\left(\frac{1}{2}\pi x_1\right)\sin\left(\frac{1}{2}\pi x_2\right)+x_2 +\mathcal{N}(0,\bar{\varepsilon}_2) \), \\
       & \(f_3^{(e)}(\mathbf{x})=x_1\sin\left(\frac{1}{2}\pi x_2\right)-x_1-x_2+\mathcal{N}(0,\bar{\varepsilon}_3) \), \\
       & \(\mathbf{x}_{min}=\{0,0\}^T,\quad \mathbf{x}_{max}=\{1,1\}^T\),\\
       &\(\bm{\theta}_{min}=\{ 0, 0\}^T,\quad\bm{\theta}_{max}=\{2,2 \}^T ,\quad\bm{\theta}_{true}=\{1,1 \}^T\), \\
       3 & \( f_1^{(s)}(\mathbf{x},\bm{\theta})=\left(1+\theta_1^3\sum_{i=3}^5(x_i-\frac{1}{2})^2\right)\cos\left(\frac{1}{2}\pi(x_1+\theta_2)\right)\cos\left(\frac{1}{2}\pi x_2\right)+2\theta_2\), \\
       & \( f_2^{(s)}(\mathbf{x},\bm{\theta})=\left(1+\theta_1^3\sum_{i=3}^5(x_i-\frac{1}{2})^2\right)\cos\left(\frac{1}{2}\pi(x_1+\theta_2)\right)\sin\left(\frac{1}{2}\pi x_2\right) - 1.5\theta_2\), \\
       & \( f_3^{(s)}(\mathbf{x},\bm{\theta})=\left(1+\theta_1^3\sum_{i=3}^5(x_i-\frac{1}{2})^2\right)\sin\left(\frac{1}{2}\pi(x_1+\theta_2)\right) + \frac{1}{2}\theta_2\), \\      
       & \( f_1^{(e)}(\mathbf{x})=\frac{6}{5}\left(1 + \sum_{i=3}^5(x_i-\frac{1}{2})^2\right)\cos\left(\frac{1}{2}\pi(x_1)\right)\cos\left(\frac{1}{2}\pi x_2\right)+\mathcal{N}(0,\bar{\varepsilon}_1)\), \\
       & \( f_2^{(e)}(\mathbf{x})=\frac{4}{5}\left(1+\sum_{i=3}^5(x_i-\frac{1}{2})^2\right)\cos\left(\frac{1}{2}\pi(x_1)\right)\sin\left(\frac{1}{2}\pi x_2\right)+\mathcal{N}(0,\bar{\varepsilon}_2) \), \\
       & \( f_3^{(e)}(\mathbf{x})=-\left(1+\sum_{i=3}^5(x_i-\frac{1}{2})^2\right)\sin\left(\frac{1}{2}\pi(x_1)\right)+\mathcal{N}(0,\bar{\varepsilon}_3)\), \\
       & \(\mathbf{x}_{min}=\{0,0,0,0,0\}^T,\quad \mathbf{x}_{max}=\{1,1,1,1,1\}^T\),\\
       &\(\bm{\theta}_{min}=\{ 0, -0.5, -0.5\}^T,\quad\bm{\theta}_{max}=\{2,0.5,0.5 \}^T ,\quad\bm{\theta}_{true}=\{1,0,0 \}^T\). \\
    \bottomrule
  \end{tabular}
\end{table}

To test the performance of the developed calibration model on the above test problems, we use a set of \(n_{test}\) validation metrics \(\mathbf{X}^{(test)}=\left\{\mathbf{X}^{(test)}_1,\ldots,\mathbf{X}^{(test)}_{n_{test}} \right\}^T,\quad\mathbf{Y}^{(test)}=\left\{Y^{(test)}_1,\ldots,Y^{(test)}_{n_{test}} \right\}^T\) to generate the following set of three metrics. First, to test the predictive performance of the model we define the NRMSE as 
\begin{equation}
\gamma_{MSE}\left(\mathbf{X}^{(t)},\mathbf{Y}^{(t)}\right) = \frac{1}{\text{std}\left(\textbf{Y}^{(t)}\right)}\sqrt{\frac{1}{n_{t}}\sum_{i=1}^{n_{t}}\left(Y_i^{(t)} -\mu\left(\mathbf{X}^{(t)}_i\right)\right)^2},
\end{equation}
where \(\text{std}(\cdot)\) is the standard deviation. The metric \(\gamma_{MSE}(\cdot)\) provides an estimation for central predictive performance of a data-driven model. However, an advantage of a Bayesian model is that prediction comes in the form of a distribution. Consequently, to asses the accuracy of the posterior predictive distributions we use the NIS given as
\begin{align}
    \gamma_{NIS}\left(\mathbf{X}^{(t)},\mathbf{Y}^{(t)} \right) = &
    \frac{1}{\text{std}\left(\textbf{Y}^{(t)}\right)}\Bigg( \frac{1}{n_{test}}\sum_{i=1}^{n_{t}}\left(U\left(\textbf{X}^{(t)}\right)-L\left(\textbf{X}^{(t)}\right) \right)\nonumber\\
    & \quad +\frac{2\left( L\left(\textbf{X}^{(t)}\right)-Y_i^{(t)} \right) }{1-\alpha}\mathbf{1}_{Y_i^{(t)}<L\left(\textbf{X}^{(t)}\right)} \nonumber \\
    &\quad +\frac{2\left(Y_i^{(t)}-U\left(\textbf{X}^{(t)}\right) \right)}{1-\alpha}\mathbf{1}_{Y_i^{(t)}>U\left(\textbf{X}^{(t)}\right)}\Bigg)
\end{align}
where \(\alpha\) is the confidence interval at which we evaluate performance. Here we use \(\alpha=95\%\) so that \( L\left(\textbf{X}^{(t)}\right) = \mu\left(\textbf{X}^{(t)}\right)-1.96 s\left(\textbf{X}^{(t)}\right) \) and \( U\left(\textbf{X}^{(t)}\right) = \mu\left(\textbf{X}^{(t)}\right) + 1.96 s\left(\textbf{X}^{(t)}\right) \), respectively. 

Because the posterior distributions are Gaussian but problem dependent in scale, direct comparison across calibration problems is difficult. For each calibration trial we use the true calibration parameters \(\bm{\theta}_{true}\) to compute the standardized residual \(u_i=\frac{\hat{\theta}_i - \theta_{true}}{\hat{\sigma}_i}\) where \(\hat{\theta}_i\) and  \(\hat{\sigma}_i\) denote the posterior mean and posterior standard deviation obtained for the \(i^{th}\)  calibration experiment. If the Gaussian approximation provides a well-calibrated description of the posterior uncertainty, then the standardized residuals should follow a standard normal distribution. We therefore compare their empirical cumulative distribution function with the standard normal cumulative distribution function using the Wasserstein-1 distance \cite{ferson2008} as
\begin{equation}
    \gamma_d = \int_{-\infty}^{\infty} \left| \hat{F}_u(z) - \Phi(z)\right|\text{d}z,
\end{equation}
where \(\Phi(\cdot)\) denotes the standard normal cumulative distribution function and 
\begin{equation}
    \hat{F}_u(z) = \frac{1}{M} \sum_{i=1}^M \mathbf{1}(u_i\leq z),
\end{equation}
is the empirical cumulative distribution function calculated over \(M\) repeated experiments. 

\begin{figure}[ht]
\centering
\begin{tikzpicture}
\node[inner sep=0pt] (F) at (25mm,15mm){\includegraphics[width=46mm]{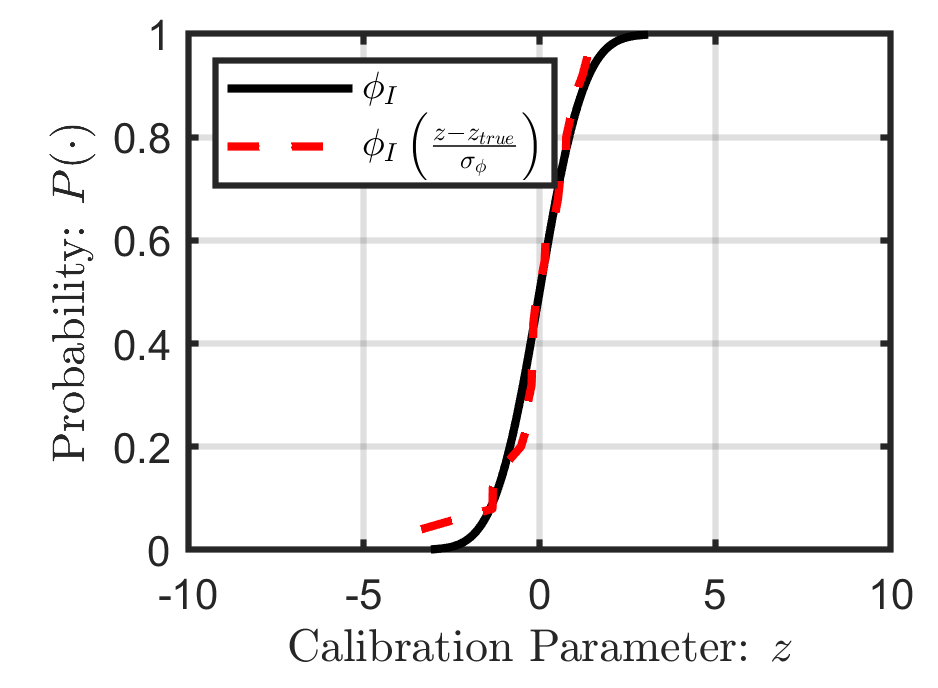}};
\node[inner sep=0pt] (F) at (72mm,15mm){\includegraphics[width=46mm]{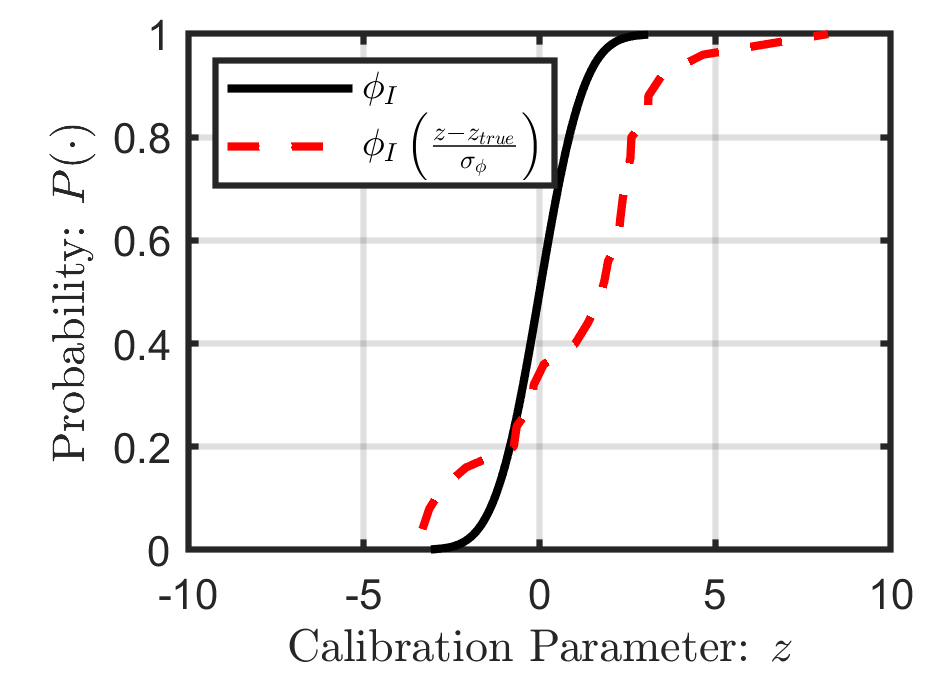}};
\node[inner sep=0pt] (F) at (119mm,15mm){\includegraphics[width=46mm]{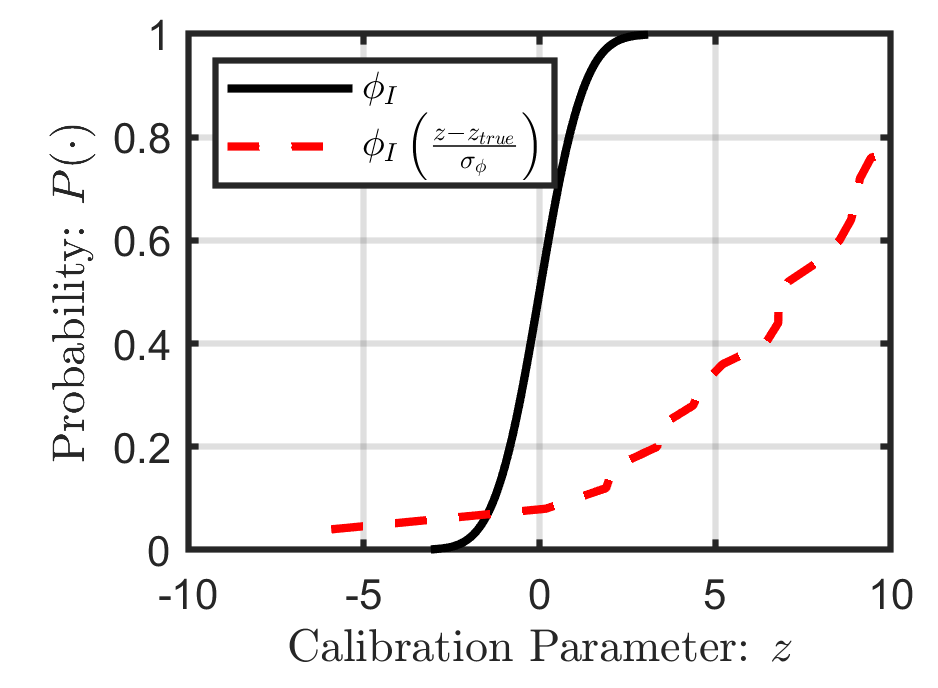}};

\draw[ line width=0.2mm, fill = white] (43.5mm,28.5mm) --++ (0,4mm) --++ (4mm,0mm) --++ (0mm,-4mm) -- cycle;
\node[mynode, anchor=center, align=center, text width = 4mm] at (45.5mm,30.5mm) {A};

\draw[ line width=0.2mm, fill = white] (90.5mm,28.5mm) --++ (0,4mm) --++ (4mm,0mm) --++ (0mm,-4mm) -- cycle;
\node[mynode, anchor=center, align=center, text width = 4mm] at (92.5mm,30.5mm) {B};

\draw[ line width=0.2mm, fill = white] (137.5mm,28.5mm) --++ (0,4mm) --++ (4mm,0mm) --++ (0mm,-4mm) -- cycle;
\node[mynode, anchor=center, align=center, text width = 4mm] at (139.5mm,30.5mm) {C};

\draw[ line width=0.2mm, fill = white, white] (38.3mm,-1mm) --++ (0,2mm) --++ (4mm,0mm) --++ (0mm,-2mm) -- cycle;

\draw[ line width=0.2mm, fill = white, white] (85.3mm,-1mm) --++ (0,2mm) --++ (4mm,0mm) --++ (0mm,-2mm) -- cycle;

\draw[ line width=0.2mm, fill = white, white] (132.3mm,-1mm) --++ (0,2mm) --++ (4mm,0mm) --++ (0mm,-2mm) -- cycle;

\end{tikzpicture}
\caption{Comparison of normalized predictive communicative density functions of calibration parameters over 25 trials. A) example of a strong adherence observed for \(\theta_3\) using \(\bar{\varepsilon} = 0.25\), and constant nugget parameter, B) moderate adherence observed for  \(\theta_1\) using \(\bar{\varepsilon} = 0.05\), and a flexible nugget parameter, and C) poor adherence observed for \(\theta_1\) using \(\bar{\varepsilon} = 0.05\), and a flexible nugget parameter.}
\label{u_pool_fig}
\end{figure}

Figure~\ref{u_pool_fig} illustrates the normalized cumulative distributions for three representative cases from Test Problem~3. Panel~A shows a well-identified parameter \(\theta_3\) obtained with \(\bar{\varepsilon} = 0.25\), and a constant nugget parameter, yielding \( \gamma_d=0.175\) which indicates relatively strong agreement with the standard normal reference distribution (solid black). Panel~C presents an example of poor agreement for \(\theta_1\) obtained with \(\bar{\varepsilon} = 0.05\) and a flexible nugget parameter, resulting in \( \gamma_d=7.30 \). Finally, Panel~B shows an intermediate case for \(\theta_2\), obtained with \(\bar{\varepsilon} = 0.10\) and the smallest admissible nugget parameter, where  \( \gamma_d=1.58\)  reflects moderate agreement.

\section{Identifiability with Single and Multiple Responses and Experimental Uncertainty}
\label{additiona_results_diff_outputs_appx}
\begin{figure}[t]
\centering
\begin{tikzpicture}
\node[inner sep=0pt] (F) at (20mm,28mm){\includegraphics[width=51mm]{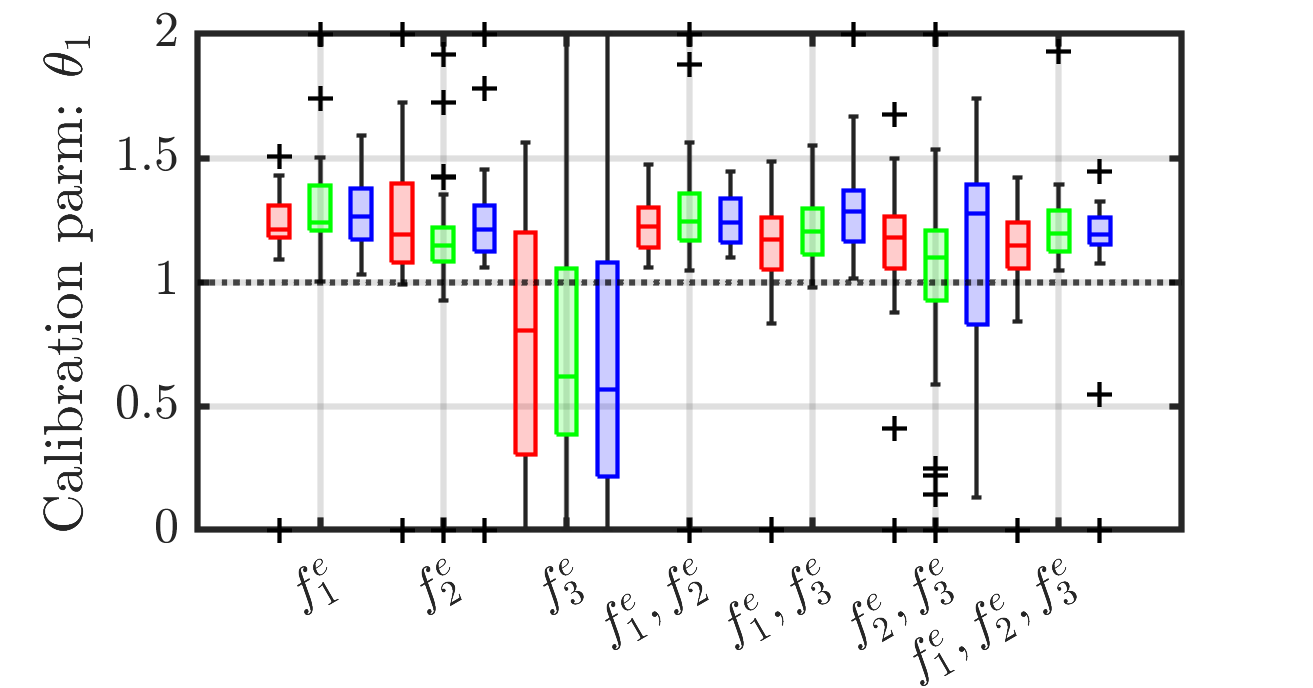}};
\node[inner sep=0pt] (F) at (67mm,28mm){\includegraphics[width=51mm]{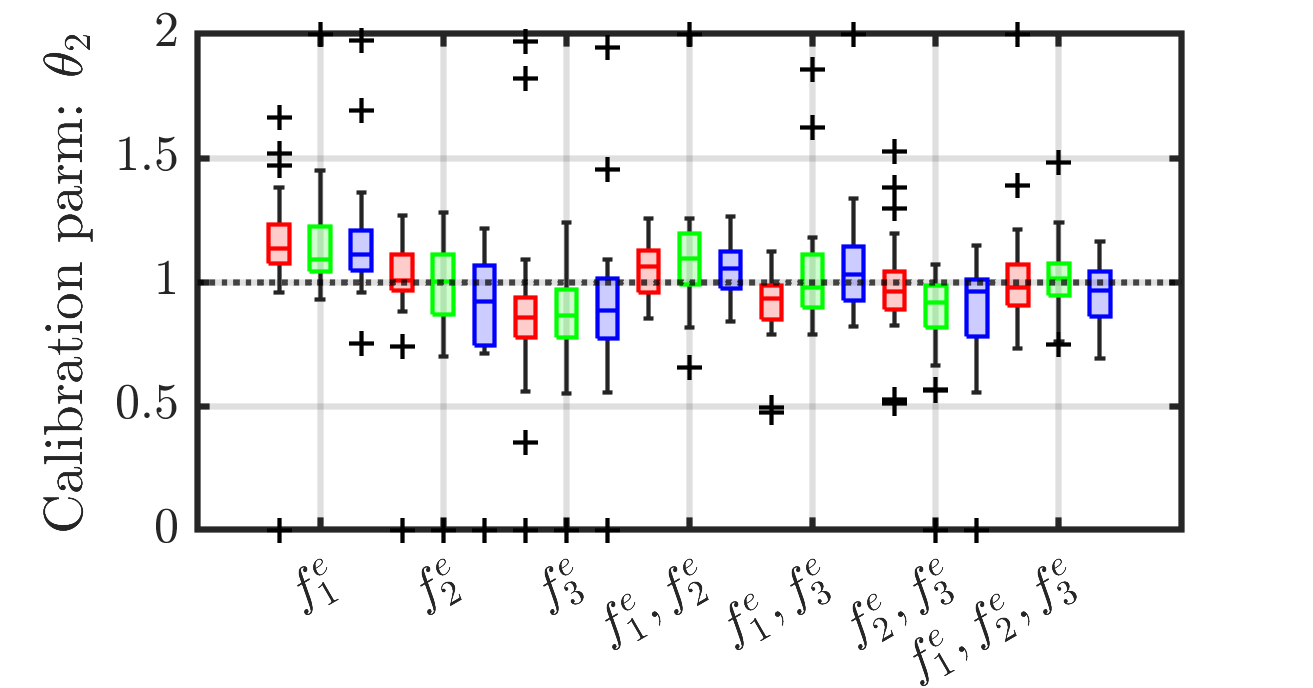}};
\node[inner sep=0pt] (F) at (108mm,30mm){\includegraphics[width=25mm]{Figures/legend2.png}};

\node[inner sep=0pt] (F) at (20mm,0mm){\includegraphics[width=51mm]{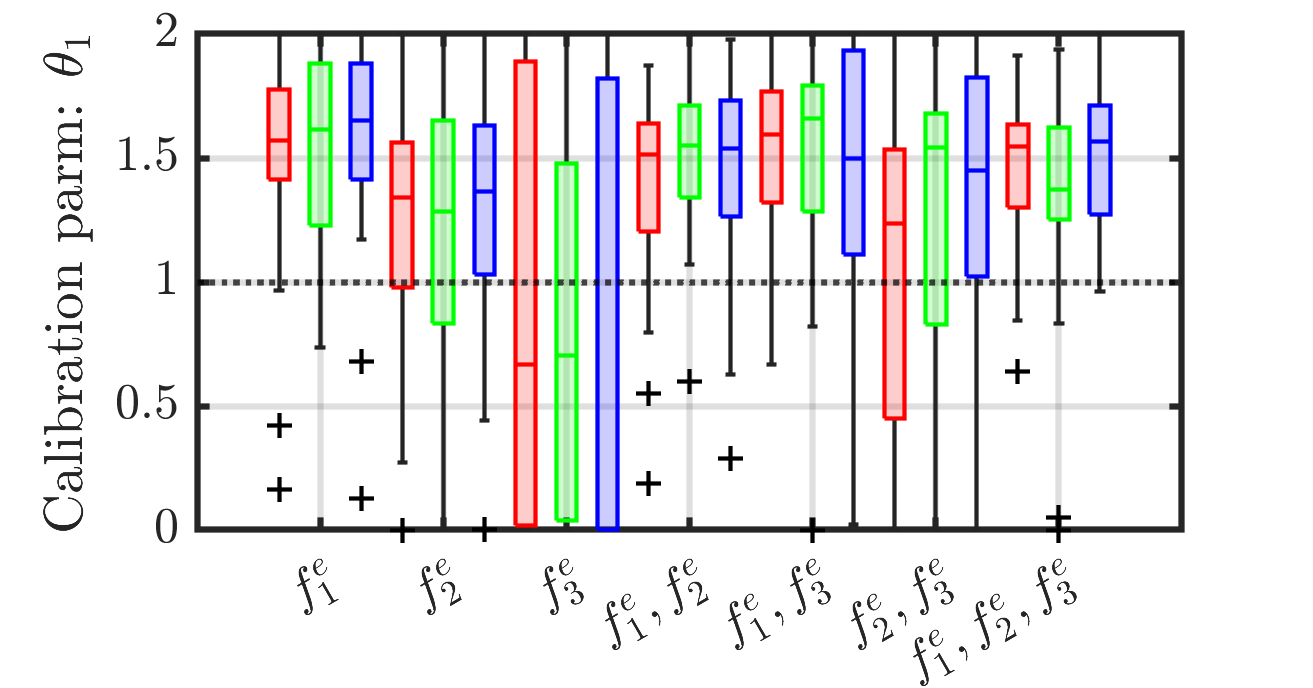}};
\node[inner sep=0pt] (F) at (67mm,0mm){\includegraphics[width=51mm]{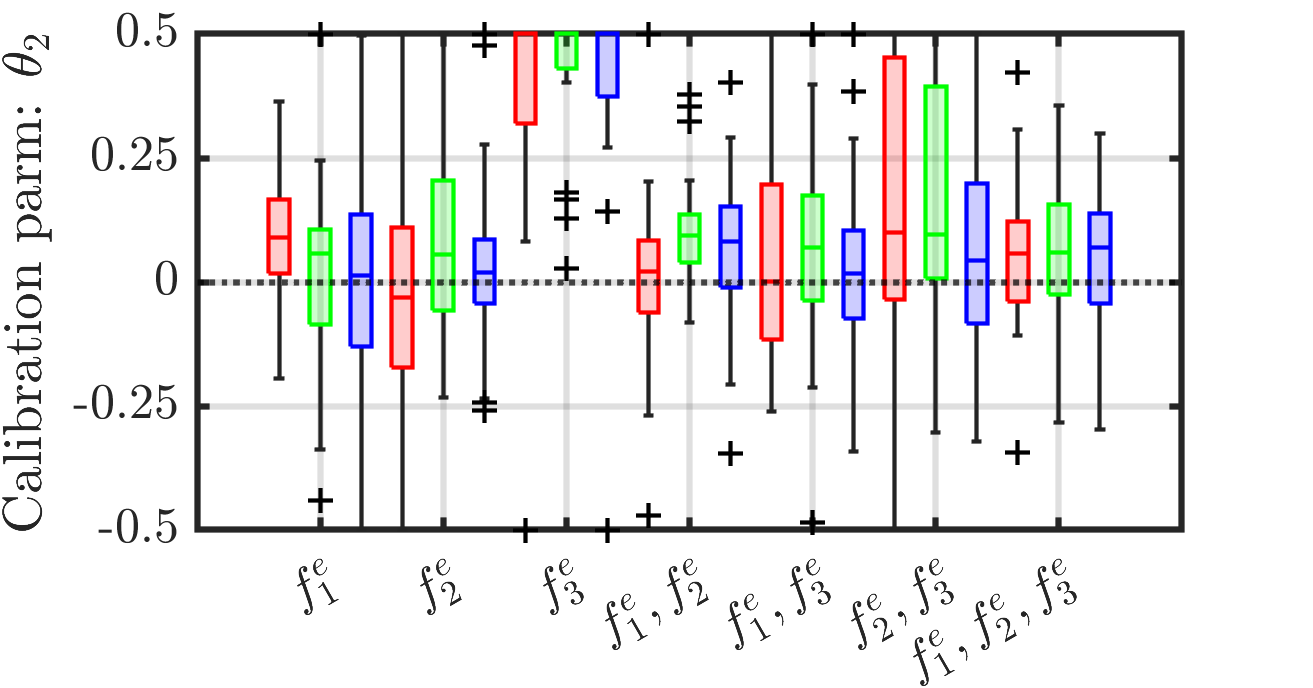}};
\node[inner sep=0pt] (F) at (114mm,0mm){\includegraphics[width=51mm]{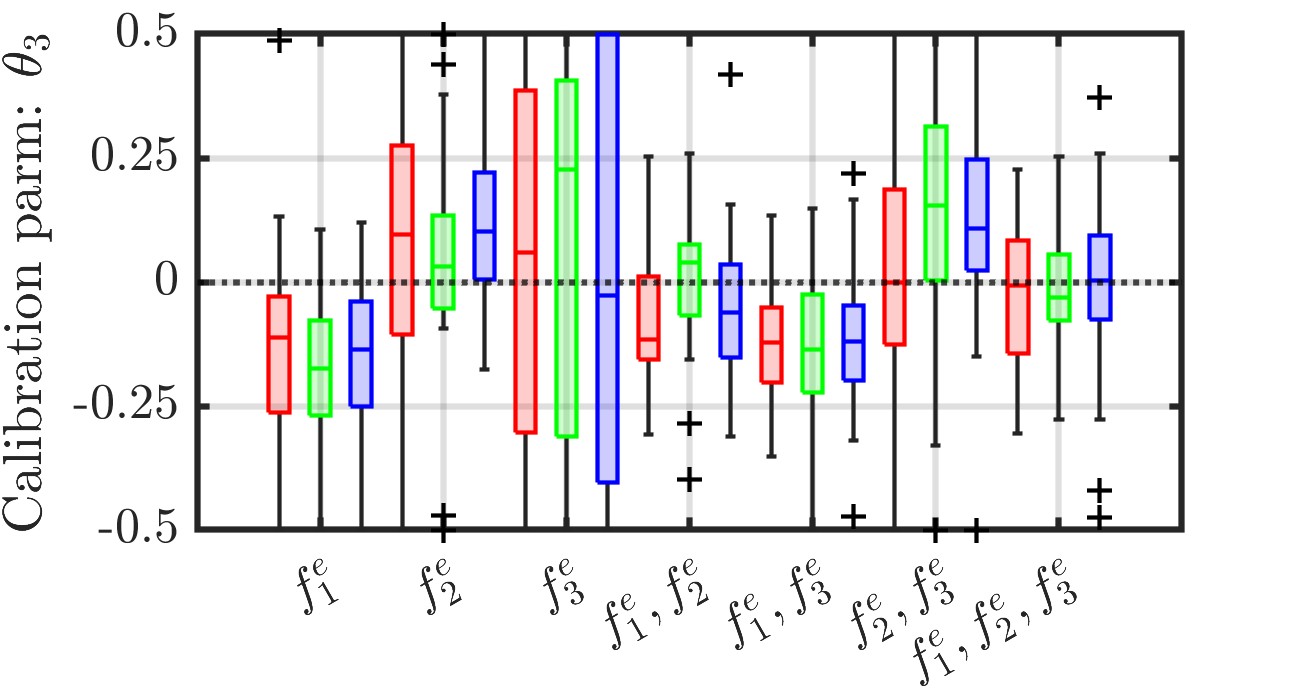}};

\draw[ line width=0.2mm, fill = white] (2mm,38mm) --++ (0,4mm) --++ (4mm,0mm) --++ (0mm,-4mm) -- cycle;
\node[mynode, anchor=center, align=center, text width = 4mm] at (2mm+2mm,38mm+2mm) {A};

\draw[ line width=0.2mm, fill = white] (49.25mm,38mm) --++ (0,4mm) --++ (4mm,0mm) --++ (0mm,-4mm) -- cycle;
\node[mynode, anchor=center, align=center, text width = 4mm] at (49.25mm+2mm,38mm+2mm) {B};

\draw[ line width=0.2mm, fill = white] (2mm,10mm) --++ (0,4mm) --++ (4mm,0mm) --++ (0mm,-4mm) -- cycle;
\node[mynode, anchor=center, align=center, text width = 4mm] at (2mm+2mm,10mm+2mm) {C};

\draw[ line width=0.2mm, fill = white] (49.25mm,10mm) --++ (0,4mm) --++ (4mm,0mm) --++ (0mm,-4mm) -- cycle;
\node[mynode, anchor=center, align=center, text width = 4mm] at (49.25mm+2mm,10mm+2mm) {D};

\draw[ line width=0.2mm, fill = white] (96mm,10mm) --++ (0,4mm) --++ (4mm,0mm) --++ (0mm,-4mm) -- cycle;
\node[mynode, anchor=center, align=center, text width = 4mm] at (96mm+2mm,10mm+2mm) {E};

\end{tikzpicture}
\caption{Calibration results for varying numbers of outputs (\(\bar{\varepsilon}=0.0\); 25 runs; ‘+’ denotes outliers). A,B) parameter estimates for problem 2; C–E) parameter estimates for problem 3. Increasing the number of outputs selectively improves identifiability, with anisotropic gains across parameters.}
\label{exp_num_out_results_n0}
\end{figure}

\begin{figure}[t]
\centering
\begin{tikzpicture}
\node[inner sep=0pt] (F) at (20mm,28mm){\includegraphics[width=51mm]{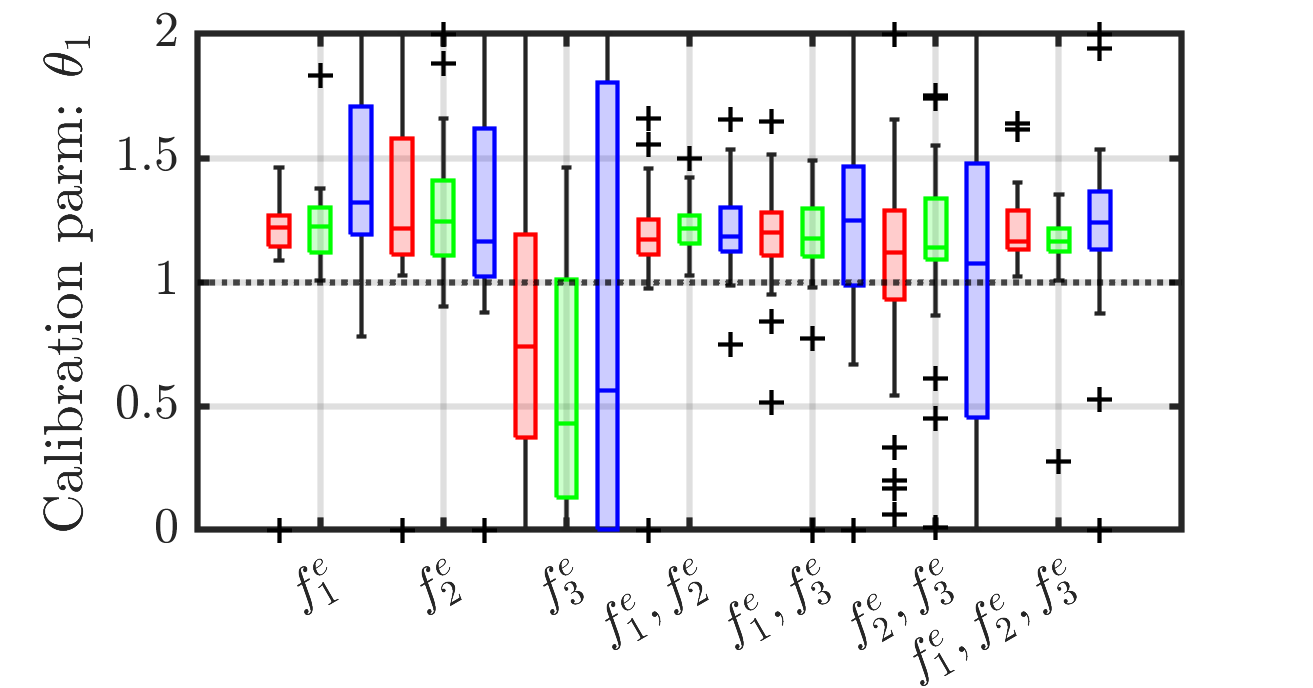}};
\node[inner sep=0pt] (F) at (67mm,28mm){\includegraphics[width=51mm]{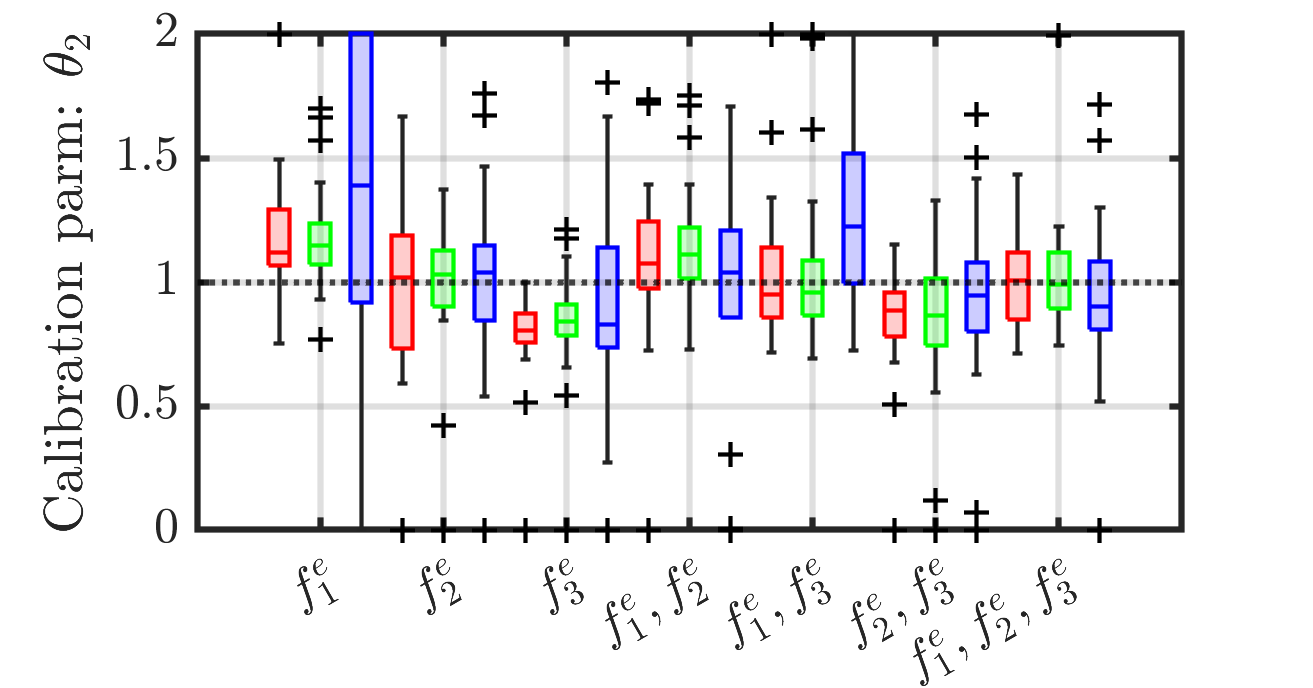}};
\node[inner sep=0pt] (F) at (108mm,30mm){\includegraphics[width=25mm]{Figures/legend2.png}};

\node[inner sep=0pt] (F) at (20mm,0mm){\includegraphics[width=51mm]{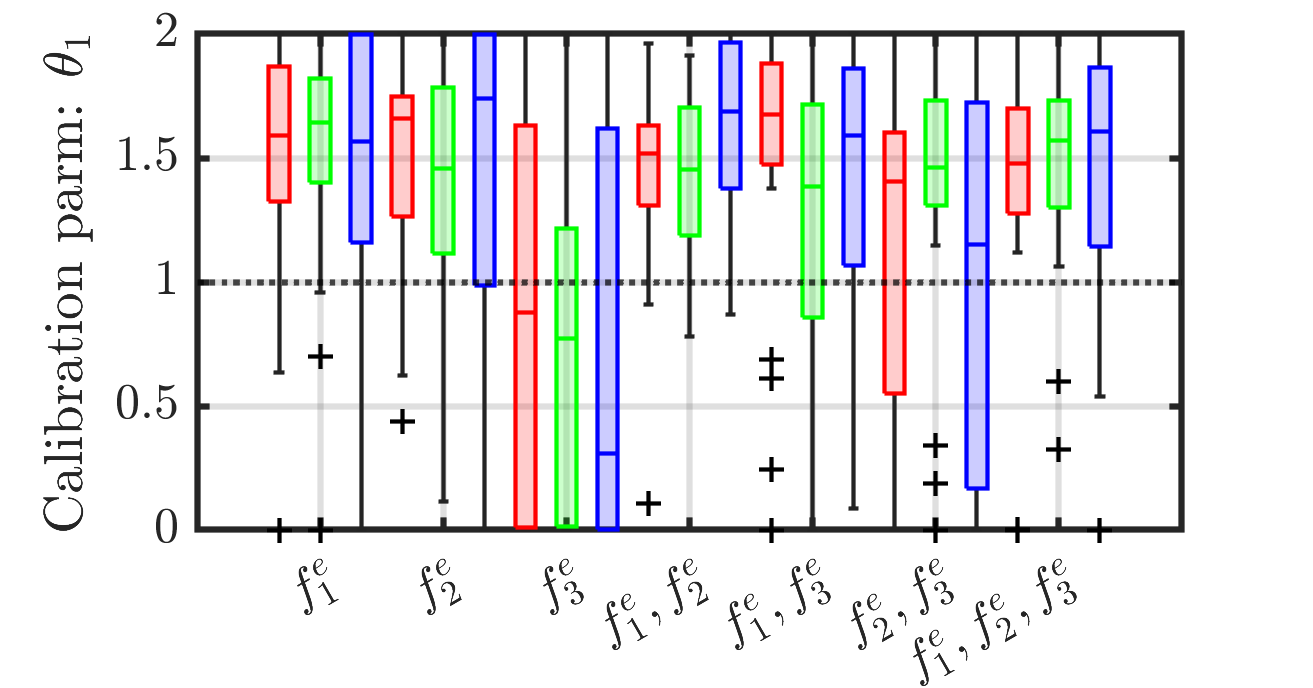}};
\node[inner sep=0pt] (F) at (67mm,0mm){\includegraphics[width=51mm]{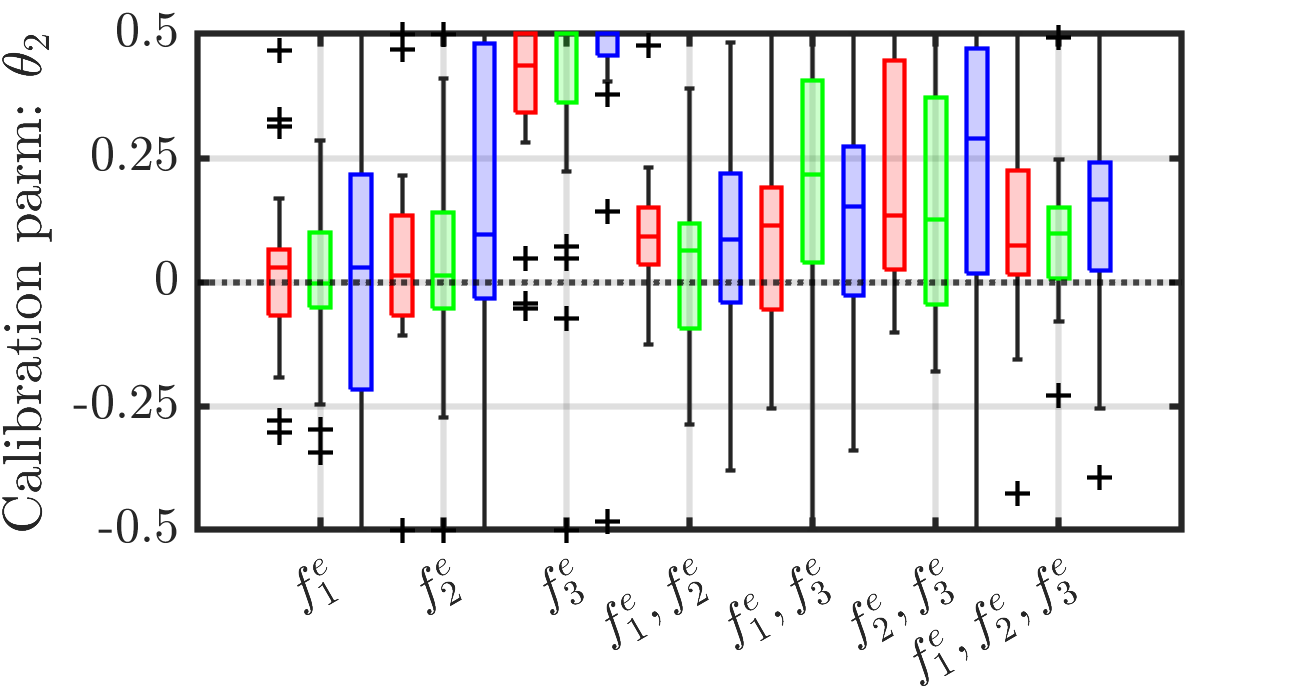}};
\node[inner sep=0pt] (F) at (114mm,0mm){\includegraphics[width=51mm]{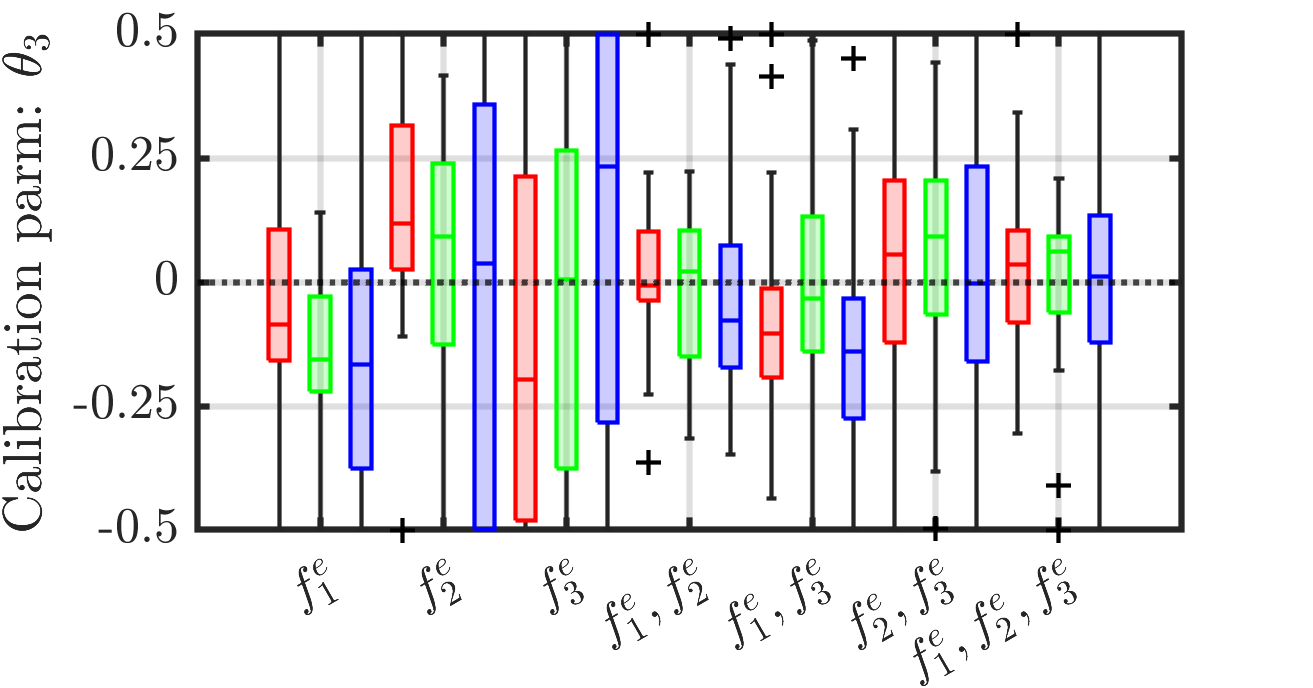}};

\draw[ line width=0.2mm, fill = white] (2mm,38mm) --++ (0,4mm) --++ (4mm,0mm) --++ (0mm,-4mm) -- cycle;
\node[mynode, anchor=center, align=center, text width = 4mm] at (2mm+2mm,38mm+2mm) {A};

\draw[ line width=0.2mm, fill = white] (49.25mm,38mm) --++ (0,4mm) --++ (4mm,0mm) --++ (0mm,-4mm) -- cycle;
\node[mynode, anchor=center, align=center, text width = 4mm] at (49.25mm+2mm,38mm+2mm) {B};

\draw[ line width=0.2mm, fill = white] (2mm,10mm) --++ (0,4mm) --++ (4mm,0mm) --++ (0mm,-4mm) -- cycle;
\node[mynode, anchor=center, align=center, text width = 4mm] at (2mm+2mm,10mm+2mm) {C};

\draw[ line width=0.2mm, fill = white] (49.25mm,10mm) --++ (0,4mm) --++ (4mm,0mm) --++ (0mm,-4mm) -- cycle;
\node[mynode, anchor=center, align=center, text width = 4mm] at (49.25mm+2mm,10mm+2mm) {D};

\draw[ line width=0.2mm, fill = white] (96mm,10mm) --++ (0,4mm) --++ (4mm,0mm) --++ (0mm,-4mm) -- cycle;
\node[mynode, anchor=center, align=center, text width = 4mm] at (96mm+2mm,10mm+2mm) {E};

\end{tikzpicture}
\caption{Calibration results for varying numbers of outputs (\(\bar{\varepsilon}=0.05\); 25 runs; ‘+’ denotes outliers). A,B) parameter estimates for problem 2; C–E) parameter estimates for problem 3. Increasing the number of outputs selectively improves identifiability, with anisotropic gains across parameters.}
\label{exp_num_out_results_n05}
\end{figure}

\begin{figure}[t]
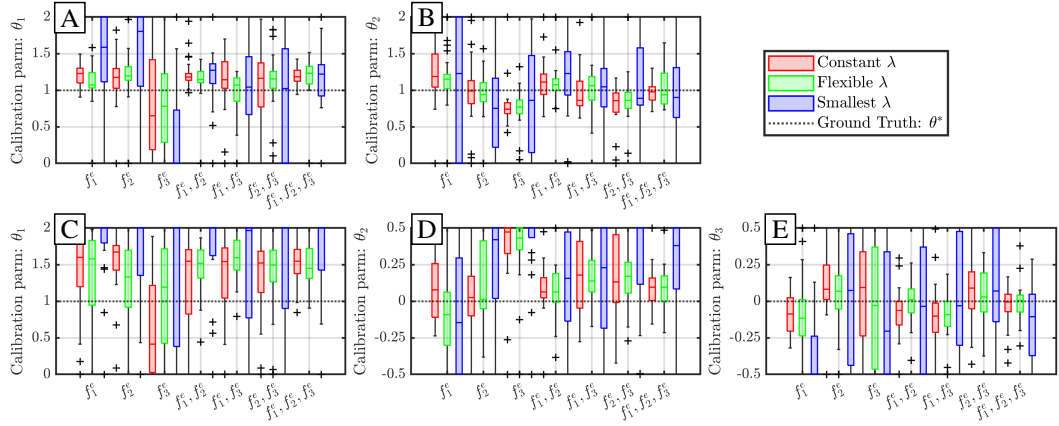

\centering
\begin{tikzpicture}
\node[inner sep=0pt] (F) at (20mm,28mm){\includegraphics[width=51mm]{Figures/problem1_theta1_diff_out_n20.png}};
\node[inner sep=0pt] (F) at (67mm,28mm){\includegraphics[width=51mm]{Figures/problem1_theta2_diff_out_n20.png}};
\node[inner sep=0pt] (F) at (108mm,30mm){\includegraphics[width=25mm]{Figures/legend2.png}};

\node[inner sep=0pt] (F) at (20mm,0mm){\includegraphics[width=51mm]{Figures/problem2_theta1_diff_out_n20.png}};
\node[inner sep=0pt] (F) at (67mm,0mm){\includegraphics[width=51mm]{Figures/problem2_theta2_diff_out_n20.png}};
\node[inner sep=0pt] (F) at (114mm,0mm){\includegraphics[width=51mm]{Figures/problem2_theta3_diff_out_n20.png}};

\draw[ line width=0.2mm, fill = white] (2mm,38mm) --++ (0,4mm) --++ (4mm,0mm) --++ (0mm,-4mm) -- cycle;
\node[mynode, anchor=center, align=center, text width = 4mm] at (2mm+2mm,38mm+2mm) {A};

\draw[ line width=0.2mm, fill = white] (49.25mm,38mm) --++ (0,4mm) --++ (4mm,0mm) --++ (0mm,-4mm) -- cycle;
\node[mynode, anchor=center, align=center, text width = 4mm] at (49.25mm+2mm,38mm+2mm) {B};

\draw[ line width=0.2mm, fill = white] (2mm,10mm) --++ (0,4mm) --++ (4mm,0mm) --++ (0mm,-4mm) -- cycle;
\node[mynode, anchor=center, align=center, text width = 4mm] at (2mm+2mm,10mm+2mm) {C};

\draw[ line width=0.2mm, fill = white] (49.25mm,10mm) --++ (0,4mm) --++ (4mm,0mm) --++ (0mm,-4mm) -- cycle;
\node[mynode, anchor=center, align=center, text width = 4mm] at (49.25mm+2mm,10mm+2mm) {D};

\draw[ line width=0.2mm, fill = white] (96mm,10mm) --++ (0,4mm) --++ (4mm,0mm) --++ (0mm,-4mm) -- cycle;
\node[mynode, anchor=center, align=center, text width = 4mm] at (96mm+2mm,10mm+2mm) {E};

\end{tikzpicture}
\caption{Calibration results with respect the different outputs for a constant degree of experimental uncertainty of \(\bar{\varepsilon} = 0.25\) over 25 repeated experiments with `+' indicating outliers. A,B) calibration parameter estimation for problem 2, and C-E) calibration parameter estimation for problem 3.}
\label{exp_num_out_results_n20}
\end{figure}

We present additional calibration results for Problems 2 and 3 across varying levels of experimental uncertainty, including \(\bar{\bm{\varepsilon}}=0\), small \(\bar{\bm{\varepsilon}}=0.05\), and large noise \(\bar{\bm{\varepsilon}}=0.25\). The corresponding results are shown in Figs.~\ref{exp_num_out_results_n0},~\ref{exp_num_out_results_n05},~\ref{exp_num_out_results_n20} and Tbls.~\ref{exp_num_out_results_n0_tbl},~\ref{exp_num_out_results_n05_tbl},~\ref{exp_num_out_results_n20_tbl}. Consistent with Sec.~\ref{results}, the Min treatment yields unstable parameter estimates but improved distance metrics relative to Const and Flex.

\begin{table}[t]
\caption{Calibration performance for Problems 2 and 3 at fixed experimental variance (\(\bar{\varepsilon}=0\)) across different output sets. Const and Flex yield the best point-wise predictive accuracy, while Min achieves the best identifiability (\( \gamma_d\)). Bold indicates the best value per metric and output set.}
\label{exp_num_out_results_n0_tbl}
\centering

\setlength{\tabcolsep}{1.8pt}
\renewcommand{\arraystretch}{1}

{\scriptsize
\begin{tabular}{cc cccc cccc cccc cccc cccc ccc}
\hline
&& \multicolumn{11}{c}{\textbf{Problem 2}} && \multicolumn{11}{c}{\textbf{Problem 3}} \\

\cline{3-13} \cline{15-25}

&& \multicolumn{3}{c}{\(\gamma_{MSE}\)} && \multicolumn{3}{c}{\(\gamma_{NIS}\)} && \multicolumn{3}{c}{\(\mathbb{E}(\gamma_{d})\)} && \multicolumn{3}{c}{\(\gamma_{MSE}\)} && \multicolumn{3}{c}{\(\gamma_{NIS}\)} && \multicolumn{3}{c}{\(\mathbb{E}(\gamma_{d})\)}\\

\cline{3-5} \cline{7-9} \cline{11-13} \cline{15-17} \cline{19-21} \cline{23-25}

\( \bar{\varepsilon}\) && Const & Flex & Min && Const & Flex & Min && Const & Flex & Min && Const & Flex & Min && Const & Flex & Min && Const & Flex & Min \\
\hline
\( f_1^{(e)}\)          && 0.368 & 0.358 & \textbf{0.353} && 14.42 & 14.55 & \textbf{14.18} && 0.810 & 0.883 & \textbf{0.764} && 0.560 & 0.572 & \textbf{0.554} && 27.14 & 28.12 & \textbf{26.40} && 3.359 & 3.461 & \textbf{2.679} \\
\( f_2^{(e)}\)          && \textbf{0.363} & 0.371 & 0.379 && \textbf{14.39} & 14.40 & 14.90 && \textbf{1.541} & 3.001 & 2.439 && 0.552 & 0.540 & \textbf{0.531} && \textbf{25.66} & 26.48 & 26.05 && 3.386 & 2.618 & \textbf{2.615} \\
\( f_3^{(e)}\)          && 0.353 & \textbf{0.352} & 0.362 && \textbf{14.57} & 14.63 & 15.84 && 12.89 & 14.16 & \textbf{11.47} && \textbf{0.564} & 0.590 & 0.619 && \textbf{29.40} & 29.87 & 31.17 && \textbf{1.506} & 2.195 & 1.588 \\
\( f_{1,2}^{(e)}\)      && 0.297 & 0.298 & \textbf{0.286} && 12.36 & 11.80 & \textbf{11.61} && 2.467 & 2.025 & \textbf{1.722} && \textbf{0.479} & 0.504 & 0.480 && \textbf{23.61} & 25.72 & 23.62 && 3.491 & 4.038 & \textbf{2.692} \\
\( f_{1,2}^{(e)}\)      && \textbf{0.289} & 0.290 & 0.290 && 12.09 & \textbf{11.54} & 11.81 && 6.667 & \textbf{4.098} & 5.190 && 0.529 & 0.518 & \textbf{0.495} && 26.74 & 26.19 & \textbf{24.43} && 3.535 & 3.613 & \textbf{2.172} \\
\( f_{2,3}^{(e)}\)      && \textbf{0.284} & 0.286 & 0.305 && 12.33 & \textbf{12.13} & 12.80 && \textbf{5.462} & 6.008 & 8.662 && 0.517 & 0.517 & \textbf{0.513} && 27.23 & 27.37 & \textbf{25.53} && 3.196 & 3.302 & \textbf{2.693} \\
\( f_{1,2,3}^{(e)}\)    && 0.127 & \textbf{0.126} & 0.136 && 7.041 & \textbf{6.973} & 7.554 && \textbf{3.286} & 4.630 & 6.398 && 0.369 & 0.357 & \textbf{0.353} && 20.64 & 20.78 & \textbf{20.10} && 2.912 & 3.101 & \textbf{2.423} \\
\hline
\end{tabular}
}
\end{table}

\begin{table}
\caption{Calibration performance for Problems 2 and 3 at fixed experimental variance (\(\bar{\varepsilon}=0.05\)) across different output sets. Const and Flex yield the best point-wise predictive accuracy, while Min achieves the best identifiability (\( \gamma_d\)). Bold indicates the best value per metric and output set.}
\label{exp_num_out_results_n05_tbl}
\centering

\setlength{\tabcolsep}{1.8pt}
\renewcommand{\arraystretch}{1}

{\scriptsize
\begin{tabular}{cc cccc cccc cccc cccc cccc ccc}
\hline
&& \multicolumn{11}{c}{\textbf{Problem 2}} && \multicolumn{11}{c}{\textbf{Problem 3}} \\

\cline{3-13} \cline{15-25}

&& \multicolumn{3}{c}{\(\gamma_{MSE}\)} && \multicolumn{3}{c}{\(\gamma_{NIS}\)} && \multicolumn{3}{c}{\(\mathbb{E}(\gamma_{d})\)} && \multicolumn{3}{c}{\(\gamma_{MSE}\)} && \multicolumn{3}{c}{\(\gamma_{NIS}\)} && \multicolumn{3}{c}{\(\mathbb{E}(\gamma_{d})\)}\\

\cline{3-5} \cline{7-9} \cline{11-13} \cline{15-17} \cline{19-21} \cline{23-25}

\( \bar{\varepsilon}\) && Const & Flex & Min && Const & Flex & Min && Const & Flex & Min && Const & Flex & Min && Const & Flex & Min && Const & Flex & Min \\
\hline
\( f_1^{(e)}\)          && 0.398 & \textbf{0.387} & 0.433 && 16.57 & \textbf{16.17} & 18.34 && 0.831 & \textbf{0.776} & 0.974 && \textbf{0.551} & 0.556 & 0.615 && 27.01 & \textbf{26.84} & 29.87 && 2.522 & 3.066 & \textbf{1.912} \\
\( f_2^{(e)}\)          && \textbf{0.415} & 0.420 & 0.435 && \textbf{16.57} & 16.72 & 17.74 && \textbf{2.029} & 2.055 & 2.468 && 0.618 & \textbf{0.564} & 0.646 && 29.56 & \textbf{28.01} & 32.24 && 3.273 & 3.228 & \textbf{2.226} \\
\( f_3^{(e)}\)          && \textbf{0.358 }& 0.369 & 0.428 && \textbf{14.66} & 15.17 & 19.99 && 6.156 & \textbf{5.941} & 6.926 && 0.617 & \textbf{0.591} & 0.637 && 31.14 & \textbf{29.76} & 32.70 && 2.593 & 1.884 & \textbf{1.158} \\
\( f_{1,2}^{(e)}\)      && 0.363 & \textbf{0.348} & 0.408 && 15.22 & \textbf{14.74} & 16.53 && 2.219 & \textbf{2.167} & 2.294 && 0.521 & \textbf{0.506} & 0.548 && 26.33 & \textbf{25.46} & 28.44 && \textbf{2.535} & 3.087 & 2.955 \\
\( f_{1,2}^{(e)}\)      && \textbf{0.326} & 0.349 & 0.392 && \textbf{14.70} & 15.91 & 18.21 && \textbf{2.539} & 3.403 & 5.513 && 0.546 &\textbf{ 0.543} & 0.568 && 28.47 & \textbf{27.77} & 28.92 && 2.738 & 2.866 & \textbf{1.642} \\
\( f_{2,3}^{(e)}\)      && \textbf{0.339} & 0.373 & 0.404 && \textbf{14.93} & 15.61 & 17.42 && \textbf{4.285} & 4.323 & 4.467 && 0.543 & \textbf{0.525} & 0.558 && 27.59 & \textbf{27.12} & 29.27 && 2.659 & 3.126 & \textbf{2.235} \\
\( f_{1,2,3}^{(e)}\)    && \textbf{0.183} & 0.195 & 0.272 && \textbf{10.81} & 11.00 & 14.30 && 2.456 & \textbf{2.274} & 6.558 && \textbf{0.376} & 0.393 & 0.419 && \textbf{21.31} & 22.45 & 23.05 && 2.753 & 3.080 & \textbf{1.789} \\
\hline
\end{tabular}
}
\end{table}

\begin{table}[t]
\caption{Calibration performance for Problems 2 and 3 at fixed experimental variance (\(\bar{\varepsilon}=0.25\)) across different output sets. Const and Flex yield the best point-wise predictive accuracy, while Min achieves the best identifiability (\( \gamma_d\)). Bold indicates the best value per metric and output set.}
\label{exp_num_out_results_n20_tbl}
\centering

\setlength{\tabcolsep}{1.8pt}
\renewcommand{\arraystretch}{1}

{\scriptsize
\begin{tabular}{cc cccc cccc cccc cccc cccc ccc}
\hline
&& \multicolumn{11}{c}{\textbf{Problem 2}} && \multicolumn{11}{c}{\textbf{Problem 3}} \\

\cline{3-13} \cline{15-25}

&& \multicolumn{3}{c}{\(\gamma_{MSE}\)} && \multicolumn{3}{c}{\(\gamma_{NIS}\)} && \multicolumn{3}{c}{\(\mathbb{E}(\gamma_{d})\)} && \multicolumn{3}{c}{\(\gamma_{MSE}\)} && \multicolumn{3}{c}{\(\gamma_{NIS}\)} && \multicolumn{3}{c}{\(\mathbb{E}(\gamma_{d})\)}\\

\cline{3-5} \cline{7-9} \cline{11-13} \cline{15-17} \cline{19-21} \cline{23-25}

\( \bar{\varepsilon}\) && Const & Flex & Min && Const & Flex & Min && Const & Flex & Min && Const & Flex & Min && Const & Flex & Min && Const & Flex & Min \\
\hline
\( f_1^{(e)}\)          && \textbf{0.488} & 0.493 & 0.676 && \textbf{18.99} & 19.73 & 28.71 && 1.028 & 1.287 & \textbf{0.336} && \textbf{0.563} & 0.589 & 0.794 && 28.53 & \textbf{27.23} & 40.56 && 1.819 & 1.869 & \textbf{0.212} \\
\( f_2^{(e)}\)          && 0.541 & \textbf{0.521} & 0.653 && 20.79 & \textbf{19.05} & 27.57 && 1.332 & 1.904 & \textbf{0.276} && \textbf{0.590} & 0.604 & 0.799 && 30.24 & \textbf{28.68} & 41.81 && 1.832 & 1.820 & \textbf{0.514} \\
\( f_3^{(e)}\)          && 0.456 & \textbf{0.438} & 0.631 && 19.01 & \textbf{17.76} & 26.40 && 2.785 & 2.887 & \textbf{0.382} && \textbf{0.609} & 0.611 & 0.783 && \textbf{28.84} & 29.13 & 38.64 && 1.987 & 2.059 & \textbf{0.269} \\
\( f_{1,2}^{(e)}\)      && \textbf{0.455} & 0.478 & 0.690 && \textbf{18.34} & 22.10 & 31.67 && 1.413 & 1.137 & \textbf{0.926} && 0.522 & \textbf{0.512} & 0.819 && \textbf{24.78} & 24.80 & 44.39 && 1.648 & 1.658 & \textbf{0.510} \\
\( f_{1,2}^{(e)}\)      && \textbf{0.431} & 0.446 & 0.671 && \textbf{18.58} & 19.34 & 30.26 && \textbf{1.323} & 1.537 & 1.490 && 0.594 & \textbf{0.570} & 0.818 && 27.89 & \textbf{27.53} & 44.94 && 2.193 & 2.296 & \textbf{0.483} \\
\( f_{2,3}^{(e)}\)      && \textbf{0.431} & 0.447 & 0.696 && \textbf{18.12} & 19.21 & 29.74 && 2.566 & 1.994 & \textbf{0.961} && 0.587 & \textbf{0.561} & 0.824 && 26.81 & \textbf{26.09} & 45.49 && 2.333 & 1.645 & \textbf{0.653} \\
\( f_{1,2,3}^{(e)}\)    && \textbf{0.300} & 0.310 & 0.571 && 15.74 & \textbf{14.57} & 19.35 && \textbf{0.986} & 1.573 & 2.404 && \textbf{0.425} & 0.509 & 0.671 && \textbf{21.12} & 27.29 & 39.08 && 1.744 & 2.569 & \textbf{0.585} \\
\hline
\end{tabular}
}
\end{table}

\section{Data Sources for Coin Cell Electrode Experiments}
\label{data_gen_sec}
For the validation problem in Sec.~\ref{results} we used an experimental setup that includes multi-physics-based equations and physical experiments of the electrodes of coin cell batteries. In this section of the Appendix, we will present how the data was generated.

\subsection{Physical Electrode Experiments}
\label{battery_experiments}
A positive electrode slurry was prepared using the active material \(LiFePO_4\) (LFP, Aleees), conductive additive carbon black (C-Nergy Super \(C65\), Timcal) and polyvinylidene fluoride (PVDF) binder (Solef, Solvay Specialty Polymers) in a \(90:5:5\) ratio. These were dispersed in N-methyl pyrrolidone (NMP, anhydrous \(99.5\%\), Sigma Aldrich), with a solids content of \(50\) wt\(\%\). PVDF was initially dissolved in NMP to give a 8 wt\% solution, into a portion of which the carbon black was added and mixed in a planetary mixer (Thinky ARE-250, Intertronics) for 1 min at 500 rpm and \(5\) mins at \(2000\) rpm. The LFP was then added, alongside additional NMP to achieve the desired solids content, and mixed for 1 minute at \(500\) rpm and \(10\) minutes at \(2000\) rpm with a further \(3\) minutes degassing mix at \(2200\) rpm. The electrode was then coated using a calibrated doctor blade (Elcometer \(3600/4\)) and draw down coater (Elcometer 4340), and dried on a hot plate at \(80\)\degree C, giving an initial dry thickness of \(120 \mu\)m. A rectangular electrode region of \(50 cm^2\) was cut and calendered to a thickness of \(100 \mu\)m, giving a porosity of approximately \(50\%\), using a benchtop calendering machine (MSK-HRP MR100DC, MTI).

Positive electrode discs of \(15\)mm diameter were cut and dried at \(80\)\degree C under vacuum for a minimum of \(8\) hours, alongside CR2032 coin cells parts (Hohsen), including a top, bottom, spring and \(1\) mm spacer. The coin cell gasket and a \(19\)mm diameter separator disc (Celgard 2325) were dried at \(60\)\degree C for the same period. Coin cells were assembled under an argon atmosphere with \( 100 \mu\)L  of an electrolyte consisting of 1M \(LiPF_6\) in a solvent mixture of \(30:70\) wt/wt ethylene carbonate:ethyl methyl carbonate with \(2\) wt\(\%\) vinylene carbonate, with a \(16\) mm diameter Li chip serving as the negative electrode. Coin cells were allowed to rest for \(12\) hours after assembly to allow for electrolyte wetting, and subsequently formation was carried out with \(2\) cycles of constant-current constant-voltage (CC/CV) charging and constant-current (CC) discharging at a nominal rate of \(0.05\)C, based on a presumed LFP capacity of \(155\) mAh/g. The C-rate specifies the charge or discharge current relative to the battery capacity, with 1C corresponding to a full charge or discharge in one hour. There was a \(10\) minute rest between charge and discharge, and a 1 hour rest after discharge before the next cycle. These were followed by a single cycle each at \(0.1\)C and \(0.2\)C, with the same rest times. Discharge rate testing was then carried out, with all charges carried out with CC/CV charge at \(0.1\)C, followed by variable discharge rates of \(0.1\)C, \(0.2\)C, \(0.5\)C, \(1\)C, \(2\)C and \(4\)C, each of which rates were tested \(4\) times sequentially. In all cases, the CV current limit was equal to the C-rate divided by \(10\). 

Cell testing was carried out in triplicate, and the value of the initial cell potential, potential plateau and discharge capacity were recorded for the second cycle at each C-rate. A mean for each value was calculated from the repeats, where the plateau potential for each test was taken as the mean of \(3\) evenly-spaced points along the plateau region in the discharge profile.

\subsection{Electrochemical simulations of lithium-ion batteries}
\label{electrochemical_simulations}
The simulation data sources are constructed through a finite element model that captures lithium diffusion within the LFP active material, lithium-ion transport, conduction and migration in the liquid electrolyte, and interfacial reaction kinetics between the active material and electrolyte. The electrode is treated as a homogenised porous medium, in which solid-state diffusion, electronic conduction and ionic transport are scaled by electrode porosity using Bruggeman-type relations. Discharge was simulated by applying a constant current density normal to the cathode-side current collector. At the anode-side boundary, a prescribed lithium-ion flux and zero electric potential were imposed to represent a lithium-metal foil. The cathode was initialized in a near-zero, spatially uniform lithiation state.

The model was implemented in COMSOL Multiphysics \(v6.4\). A detailed description of the modeling framework, including the governing equations, boundary conditions and parameter values, is provided in the author’s previous work on LFP modeling  \cite{Tredenick2025}. Key model input parameters were varied across 383 simulations using a Sobol sequence, including C-rate (values as outlined in Appx.~\ref{battery_experiments}), active material particle size, solid-state lithium diffusion coefficient, electrode porosity, charge-transfer reaction rate constant, electrode electronic conductivity and lithium-foil exchange current density.

\section{Discussion on Limitations and Challenges}
\label{discussion}
We discuss key limitations of the proposed framework and outline directions for future work. 

\begin{enumerate}
    \item {\textbf{Identifiably:} Identifiability is the central objective of the proposed method, yet it admits multiple interpretations. One view relates identifiability to posterior concentration (i.e., the spread of the calibration parameter distribution), while another concerns how well the posterior captures the true parameter values. In this work, we adopt the latter perspective. Although we observe that geometric regularization can lead to more concentrated posteriors and improved empirical identifiability, the resulting distributions do not always accurately recover the ground truth, as reflected by the area metric \( \gamma_d(\cdot) \). This suggests that identifiability remains fundamentally ill-posed under model discrepancy, and posterior estimates should not be interpreted as definitive ground truth. Nonetheless, the framework provides improved insight into plausible parameter regimes and supports more informed scientific interpretation.} 
    \item{\textbf{Assumption of Normality:} Our analysis relies on a local Gaussian approximation of the posterior induced by the expected Fisher information. This approximation may break down in the presence of strong non-Gaussian structure (e.g., multi-modality or heavy tails), where posterior geometry is not well captured by second-order information. In the experiments considered, the approximation appears adequate, as indicated by small deviations under the metric \(\gamma_d(\cdot)\).}
    \item{\textbf{Scalability and Stability:} The proposed model involves a potentially large number of calibration parameters and outputs, leading to computational costs that scale cubically with the number of samples, \(\mathcal{O}(n^3)\) (i.e., the computational cost associated with inversion of the covariance matrix). This limits practical applicability to problems with approximately \(n < 10^4\) samples. While scalability could be improved through techniques such as inducing point methods \cite{titsias2009}, their impact on identifiability is not yet well understood and warrants further investigation.}
    \item{\textbf{Prior Knowledge:} As a Bayesian method, performance depends on the specification of prior distributions. While weakly informative priors are often appropriate for calibration parameters, practical challenges arise in defining feasible ranges for design variables that are consistent with both simulation and experimental constraints. This highlights the importance of incorporating domain knowledge and maintaining close collaboration between modeling  and experimental experts to ensure meaningful and efficient experimental design.}
    \item{\textbf{Heteroscedastic Noise:} The current formulation assumes homoscedastic noise within experimental observations, which may be restrictive in practice. Extending the framework to heteroscedastic settings, for example, by placing a Gaussian process prior over the noise variance \cite{Binois2019} or using deep kernel learning approaches \cite{wilson2016}, could improve predictive accuracy and potentially enhance identifiability. However, these extensions introduce additional hyperparameters. Although the Fisher information exhibits an approximately block-diagonal structure (Appx.~\ref{fish_gen_sec}), the coupling between noise and calibration parameters is only weakly suppressed. As a result, the increased parameterization can degrade identifiability in practice. This trade-off between improved noise modeling and parameter identifiability remains unresolved.}
    \item{\textbf{Unique Experimental Outputs:} The proposed framework benefits from diversity in experimental designs across outputs. In many practical settings, however, multiple outputs are observed at identical input locations, reducing the effective diversity of the data. For example, in the battery calibration problem, a single electrode would be manufactured and tested on all three outputs (i.e., we would measure plateau voltage, capacity, and initial voltage for the same input \( \textbf{X}\)). While posterior geometric regularization can partially mitigate this limitation, identifiability is further improved when both input diversity and appropriate regularization are present. Designing experiments that balance these factors remains an important practical consideration.}
\end{enumerate}
These limitations primarily arise from the geometric nature of posterior identifiability under model discrepancy, for which the present works provides a structured perspective for further study.

\section{Computational Resources}
\label{computational_resources}
All calibration experiments were run on the UCD Sonic high-performance computing cluster using GPU-enabled jobs distributed across 16 nodes. Individual runs required from seconds to approximately 20 minutes of wall-clock time, depending primarily on the number of training samples. The dominant cost arises from multi-start optimization of the negative log-likelihood, which involves repeated covariance matrix inversions with \(\mathcal{O}(n^3)\) complexity. 

We used Matlab’s \texttt{fmincon} with 25 random initializations (BFGS-based updates). Problem sizes ranged from 25 to 315 training samples. The total compute across all reported experiments amounts to several GPU-hours. Memory usage remained within standard node limits.

Additional battery calibration experiments were conducted on a local workstation (Dell XPS, 64 GB RAM) using a dataset containing 1,089 observations. These runs required approximately 70 minutes per calibration. Beyond the experiments reported in the paper, additional exploratory and preliminary calibration runs were performed to evaluate alternative model configurations and hyperparameter settings. These required additional compute beyond the reported experiments but remained comparable in scale to the final experimental workload and did not substantially increase overall computational requirements.

\newpage

\end{document}